\documentclass[12pt]{article}

\usepackage{amssymb}
\usepackage{graphicx,epstopdf}
\usepackage{amsthm}
\usepackage{mathrsfs}
\usepackage{array,multirow}
\usepackage{tikz}
\usepackage{subfig}
\usepackage{caption}
\usepackage[square,numbers]{natbib}
\usepackage{multicol}
\usepackage{paralist}
\usepackage{url}
\usepackage{amsmath}
\usepackage{amsfonts}
\usepackage{color}
\usepackage{afterpage}
\usepackage{float}
\usepackage{soul}
\usepackage[boxruled, linesnumbered]{algorithm2e}
\SetKwInput{KwData}{Input}
\usepackage{hyperref}
\usepackage{verbatim}
\usepackage{array}
\usepackage{tabularx}
\usepackage{enumitem}
\usepackage{wrapfig}

\allowdisplaybreaks
\makeatletter
\newcommand{\nosemic}{\renewcommand{\@endalgocfline}{\relax}}
\newcommand{\dosemic}{\renewcommand{\@endalgocfline}{\algocf@endline}}
\let\oldnl\nl
\newcommand{\nonl}{\renewcommand{\nl}{\let\nl\oldnl}}
\makeatother

\newcommand{\MM}{\mathcal{S}}

\newcommand{\reals}{\mathbb{R}}

\newcommand{\Ctau}{C_{\tau}}

\newcommand{\EE}{\mathbb{E}}
\newcommand{\PP}{\mathbb{P}}

\newcommand{\umod}[1]{ \left\lVert #1 \right\rVert}
\newcommand{\umodnot}[1]{\| #1 \|}

\newcommand{\diag}{\mbox{diag}}

\newtheorem{remark}{\n{Remark}}[section]
\theoremstyle{definition}
\newtheorem{theorem}{\n{Theorem}}[section]

\newcommand{\vertiii}[1]{{\left\vert\kern-0.25ex\left\vert\kern-0.25ex\left\vert #1 
    \right\vert\kern-0.25ex\right\vert\kern-0.25ex\right\vert}}

\font\n=cmcsc10

\long\def\ignore#1{}

\usepackage{url} 

\newcommand{\blind}{1}

\begin{document}

\def\spacingset#1{\renewcommand{\baselinestretch}%
{#1}\small\normalsize} \spacingset{1}


\if1\blind
{
  \title{\bf Scalable community detection in massive networks via predictive assignment}
  \author{Subhankar Bhadra\hspace{.2cm}\\
    Department of Statistics, North Carolina State University\\
    Marianna Pensky\\
    Department of Mathematics, University of Central Florida\\
    Srijan Sengupta\\
    Department of Statistics, North Carolina State University
    }
  \maketitle
} \fi

\if0\blind
{
  \bigskip
  \bigskip
  \bigskip
  \begin{center}
    {\LARGE\bf Scalable community detection in massive networks via predictive assignment}
\end{center}
  \medskip
} \fi

\bigskip
\begin{abstract}
Massive network datasets are becoming increasingly common in scientific applications.
Existing community detection methods encounter significant computational challenges for such massive networks due to two reasons. First, the full network needs to be stored and analyzed on a single server, leading to high memory costs. Second, existing methods typically use matrix factorization or iterative optimization using the full network, resulting in high runtimes. We propose a strategy called \textit{predictive assignment} to enable computationally efficient community detection while ensuring statistical accuracy. The core idea is to avoid large-scale matrix computations by breaking up the task into a smaller matrix computation plus a large number of vector computations that can be carried out in parallel. 
Under the proposed method, community detection is carried out on a small subgraph to estimate the relevant model parameters. Next, each remaining node is assigned to a community based on these estimates. We prove that predictive assignment achieves strong consistency under the stochastic blockmodel and its degree-corrected version.
We also demonstrate the empirical performance of predictive assignment on simulated networks and two large real-world datasets:  DBLP (Digital Bibliography \& Library Project), a computer science bibliographical database, and the Twitch Gamers Social Network.
\end{abstract}

\noindent%
{\it Keywords:}  Spectral Clustering, Bias-adjusted Spectral Clustering, Stochastic Block Model, Degree Corrected Stochastic Block Model, Computational Efficiency,  {Scalable Inference}
\vfill

\newpage
\spacingset{1.9} 
\section{Introduction}
\label{sec:intro}

Community structure is a common feature of networks,
where the nodes in a network belong to clusters or communities that
exhibit similar behavior \citep{fortunato2010community, 
 yanchenko2021generalized}.
Numerous community detection methods have been developed and studied in the statistics literature, e.g., spectral methods \citep{jin2015fast, rohe2011spectral, sengupta2015spectral}, modularity based methods \citep{bickel2009nonparametric, zhao2012consistency}, and likelihood based methods \citep{amini2013, senguptapabm}.
These community detection methods are statistically sound, with rigorous theoretical guarantees, making them valuable tools for network analysis.
However, applying these existing methods becomes computationally challenging in many scientific fields where massive networks are becoming increasingly common, e.g., epidemic modeling \citep{venkatramanan2021forecasting}, brain networks \citep{penny2011statistical, roncal2013migraine}, online social networks \citep{guo2022safer, leskovec2012learning}, and biomedical text networks \citep{gurtcheff2003complications,komolafe2022scalable}.

How serious is this problem?  To illustrate this, we report a brief computational experiment. 
Consider an undirected network of $n$ nodes with no self-loops, represented by an adjacency matrix $A \in \{0,1\}^{n \times n}$, where $A_{i,j} \sim \text{Bernoulli}(P_{i,j})$ for $1\leq i < j \leq n$. 
Suppose the network has $K$ communities, where $K$ is known, with membership vector $c = \{c_i\}_{i=1}^n$ and membership matrix $M \in \{0,1\}^{n \times K}$, where $M_{i,j} = \mathbb{I}(c_i = j)$.
Under the stochastic block model (SBM) \citep{holland1983stochastic}, we set $P = M\Omega M^T$, where $\Omega \in \mathbb{R}^{K \times K}$ defines block interactions:
$$
\Omega_{rs} = \frac{\alpha Kh}{h+(K-1)}\, I(r=s) + \frac{\alpha K}{h+(K-1)}\, I(r \neq s), \quad r,s \in \{1, \dots, K\},
$$
with density parameter $\alpha = 0.01$ and homophily factor $h=3$. We generated balanced SBMs (each community has $n/K$ nodes) under five scenarios:  
(i) $n=20000, K=10$,  
(ii) $n=50000, K=15$,  
(iii) $n=100000, K=20$,  
(iv) $n=150000, K=20$,  
(v) $n=200000, K=20$.  
For each scenario, we generated 30 networks and performed community detection using spectral clustering and bias-adjusted spectral clustering \citep{rohe2011spectral, sussman2012consistent, lei2019}.

{\spacingset{1.4}

\begin{table}[h]
\centering

{\footnotesize   
\begin{tabular}{|c|c|ccc|ccc|}
\hline
 
   & & 
      \multicolumn{3}{|c|}{Spectral Clustering} & 
   \multicolumn{3}{|c|}{Bias-adjusted Spectral Clustering} \\
   \hline
   $n$ & $K$ & Error (\%) &  Memory (Mb.) & Runtime (min) 
   & Error (\%) &  Memory (Mb.) & Runtime (min)\\
  \hline \hline
   $20000$ & $10$ & 
   $0.0 \pm 0.0$ & $98.5$ & $0.87$ &
   $0.0 \pm 0.0$ & $8079.6$ & $2.71$ \\ 
   \hline
  $50000$ & $15$ & 
  $0.0 \pm 0.0$ & $555$ & $3.89$ &
  \multicolumn{3}{|c|}{Memory Overload} \\   
  \hline
    $100000$ & $20$ & $0.0 \pm 0.0$ & $1907$ & $12.90$ &
  \multicolumn{3}{|c|}{Memory Overload}\\ 
   \hline
  $150000$ & $20$ & $0.2 \pm 1.4$ & $4284$ & $19.92$ &
  \multicolumn{3}{|c|}{Memory Overload}\\ 
  \hline
  $200000$ & $20$ & $0.2 \pm 1.4$ & $7632$ & $28.31$ &
  \multicolumn{3}{|c|}{Memory Overload}\\ 
  \hline  
\end{tabular}
 } 
\captionof{table}{\footnotesize 
Computational cost of community detection, with units in parentheses, on a server with Intel Xeon(R) E5-4627 v3 processors.
Community detection errors are reported as mean $\pm$ standarad deviation in percentage.
The R functions \texttt{irlba} and \texttt{peakRAM} were used to implement spectral decomposition of the adjacency matrix and to compute the memory requirement, respectively.}
\label{memcomp}
\end{table}

}  

In Table \ref{memcomp}, we report the runtime (in minutes) and  memory usage (MB), in addition to the Hamming loss community detection error.
We observe that while spectral clustering and its bias-adjusted version are statistically very accurate, with errors close to zero,
the computational costs are rather high.
Spectral clustering takes 20 minutes for $n=150000$ and over 28 minutes for $n=200000$. For bias-adjusted spectral clustering, memory exceeds 8000 MB for $n=20000$ (exceeding the 8000 MB RAM of a typical laptop) and 16 GB for $n\geq50000$, causing a ``memory overload'' error on the server
since it exceeds the 16 GB memory allocation.
We would like to point out that the statistical literature on scalable inference tends to focus on runtime as the only measure of computational cost \citep{kleiner2014scalable, mukherjee2021two, sengupta2016subsampled}.
But in practice, the memory requirement of a statistical method is also a critical component of computational cost.
Also note that spectral clustering is one of the fastest community detection algorithms \citep{mukherjee2021two, wang2021fast}, especially with our fast implementation using \texttt{irlba}.
Other community detection algorithms, e.g., likelihood-based methods, are likely to fare worse.

In this paper, we introduce \textit{predictive assignment}, a new technique designed to scale up community detection.
The key idea is that if we can have reasonably accurate estimates of the model parameters, we can assign the nodes to communities individually, eliminating the need for clustering.
These estimates can be efficiently obtained via community detection on a small subgraph of the network, significantly reducing computational costs compared to community detection on the full network, in the spirit of randomized sketching \citep{woodruff2014sketching}.

Predictive assignment consists of three steps. 
In Step 1, 
we 
select a subsample of nodes from the network.
In Step 2, we implement a standard community detection algorithm
\begin{wrapfigure}{r}{0.5\textwidth}
{\spacingset{1.4}
\centering
 	\includegraphics[width= 0.9\linewidth]{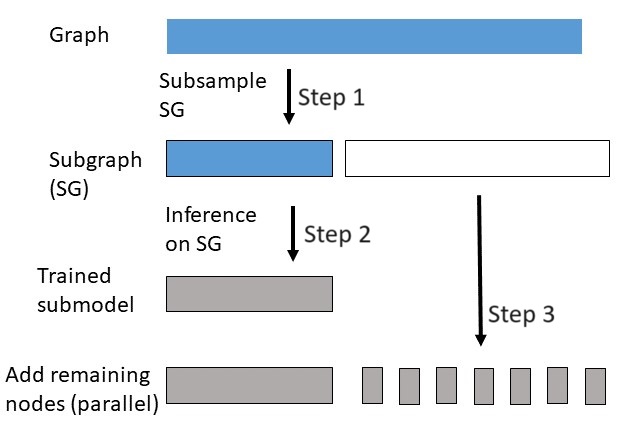}
		\caption{\footnotesize A schema of the predictive assignment algorithm. Step 1: subsample selection; Step 2: community detection from the subgraph and estimation of the structural link parameter; Step 3: assignment of the remaining nodes to communities.}
\label{predinf}
} 
\end{wrapfigure}
 (such as spectral clustering) on the subgraph formed by the subsampled nodes. When the subsample size is small compared to the full network size, this step drastically reduces runtime and memory usage. For example, if the subsample comprises $20\%$ of the nodes, an $O(n^3)$ algorithm would run approximately $125$ times faster than on the full network. 
In Step 3, we assign the remaining nodes to communities by exploiting the mathematical structure of the model. A model assumption (e.g., SBM) provides a \textit{structural link} between the community memberships of the subgraph nodes (which have already been estimated) and the community memberships of the rest of the nodes (which still need to be estimated). 
We leverage this structural link to formulate a decision rule that assigns each remaining node to its community using only vector computations. See Figure \ref{predinf} for a visual illustration.

Predictive assignment is highly versatile as it 
can accommodate any reasonably accurate community detection method in Step 2 while offering theoretical guarantees of asymptotic accuracy.
In this paper, we theoretically analyze predictive assignment under the stochastic blockmodel (defined earlier) as well as the degree corrected blockmodel (DCBM). 
In Section~\ref{sec:theo}, we prove that under certain mild assumptions, predictive assignment achieves \textit{strong} consistency in Step 3 (i.e., perfect community assignment with probability tending to 1), even when subgraph community detection in Step 2 is \textit{not} strongly consistent. Notably, strong consistency holds for any community detection method in Step 2 that meets a specific error bound, making predictive assignment highly robust. 
Since Step 2 operates on a much smaller subgraph, the overall error rate is primarily influenced by the error from Step 3.
Therefore, one could use a fast but relatively less accurate community detection method in Step 2, and still achieve high overall accuracy due to the strong consistency of predictive assignment.
In other words, the proposed technique, remarkably, can achieve higher overall accuracy than the underlying community detection method.
This phenomenon is further reinforced in our empirical results.

 The rest of the paper is organized as follows. 
 In Section \ref{sec:prior} we describe prior work on scalable community detection.
In Section \ref{sec:sbm}, we describe the methodological details of predictive assignment under the SBM and the DCBM, and in Section \ref{sec:theo}, we study its theoretical properties. 
In Section \ref{sec:sim}, we report the computational and statistical performance of predictive assignment in numerical experiments compared to standard community detection as well as existing scalable algorithms. In Section \ref{sec:data}, we illustrate the algorithm using two real-world networks:   {the Digital Bibliography \& Library Project (DBLP) database and the Twitch Gamers Social Network.} In Section \ref{dis} we conclude the paper with a discussion.
A supplementary file contains technical proofs of the theoretical results.

\subsection{Prior methodologies}
\label{sec:prior}

In related work, \citet{amini2013} developed a pseudo-likelihood approach to improve the computational efficiency of community detection, and their work was further refined by \citet{wang2021fast}.
Although these methods are highly innovative and have excellent theoretical properties,
they rely on a likelihood-based approach that is slower than spectral clustering on the full network, as demonstrated by \cite{mukherjee2021two}. 
Indeed, \citet{wang2021fast} recommend using spectral clustering on the full network as the initialization step of their algorithm.
Since our algorithm is significantly faster than spectral clustering, it is already faster than their initialization step, with subsequent steps only adding to the runtime.

Another approach to scalable community detection is distributed computation.
\citet{zhang2022distributed} proposed a distributed community detection algorithm for large networks specifically designed for block models with a \textit{grouped} community structure.
In their model assumption, the group structure overlaps with the community structure such that nodes and communities within the same group have higher link probabilities than those in different groups. While the proposed distributed algorithm in \cite{zhang2022distributed} is effective in this setting, their method is limited by this structural assumption. In contrast, our method applies to a broader class of models without requiring such constraints.

Divide-and-conquer strategies have also gained attention as a scalable alternative to direct community detection on large networks \cite{chakrabarty2023sonnet, mukherjee2021two,  wu2020distributed}. \citet{mukherjee2021two} introduced two notable algorithms: PACE (Piecewise Averaged Community Estimation) and GALE (Global Alignment of Local Estimates). 
The core idea behind both PACE and GALE is a divide and conquer strategy, where $T$ subgraphs are sampled from the given network.
Community detection is carried out on the $T$ subgraphs using some standard community detection method, and the resulting community assignments are aggregated to obtain communities for the full network.
Under PACE, this aggregation is carried out in a piecewise manner by considering each pair of nodes and averaging their estimated communities over the subgraphs where both nodes were selected. 
Under GALE, the aggregation is carried out by 
using a traversal through the subgraphs.
\citet{chakrabarty2023sonnet} proposed a divide and conquer strategy  using overlapping subgraphs, while
\citet{wu2020distributed} developed a distributed computational framework for spectral decomposition under the SBM framework.

Although these divide-and-conquer methods improve scalability, they still require matrix computations for community detection on each subgraph, leading to substantial computational overhead. In contrast, our predictive assignment approach requires matrix computations for only a single subgraph. The remaining nodes are assigned to communities individually through efficient vector-based operations, significantly reducing computational complexity.
{Morever, predictive assignment also offers stronger theoretical guarantees than the divide-and-conquer methods, as the divide-and-conquer methods can only provide convergence rates of the same order as the underlying community detection algorithm applied to the subgraphs.
For predictive assignment, we show in Section \ref{sec:theo} that the node assignment in Step 3 can yield strongly consistent community estimates even if the community detection algorithm applied on the single subgraph is weakly consistent.}
In Section \ref{sec:compare}, we provide a numerical comparison against the methods of \citet{mukherjee2021two}, highlighting the advantages of our approach in terms of both speed and accuracy.


\section{Predictive assignment}
\label{sec:sbm}

We start with some notation. Let $[n]$ denote the set $\{1, \ldots, n\}$. For any matrix $T$, we use the notation $T_{i,j}$ to denote its $(i,j)^{th}$ element and $T_{i,.}$(resp. $T_{.,i}$) to denote its $i^{th}$ row (resp. column). For index sets $\mathcal{I}, \mathcal{J}\subset [n]$, ${T_{(\mathcal{I}, \mathcal{J})}}$ denotes the $|\mathcal{I}|\times|\mathcal{J}|$ sub-matrix of $T$ containing the corresponding rows and columns. 
   Let $D$ be the diagonal matrix of node degrees, i.e., $D_{i,i} = \sum\limits_{j=1}^n A_{i,j}$, and $\Lambda=M^TM$ be the diagonal matrix of community sizes from the full network, such that $\Lambda_{k,k}$ is the number of nodes in the $k^{th}$ community.

 {\spacingset{1.4}

\begin{algorithm}[h]
    \caption{Predictive assignment algorithm under SBM and DCBM} \medskip
    \LinesNumbered
    \KwData{ Adjacency matrix $A_{n\times n}$, number of communities $K$, subgraph size $m < n$.}
    \setcounter{AlgoLine}{0}
    \begin{enumerate}
        \item Choose $\mathcal{S} \subset \{1, \ldots, n\}$ via uniform random sampling

        \item \begin{enumerate}
            \item Carry out community detection on the subgraph $A_{(\mathcal{S},\mathcal{S})}$.
     
            \item Compute the estimates $\widehat{M}_{(\MM,.)}$ and $\widehat{\mathcal{G}}_k$ for $k=1,\ldots,K$. 
            Under SBM, \\estimate $\Theta$ by
            $\widehat{\Theta}=A_{(\MM^c,\MM)}\widehat{M}_{(\MM,.)}\widehat{\Lambda}_s^{-1}$. Under DCBM, estimate $\widetilde{\Omega}$ by
            $ \widehat{\Omega} = \widehat{M}_{(\MM,.)}^TA_{(\MM,\MM)}\widehat{M}_{(\MM,.)}$.

        \end{enumerate} 

        \item Assign the remaining $(n-m)$ nodes to communities (preferably in parallel)\\
        SBM: $ 
        \widehat{c}_{i} = \underset{k = 1, \ldots, K}{\arg \min} 
    \left\lVert{a_i-\widehat{\Theta}_{.,k}} \right\rVert_2 \text{ for all } i \in\MM^c.
    $\\
    DCBM: $ \widehat{c}_i=\underset{k = 1, \ldots, K}{\arg \min}\ \left\|{\widetilde{N}}_{i,.}  - \left( \sum\limits_r \widehat{\Omega}_{k,r}\right)^{-1}\,
\left(\widehat{\Omega}_{k,1}, \ldots, \widehat{\Omega}_{k,K} \right) \right\|_2\ \text{ for all }\ i \in\MM^c.$
    \end{enumerate}
         \label{alg:quad}
     \end{algorithm}

     } 

 \subsection{Steps 1 and 2: subgraph selection and clustering}

Let $m$ ($m<n$) be the given subsample size.
In Step 1, we use some suitable sampling scheme to select a subsample of nodes $\MM \subset [n]$, where $|\MM|=m$,  and select the subgraph spanned by the nodes in $\MM$.
Let $\mathcal{G}_k = \{i\in\MM: c_i = k\}$ be the set of subgraph nodes in the $k^{th}$ community for $k=1, \ldots, K$. 
Let $\Lambda_s={M_{(\MM,.)}}^TM_{(\MM,.)}$ be the subgraph version of $\Lambda$, such that the $k^{th}$ diagonal entry of $\Lambda_s$ is $|\mathcal{G}_k|$.
We have considered uniform random sampling to select $\MM$ for our theoretical analysis in this paper.

For Step 2, any consistent community detection algorithm under the SBM and DCBM can be used for the subgraph.
We recommend the use of fast community detection methods such as spectral clustering and its variants \citep{rohe2011spectral, sussman2012consistent}.
The chosen community detection method is implemented on the subgraph adjacency matrix $A_{(\MM,\MM)}$ to obtain 
community estimates $\widehat{c}_i$ for all $i\in \MM$,
and, subsequently,  the estimates $\widehat{M}_{(\MM,.)}$ and 
$\{\widehat{\mathcal{G}}_k\}_{1 \le k \le K}$.


\subsection{Step 3: Predictive assignment of the remaining nodes}
Next, we use $\widehat{M}_{(\MM,.)}$ and 
$\{\widehat{\mathcal{G}}_k\}_{1 \le k \le K}$ from Step 2 to estimate a ``structural link'' parameter,  and estimate ${c}_i$ for  $i\in \MM^c$.

\subsubsection{Closest community approach under SBM}
Under the SBM, consider the matrix parameter
\begin{equation}
 {\Theta=P_{(\MM^c,\MM)}M_{(\MM,.)}\Lambda_s^{-1} },
\label{eq:sbmpara}
\end{equation}
and its ``plug-in'' estimator
\begin{equation}
\widehat{\Theta}=A_{(\MM^c,\MM)} \widehat{M}_{(\MM,.)} \widehat{\Lambda}_s^{-1}.
\label{eq:sbmest}
\end{equation}
Note that the estimation of  $\Theta$  via  \eqref{eq:sbmest} uses community detection results only from the subgraph.
Furthermore, this estimator can be computed efficiently since the matrix dimensions in \eqref{eq:sbmest} are much smaller than the full adjacency matrix.
Consider the $j^{th}$ node in $\MM^c$, and note that
its
connections to $\MM^c$ are given by the $j^{th}$ column vector of the matrix $A_{(\MM^c,.)}$. See Figure \ref{adj1} for a visual illustration.
Denote this column vector as $a_j$.
Then 
$$\EE(a_j)=P_{(\MM^c,.)}e_j=\Theta {M}^Te_j=\Theta_{.,c_j},$$
where $e_j$ is the $j$th column of the $n \times n$ identity matrix.
Thus, the parameter $\Theta$ has two useful properties: it governs the behavior of the non-subgraph nodes, and via \eqref{eq:sbmest} it can be estimated
using community detection results only from the subgraph.
Therefore $\Theta$ acts as the ``structural link'' between the subgraph nodes and the non-subgraph nodes.
{\spacingset{1.4}
 
 \begin{figure}[h!]
	\centering
\includegraphics[width=\linewidth]{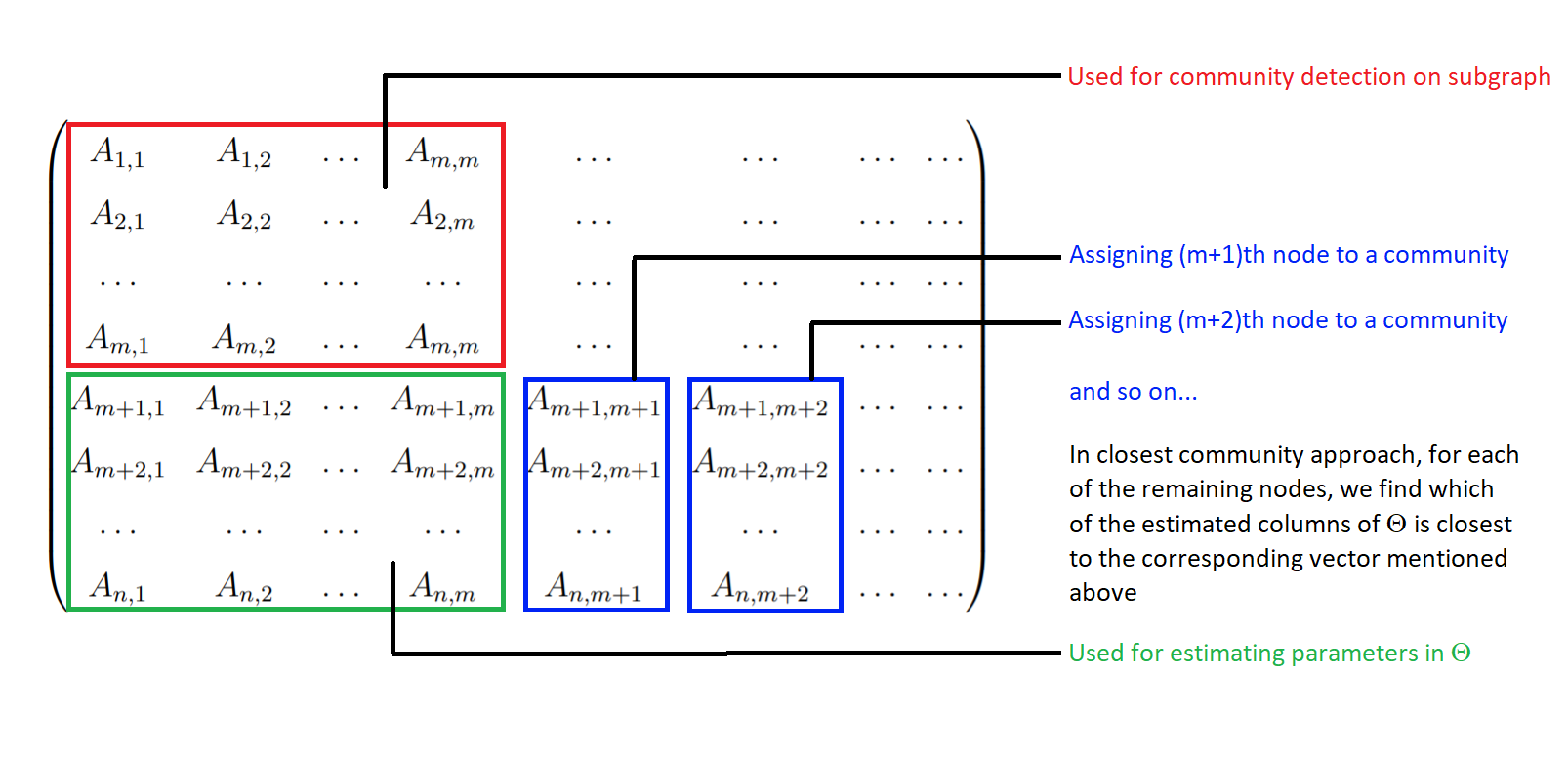}
\includegraphics[width=\linewidth]{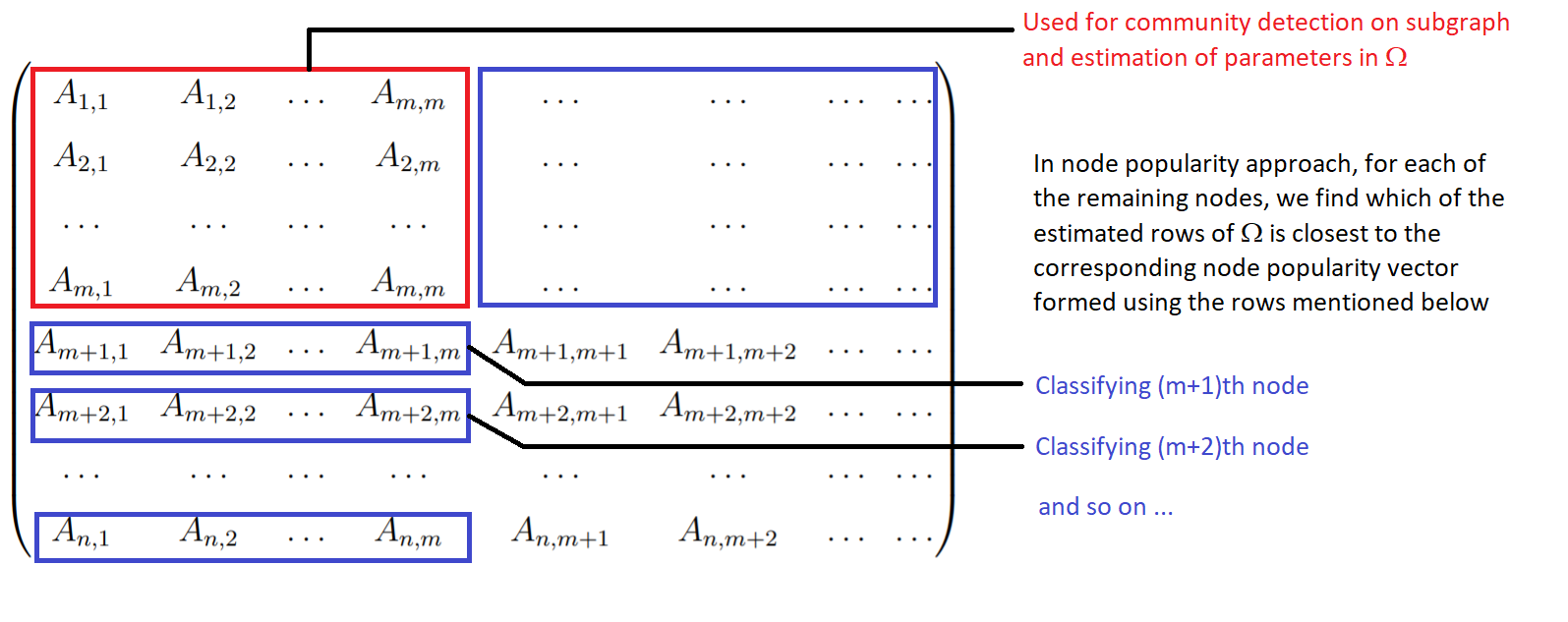}
\caption{\footnotesize Use of the different sections of the adjacency matrix under SBM (top panel) and DCBM (bottom panel).
Here we have assumed, for the sake of simplicity, that $\MM = \{1, \ldots, m\}$.
For community detection in Step 2, $A_{(\MM,\MM)}$ (red border) is utilized  under both models. 
Under the SBM, $A_{(\MM^c,\MM)}$ (green border, top panel) is used to estimate $\Theta$.
Under the DCBM, $A_{(\MM,\MM)}$ (red border, bottom panel) is used to estimate $\Omega$.
Under both models, the blue-bordered vectors are used to assign the out-of-subgraph nodes to communities one by one in Step 3.
\label{adj1}}
\end{figure}

} 

Note that $\Theta$ has $K$ unique columns, one for each community.
Consider the quantity $\left\lVert{a_j-{\Theta}_{.,k}} \right\rVert_2$, the $\ell^2$ distance between $a_j$ and the $k^{th}$ column of $\Theta$, for $k=1, \ldots, K$.
Intuitively, when $k = c_j$, $\left\lVert{a_j-{\Theta}_{.,k}} \right\rVert_2$ represents only the ``noise'', whereas when $k \neq c_j$, it represents noise plus bias. Therefore, we expect to have, in a stochastic sense,
$ \left\lVert{a_j-{\Theta}_{.,c_j}} \right\rVert_2
< \left\lVert{a_j-{\Theta}_{.,k}} \right\rVert_2$ for any $k \neq c_j$, which implies, heuristically speaking,
 $   c_j = \underset{k = 1, \ldots, K}{\arg \min} 
    \left\lVert{a_j-{\Theta}_{.,k}} \right\rVert_2.$
If we had access to $\Theta$, we could assign the $j^{th}$ node to its community by simply finding the column of ${\Theta}$ closest to  $a_j$ (hence the name \textit{closest community}).
Since we do not observe ${\Theta}$, we use its estimate from \eqref{eq:sbmest} as a proxy.
Formally, the assignment rule is 
\begin{equation}
    \widehat{c}_{j} = \underset{k = 1, \ldots, K}{\arg \min} 
    \left\lVert{a_j-\widehat{\Theta}_{.,k}} \right\rVert_2 \text{ for all } j \in\MM^c.
    \label{eq:SBM_cc}
\end{equation}
In the top panel of Figure \ref{adj1}, we provide a visual schematic of  how the different parts of the adjacency matrix are utilized in this method.
From a statistical perspective, the success of this strategy hinges on how accurately we can estimate $\Theta$.
In Section \ref{sec:theo}, we prove that the estimator \eqref{eq:sbmest} is indeed sufficiently accurate.

\subsubsection{Node popularity approach under DCBM}
The DCBM has 
 additional node-specific degree parameters $\mathbf{\theta} = (\theta_1,\ldots,\theta_n)$ such that $P=\diag(\mathbf{\theta}) M\Omega M^T\diag(\mathbf{\theta})$.
Therefore, $\EE(a_j)$ for $j\in\MM^c$ involves the degree parameter $\theta_j$, which cannot be estimated from the output of Step 2, which means that the closest community approach no longer works under the DCBM.
We propose an alternative approach for predictive assignment based on the concept of node popularity introduced by \cite{senguptapabm}.
The node popularity of the $i^{th}$ node with respect to the $k^{th}$ community is defined as the number of edges between the node and the community, i.e.,
$N_{i,k} = \sum\limits_{j=1}^n A_{i,j} \mathbb{I}(c_j=k)$. 
If $d_i = \sum\limits_{j=1}^n A_{i,j}$ is the degree of the $i$th node, then
we have 
\begin{equation}
    \frac{\EE(N_{i,k})}{\EE(d_{i})} 
    =
    \frac{\sum\limits_{j=1}^n  \theta_i\Omega_{c_i,k}\theta_j\mathbb{I}(c_j=k)}{\sum\limits_{r=1}^K\sum\limits_{j=1}^n  \theta_i\Omega_{c_i,r}\theta_j\mathbb{I}(c_j=r)}
    = 
    \frac{\Omega_{c_i,k}\sum\limits_{j=1}^n \theta_j\mathbb{I}(c_j=k)}{\sum\limits_{r=1}^K \Omega_{c_i,r} \left(\sum\limits_{j=1}^n \theta_j\mathbb{I}(c_j=r)\right)}.
\label{eq:dcbm1}
\end{equation}
The node-popularity-to-degree ratio within the subgraph is given by
\begin{equation}
    \label{eq:dcbm2}
     {\widetilde{N}}_{i,k} = 
      \sum\limits_{j \in \MM} A_{i,j} \mathbb{I}(\widehat{c}_j=k)
  \left/  \sum_{j \in \MM} A_{i,j} \right.=\left.\sum\limits_{j 
  \in \widehat{\mathcal{G}}_k} A_{i,j}\right/\sum\limits_{j \in \MM} A_{i,j},
\end{equation}
and, the subgraph analogue of the quantity on the right-hand side of \eqref{eq:dcbm1} is given by
\begin{equation}
    \frac{\Omega_{c_i,k}\sum\limits_{j\in\MM} \theta_j\mathbb{I}(c_j=k)}{\sum\limits_{r=1}^K \Omega_{c_i,r} \left(\sum\limits_{j\in\MM}\theta_j\mathbb{I}(c_j=r)\right)} = \frac{\Omega_{c_i,k}\Gamma_k}{\sum\limits_{r=1}^K \Omega_{c_i,r} \Gamma_r},
    \text{where }
    \Gamma_{k}=\sum\limits_{u\in \mathcal{G}_k} \theta_u,\ k=1,2,\ldots, K.
    \label{gammadef}
\end{equation}
Note that, the $\Gamma_k$'s defined in \eqref{gammadef} are random variables. To estimate the quantities in \eqref{gammadef} from the subgraph, observe that $\frac{\Omega_{j,k}\Gamma_k}{\sum\limits_{r=1}^K \Omega_{j,r} \Gamma_r}$ can be written as $\frac{\widetilde{\Omega}_{j,k}}{\sum\limits_{r=1}^K \widetilde{\Omega}_{j,r}}$, where $\widetilde{\Omega}$ is defined as
\begin{equation}
    \widetilde{\Omega} = M_{(\MM,.)}^TP_{(\MM,\MM)}M_{(\MM,.)},
    \quad \widetilde{\Omega}_{j,k} = \sum_{v\in\mathcal{G}_j}\sum_{u\in\mathcal{G}_k}P_{v,u} = \Gamma_j\Omega_{j,k}\Gamma_k, \ \  j,k\in[K].
    \label{eq:dcbmpar}
\end{equation}
Thus, under the DCBM, $\widetilde{\Omega}$ acts as the ``structural link'' between $\MM$ and $\MM^c$.
We can estimate $\widetilde{\Omega}$ from the output of Step 2 as follows:
\begin{equation}
    \widehat{\Omega} = \widehat{M}_{(\MM,.)}^TA_{(\MM,\MM)}\widehat{M}_{(\MM,.)}.
    \label{eq:dcbmest}
\end{equation}
Then, the community assignment rule is
\begin{equation}
    \label{eq:dcbm3}
    \widehat{c}_i=\underset{k = 1, \ldots, K}{\arg \min}\ \left\|{\widetilde{N}}_{i,.}  - \left( \sum\limits_r \widehat{\Omega}_{k,r}\right)^{-1}\,
\left(\widehat{\Omega}_{k,1}, \ldots, \widehat{\Omega}_{k,K} \right) \right\|_2\ \text{ for all }\ i \in\MM^c.
\end{equation}

The bottom panel of Figure \ref{adj1} shows how the different sections of the adjacency matrix are utilized under the node popularity approach.
The use of the submatrix in Step 2 is identical to the closest community approach.
In Step 3, the blue-bordered vectors are used one by one to assign the remaining nodes to communities.
A key difference from the closest community approach is that here we  never use the $(n-m)\times(n-m)$ sub-matrix $A_{(\MM^c,\MM^c)}$.


We conclude this section with some methodological remarks.

\begin{remark}
{\rm 
    \textbf{Algorithmic randomness:}
The random subsampling in Step 1 introduces variability in the estimated subgraph communities, which in turn affects the final community assignments.
Theorems 3.1 and 3.2 show that the impact of this algorithmic randomness is negligible as the subgraph retains the necessary properties of the full network with a high probability.
One potential strategy to further mitigate the effects of this randomness would be to implement multiple independent runs of the algorithm and then aggregate the results through majority voting. While such an extension would increase computational cost, it may still be more efficient than running full-network community detection while improving robustness. We leave a detailed exploration of this approach for future work.
    }
\end{remark}

\begin{remark}
{\rm 
\textbf{Out-of-sample extensions of graph embeddings:}
    Predictive assignment is conceptually similar to  out-of-sample extensions of graph embeddings \citep{bengio2003out, levin2021limit}. \citet{bengio2003out} introduced a framework that interprets these embeddings as eigenfunctions of data-dependent kernels, enabling the extension of learned mappings to new data without recomputing the entire eigendecomposition. Similarly, \citet{levin2021limit} developed out-of-sample extension methods for incorporating new vertices into existing graph embeddings within the Random Dot Product Graph (RDPG) model, providing theoretical guarantees.
    A key distinction, however, is that predictive assignment directly estimates community memberships rather than extending a continuous embedding. Our method exploits model-based structural relationships  to derive a decision rule for assigning the remaining nodes to communities, whereas out-of-sample graph embedding methods typically extend node positions in an embedding space, which may then be used for clustering or classification.
    }
\end{remark}

\begin{remark}
{\rm
\textbf{Semi-supervised community detection:}
Suppose there exists a set of labeled nodes $\mathcal{L}$ whose true communities $\{c_j\}_{j\in\mathcal{L}}$ are known,  while the remaining nodes in $\mathcal{U}$ have unknown community labels. 
The goal of semi-supervised community detection is to estimate the community membership of a new node $i\notin (\mathcal{L} \cup \mathcal{U})$ given its connections to $\mathcal{L} \cup \mathcal{U}$.
    \citet{jiang2023semisupervised} proposed the Anglemin+ algorithm to address this problem. The algorithm first applies a community detection method to the submatrix $A_{(\mathcal{U},\mathcal{U})}$ to estimate $\{\widehat{c}_j\}_{j\in\mathcal{U}}$. For a new node $i$, define $x\in\mathbb{R}^{2K}$ as
    $x = (x_{\mathcal{L}}, x_{\mathcal{U}})^\top$,
    where
    $$
    x_{\mathcal{L}} =  \left(\sum_{j\in \mathcal{L} \cap \mathcal{G}_1} A_{i,j},\,\ldots, \sum_{j\in \mathcal{L} \cap \mathcal{G}_K} A_{i,j}  \right),
    \text{ and }
            x_{\mathcal{U}} =  \left( \sum_{j\in \mathcal{U}\cap \widehat{\mathcal{G}}_1} A_{i,j},\,\ldots, \sum_{j\in \mathcal{U}\cap \widehat{\mathcal{G}}_K} A_{i,j} \right).
    $$
    Similarly, for each community $k \in \{1, \ldots, K\}$, define $v_k \in \mathbb{R}^{2K}$ as  
        $v_k = (v_{k,\mathcal{L}},v_{k,\mathcal{U}})^\top$,
    where
    $$
   v_{k,\mathcal{L}} = \left(
   \sum_{\substack{i\in\mathcal{L}\cap \mathcal{G}_k, \\j\in \mathcal{L}\cap \mathcal{G}_1}} A_{i,j} ,\,\ldots, \sum_{\substack{i\in\mathcal{L}\cap \mathcal{G}_k, \\j\in \mathcal{L}\cap \mathcal{G}_K}} A_{i,j}\right),
            v_{k,\mathcal{U}} = \left(
            \sum_{\substack{i\in\mathcal{L}\cap \mathcal{G}_k,\\j\in \mathcal{U}\cap \widehat{\mathcal{G}}_1}} A_{i,j} ,\,\ldots, \sum_{\substack{i\in\mathcal{L}\cap \mathcal{G}_k,\\j\in \mathcal{U}\cap \widehat{\mathcal{G}}_K}} A_{i,j}\right).
    $$
 The new node is assigned to the community  which minimizes the angle between $x$ and $v_k$.

To compare this with predictive assignment, note that predictive assignment does not assume the knowledge of a labeled set $\mathcal{L}$ whose true communities are known. Therefore, we should consider $\mathcal{L} = \emptyset$ in the context of predictive assignment.
However, since the definition of $v_k$ depends on the true communities, Anglemin+ cannot be applied when $\mathcal{L} = \emptyset$.
    
In order to construct a connection between predictive assignment and semi-supervised community detection, one could consider an extension of Anglemin+ that replaces the true community labels in the definitions of $x_{\mathcal{L}}$ and $v_{k,\mathcal{L}}$ with their \textit{estimated} versions, i.e.,
$$
\hat{x}_{\mathcal{L}} =  \left(\sum_{j\in \mathcal{L} \cap \widehat{\mathcal{G}}_1} A_{i,j},\,\ldots, \sum_{j\in \mathcal{L} \cap \widehat{\mathcal{G}}_K} A_{i,j}  \right),
\hat{v}_{k,\mathcal{L}} = \left(\sum_{i\in\mathcal{L}\cap \widehat{\mathcal{G}}_k}\sum_{j\in \mathcal{L}\cap \widehat{\mathcal{G}}_1} A_{i,j} ,\,\ldots, \sum_{i\in\mathcal{L}\cap \widehat{\mathcal{G}}_k}\sum_{j\in \mathcal{L}\cap \widehat{\mathcal{G}}_K} A_{i,j}\right),
$$
and minimizes the angle between the $K$-dimensional vectors $\hat{x}_{\mathcal{L}}$ and $\hat{v}_{k,\mathcal{L}}$, instead of the $2K$-dimensional vectors $x$ and $v_k$.
In the notation of predictive assignment, if we put $\mathcal{L}=\MM$, then $\hat{x}_{\mathcal{L}}$ becomes $\widetilde{N}_i$ (before adjustment) and $\hat{v}_{k,\mathcal{L}}$ is the $k^{th}$ row of $\widehat{\Omega}$ in \eqref{eq:dcbmest}.
While predictive assignment via node popularity minimizes the Euclidean distance between these vectors (after appropriate scaling), the extended Anglemin+ algorithm would minimize the angle between $\hat{x}_{\mathcal{L}}$ and $\hat{v}_{k,\mathcal{L}}$. 
Thus, there are two key distinctions between predictive assignment and Anglemin+.
First, predictive assignment relies entirely on estimated communities from the subgraph, whereas Anglemin+ assumes the knowledge of true community labels.
Second, the set $\MM$ in predictive assignment is chosen randomly via subsampling, while the labeled set $\mathcal{L}$ in Anglemin+ is deterministic.
}
 
\end{remark}

\begin{remark}
{\rm 
\textbf{Computational Complexity:}
Suppose that the complexity of community detection on the full network is given by $f(n^p, K)$ for some $p>1$.
Extracting the subgraph in Step 1 of predictive assignment requires $O(m^2)$ operations.
In Step 2, the complexity of subgraph-based community detection is $f(m^p, K)$.
The estimation and predictive assignment tasks have a complexity of $O(m(n-m)K)$, which dominates the Step 1 complexity of $O(m^2)$.
Therefore, the complexity of predictive assignment is $f(m^p, K) + O(m(n-m)K)$, compared to $f(n^p, K)$ for community detection on the full network.
}
\end{remark}



\section{Theoretical results}
\label{sec:theo}

 
This section describes the theoretical properties of predictive assignment under the SBM and DCBM.
We follow the definitions from Sections \ref{sec:intro} and \ref{sec:sbm}.
In particular, let $n$ be the number of nodes and $m$ be the number of subsampled nodes. 
In addition, let $n_k$ and $\mu_k$ be the size of the $k$th community in the full and 
subsampled network, respectively, and $\mu_{\min} = \min\limits_{1\leq k\leq K} \mu_k,\ \mu_{\max} = \max\limits_{1\leq k\leq K} \mu_k$.
Following the standard framework for introducing sparsity into the model \citep{senguptapabm, zhao2012consistency}, 
we assume that $\Omega = \alpha_n \Omega_0$ where $\|\Omega_0\|_\infty =1$ and  $\alpha_n$ is the sparsity parameter such that the  {expected number of edges in the network is $O(n^2 \alpha_n)$}.
For the DCBM,  following \citep{lei2015consistency}, we assume the identifiability constraint $\max\limits_{i:c_i=k}\theta_i=1$ for all $1\leq k\leq K$.
We also define $\theta_{\min} = \min\limits_{1\leq i\leq n}\theta_i,\ \theta_{\max} = \max\limits_{1\leq i\leq n}\theta_i,\ \Gamma_{\min} = \min\limits_{1\leq k\leq K}\Gamma_k$, and $\Gamma_{\max} = \max\limits_{1\leq k\leq K}\Gamma_k$ where $\Gamma_i$ is defined in \eqref{gammadef}. 
We use $\Ctau$, $C_0$,  $c_0$,  $C$, and $c$ for absolute constants independent of $m,n$ and $K$; note that $\Ctau$ can depend on $\tau$ but not on 
$m,n$ and $K$.
Here, $C_0$, $c_0$, $C$, $c$, and $\Ctau$  can take different values at different instances.

Next, we define error metrics for the different steps of predictive assignment. 
Assuming optimal label permutation,  
the \textit{average} community detection error $\Delta_{\MM}$,
 and the \textit{maximum} community-specific error $\widetilde{\Delta}_{\MM}$
 for subgraph community detection in Step 2 
are defined as
\begin{equation}
 { \Delta_{\MM} = \frac{1}{m}\sum\limits_{i\in\MM}\mathbb{I}\left({c_i\neq\widehat{c}_i}\right),
\quad \widetilde{\Delta}_{\MM}=\max\limits_{1 \le k \le K} \frac{|\mathcal{G}_k \cap \widehat{\mathcal{G}}^c_k|} {|\mathcal{G}_k|},   }
\label{eq:d1}    
\end{equation}
respectively. The   average error in the set of remaining nodes  in Step 3 of predictive assignment 
and the  overall error rate (aggregated across Steps 2 and 3) are  defined as  
\begin{equation}
\Delta_{\MM^c} = \frac{1}{n-m}\sum\limits_{i\in\MM^c}\mathbb{I}\left({c_i\neq\widehat{c}_i}\right), \quad
\quad \Delta=\frac{m\Delta_{\MM} + (n-m){\Delta}_{\MM^c}}{n},
\label{eq:d2}    
\end{equation}
respectively.
Note that since $m$ is much smaller than $n$, the overall error \eqref{eq:d2} is largely determined by ${\Delta}_{\MM^c}$.
%
Next, make the following assumptions: 

 \noindent
{\bf A1(a).}  There exists  $C_0>0$ such that 
$ (C_0\, K)^{-1} \leq \pi_k = n_k/n \leq C_0 \, K^{-1}, \  k=1,\ldots,K.$
 
 \noindent
{\bf A1(b).} Under the DCBM,
define $t_k=\sum\limits_{i=1}^n \theta_i\mathbb{I}(c_i = k)$.
Then, for some constants  $\tau >0$ and $a \in (0,1)$
\begin{equation} \label{eq:A1b}   
\min_k t_k  \geq  C_0\, n\, K^{-1}.
\end{equation}

 \noindent
{\bf A2.}  The smallest singular value of $\Omega_0$   is bounded below by a constant $\lambda > 0$.

 \noindent
{\bf A3.}
We have $m \geq  \widetilde{C}\, C_0\, K\, a^{-2}\, (\tau \log m + \log K)$, where 
 $\widetilde{C} = 4$ for the SBM and $\widetilde{C} = 20$ for the DCBM.



 \noindent
{\bf A4.} 
The sparsity parameter $ \alpha_n \geq c_0\, (\theta_{\min}\,m)^{-1}\,  K^4\,\log n$ where $c_0 > 0$ is a contant and  $\theta_{\min}=1$ in the case of the SBM.


\medskip

\noindent
Here
{\bf A1(a)} is the balanced communities assumption, which states that the community sizes are of the same order of magnitude. 
{\bf A1(b)}  controls the degree heterogeneity under the DCBM and allows variability of the node degree parameters  $\theta_i$. 
Observe that, if the $\theta_i$'s are of the same order of magnitude, {\bf A1(b)} simply follows from {\bf A1(a)}. 
%
Assumption~{\bf A2} is necessary for identifiability of communities (see, e.g., \cite{lei2015consistency}).
%
%
Assumption {\bf A3} sets  lower bounds on the subsample size, which are needed for achieving the 
required clustering accuracy in the subsample under the SBM and DCBM, respectively.
%
Assumption {\bf A4} imposes a lower bound on the sparsity of the sub-network. 
Under the SBM, we need the subsample size to satisfy $m \geq C\,\max \{ \log(m^{\tau} K), K^4 \log n \, \alpha_n^{-1}\}$. The first condition is satisfied for any reasonably large $m$ and small enough $K$, since $\log m = o(m)$ as $m \to \infty$. The second condition is only a little stronger than the well-known necessary condition $\alpha_n \geq c\,m^{-1} \log m$ for perfect community detection in an SBM.
The full network version of {\bf A4}, which requires that  
$\alpha_n\geq c_0\, n^{-1}\,  \log n$ for some $c_0 >0$, is a standard assumption in the literature, and is a kind of a necessary condition since $\alpha_n \leq c_0\, n^{-1}$ leads to impossibility of recovering communities 
\citep{lei2019, lei2015consistency}.
In {\bf A4} we have $m$ instead of $n$ since this sparsity restriction needs to be imposed on the subgraph.
Therefore, {\bf A4} seems to be close to the ``optimal'' fundamental limit under the SBM if $K$ is a constant or grows slowly with $n$.
Under the DCBM, we impose the additional condition on the degree parameters, given by
$
\theta_{\min} \geq 
c_0\, (m\, \alpha_n)^{-1}\, K^4 \log n.
$
Noting that $\theta_{\max}=1$ by the identifiability constraint for the DCBM, 
the condition states that there is a trade-off between the sparsity and the degree heterogeneity in the network.
If the network is sparse, then the degree heteregeneity in the network should be sufficiently controlled to recover communities.

We are now ready to state our theoretical results.
We first need to ensure that 
a subgraph in Step 1 inherits  analogs of the full-sample balance conditions {\bf A1(a, b)}.
The next two theorems formalize this and ensure that algorithmic randomness due to subsampling vanishes asymptotically (see Remark 2.1).
All technical proofs are in the Appendix.




\begin{theorem} \label{lem:min_max_mu_k}  
\textit{Let Assumptions {\bf A1(a)} and {\bf A3} hold. Then, 
\begin{equation}
    \PP\left\{\mu_{\min} \geq (1-a)\, (C_0 K)^{-1}\, m, \quad 
\mu_{\max} \leq (1+a)\, (K)^{-1}\,   C_0\, m  \right\}\geq 1- 2\,m ^{-\tau}.
\label{mumax_mumin}
\end{equation}}
\end{theorem}

Theorem \ref{lem:min_max_mu_k} ensures that, with high probability, 
the true community proportions of the subgraph adequately represent the true community proportions of the full network.

\begin{theorem} \label{lem:tmin_tmax}  
    \textit{Suppose that the network is generated from the DCBM as defined in Section~\ref{sec:intro}, 
and Assumptions {\bf A1(a,b)} and {\bf A3} hold. Then, for $\Gamma_k$ defined in \eqref{gammadef},  one has
    \begin{gather}
    \PP\left\{
   \Gamma_{\min} \geq (1-a)\, n^{-1}\, m\, t_k,\quad \Gamma_{\max} \leq (1+a)\, n^{-1}\, m\, t_k
    \right\}
    \geq 1- 2\, m^{-\tau}, \label{lem:tmin_tmax_eq1}\\
      \PP\left\{ \Gamma_{\min}\geq (1-a)\, (C_0\, K)^{-1}\,  m, \quad \Gamma_{\max}\leq C_0\, (1+a) \,K^{-1}\, m \right\}
      \geq 1- 2\, m^{-\tau}.   \label{lem:tmin_tmax_eq3} 
     %
    \end{gather}}
\end{theorem}

Theorem \ref{lem:tmin_tmax} ensures that, under the DCBM, 
the degree parameter-weighted community proportions of the subgraph adequately represent their  full-sample counterparts.

Next, we present three ``master" theorems that characterize the consistency of parameter estimation and the accuracy of predictive assignment under the SBM and DCBM.
These results hold independently of \textit{any} specific community detection method used in Step 2, thus demonstrating the flexibility of predictive assignment.
Suppose that $\{\widehat{c}_i:\,i\in\MM\}$ are estimated communities obtained by applying any community detection algorithm to the subgraph $A_{(\MM,\MM)}$ such that the maximum community-specific error $\widetilde{\Delta}_{\MM}$ satisfies
\begin{equation}
\PP(\widetilde{\Delta}_{\MM} \leq C_{\tau}\, \delta(n,m,K,\alpha_n)) \geq 1-  C\, m^{-\tau},
    \label{weak_cond}
\end{equation}
where $\delta(n,m,K,\alpha_n)$ may depend on $n,m,K$ and $\alpha_n$, $C_{\tau}\,\delta(n,m,K,\alpha_n)<1-\epsilon'$ for some constant $\epsilon'\in(0,1)$, and $C_{\tau}>0,\,C>0$ are constants.
In this general framework, the next three theorems provide error bounds for estimating
the link parameters $\Theta$ and $\widetilde{\Omega}$ under the SBM and the DCBM.
Theorem \ref{lem:class_weak} establishes the consistency of parameter estimation under the SBM for both weakly and strongly consistent subgraph community detection. 
Similarly, Theorem  \ref{lem:omega_est_weak}  establishes parameter estimation consistency under the DCBM for weakly and strongly consistent subgraph community detection.

\begin{theorem} \label{lem:class_weak} \textbf{(Concentration of $\widehat{\Theta}$)} 
\textit{Suppose that the network is generated from the SBM, and Assumptions {\bf A1(a), A2, A3} and {\bf A4} hold. 
If the community detection algorithm on the subgraph $A_{(\MM,\MM)}$ satisfies \eqref{weak_cond},
then 
one has 
\begin{equation} \label{eq:max_discrepancy_weak} 
\PP\left(\max_{i,\,k} |\widehat{\Theta}_{i,k}-\Theta_{i,k}| 
 \leq \Ctau\, \left(\sqrt{K m^{-1} \alpha_n\log n}
+ K\alpha_n\, \delta(n,m,K,\alpha_n) \right)  \right)  \geq 1- C\, m^{-\tau},
\end{equation}
where $\Theta$, $\widehat{\Theta}$  be defined in \eqref{eq:sbmpara} and \eqref{eq:sbmest} respectively.
Furthermore, if the community detection algorithm on the subgraph $A_{(\MM,\MM)}$ is \textit{strong consistent} with high probability, that is, $\PP(\Delta_{\MM}=0)\geq 1 - Cm^{-\tau}$, then 
\begin{equation}
\PP\left(\max_{i,\,k} |\widehat{\Theta}_{i,k}-\Theta_{i,k}| 
 \leq \Ctau\,\sqrt{ K m^{-1}\, \alpha_n\log n}
\right) \geq 1- C\, m^{-\tau}.
\end{equation}}
\end{theorem}


\medskip

\begin{theorem} (\textbf{Concentration of $\widehat{\Omega}$}) \label{lem:omega_est_weak}   
\textit{Suppose that the network is generated from the DCBM, and Assumptions {\bf A1(a,b), A2, A3} and {\bf A4} hold. 
Let $\widetilde{\Omega}, \widehat{\Omega}$  be defined in \eqref{eq:dcbmpar} and
\eqref{eq:dcbmest} respectively.
Then, 
one has  
 \begin{align}
\PP\left\{\max_k \umodnot{\widehat{\Omega}_{k,.} - \widetilde{\Omega}_{k,.}}  \leq C_{\tau} \right. 
& \left. \left(\dfrac{m^{3/2}\sqrt{\alpha_n}}{K} + \dfrac{m^2\alpha_n}{\sqrt{K}}\, \delta(n,m,K,\alpha_n)\right)(1 + K\,\delta(n,m,K,\alpha_n))\right\} \nonumber\\
&\geq 1-C\,m^{-\tau}. \label{eq:omega_gap}
\end{align}
If, in addition, clustering is strongly consistent, so that 
$\PP(\Delta_{\MM}=0)\geq 1 - Cm^{-\tau}$, then 
 \begin{equation} \label{eq:omega_gap_Del0}
\PP\left(\max_k \umodnot{\widehat{\Omega}_{k,.} - \widetilde{\Omega}_{k,.}}  \leq C_{\tau}\, m\, K^{-1/2}\, \sqrt{\log m} \right) \geq 1-C\,m^{-\tau}.
\end{equation} 
} 
\end{theorem}

Note that setting $\delta(n,m,K,\alpha_n)$ to 0 in \eqref{eq:omega_gap}, leads to  $\umodnot{\widehat{\Omega}_{k,.} - \widetilde{\Omega}_{k,.}} =O(m^{3/2}\sqrt{\alpha_n}/K)$ with probability at least $1-Cm^{-\tau}$.
The  sharper error bound in \eqref{eq:omega_gap_Del0} is obtained by more nuanced calculations.


Building on Theorems \ref{lem:class_weak} and  \ref{lem:omega_est_weak}, we now present our main result in Theorem \ref{th:step2_weak}, which establishes that predictive assignment achieves strong consistency for the nodes in ${\MM^c}$ under both the SBM and the DCBM.


\begin{theorem}\label{th:step2_weak}   
\textit{Suppose that the network is generated from the SBM or DCBM, and Assumptions {\bf A1(a,b), A2, A3} and {\bf A4} hold.
Assume that 
\begin{eqnarray}
& \lim\limits_{n \to \infty} K^3\, \delta^2(n,m,K,\alpha_n)  =0, & \mbox{for the SBM}, \label{deltapr_sbm}\\
& \lim\limits_{n \to \infty} K^3\,\delta(n,m,K,\alpha_n) =0,\quad  
& \mbox{for the DCBM}.  \label{deltapr_dcbm}
\end{eqnarray}
If the constant $c_0$ in Assumption {\bf A4} is sufficiently large
and $n\geq 2\,(1+a)\,m$ for the SBM, or $K=O(\log n)$ for the DCBM,
then for some absolute positive constant $C$, one has 
$$\PP ({\Delta}_{\MM^c}=0) \geq 1 - C\, 
m^{-\tau}.$$}
\end{theorem}

A remarkable implication of Theorem \ref{th:step2_weak} is that predictive assignment achieves strong consistency even when subgraph community detection in Step 2 is only weakly consistent, provided that $\delta(n,m,K,\alpha_n)$ satisfies \eqref{deltapr_sbm} under the SBM and the \eqref{deltapr_dcbm} under the DCBM. This highlights a key strength of the method: predictive assignment is both computationally efficient and statistically accurate. It allows the use of a fast but less precise community detection algorithm in Step 2, and even if this results in only moderate accuracy for subgraph nodes, Theorem \ref{th:step2_weak} guarantees strong consistency for the remaining nodes. Since most nodes in the full network are not part of the subgraph, overall accuracy is determined primarily by the predictive assignment step rather than subgraph community detection. Consequently, because predictive assignment is strongly consistent, the overall accuracy remains high even when subgraph community detection is only moderately accurate.

A natural question at this point is whether there is any advantage in using a strongly consistent community detection method in Step 2. 
More broadly, does employing a community detection method that achieves better accuracy than \eqref{deltapr_sbm} or \eqref{deltapr_dcbm} provide any benefits?
At first glance, the answer appears to be no, as Theorem \ref{th:step2_weak} suggests that additional accuracy is unnecessary.
However, the answer is more nuanced and hinges on the lower-bound requirement for $c_0$.
Specifically, the required magnitude of $c_0$ depends on the absolute constants such as $C_0$, $\tau$, $a$, $\lambda$, and $\epsilon'$.
    When subgraph community detection is strongly consistent, the lower bound requirement on $c_0$ is reduced compared to the weak consistency case, as shown in the proof of Theorem \ref{th:step2_weak}.
    A smaller value of $c_0$ would imply that 
    Assumption~{\bf A4} would be satisfied for smaller values of $m$, meaning a strongly consistent community detection algorithm in Step 2 can help achieve strong consistency on the full network with a lower computational cost.
Additionally, increasing $m$ allows Assumption {\bf A4} to hold for a larger $c_0$, which leads to faster convergence to perfect clustering.
Thus, increasing $m$ sharpens the rate at which strong consistency is achieved.

 Also note that condition \eqref{deltapr_dcbm} is stricter than condition \eqref{deltapr_sbm}.
 This implies that, compared to the SBM, more accurate subgraph community detection is needed under the DCBM to achieve strong consistency for predictive assignment.
 
Finally, we establish in Theorem \ref{th:step1} that strong consistency in subgraph community detection is indeed achieved by two well-known algorithms: spectral clustering under the SBM and regularized spectral clustering under the DCBM. While strong consistency for these methods is well established in the literature when applied to the full network \citep{lyzinski2014perfect, sussman2012consistent}, this theorem shows that this property extends to the case where the methods are applied to a subgraph spanning randomly subsampled nodes from the full network.

\begin{theorem}\label{th:step1}   
    \textit{Suppose that the network is generated from the SBM and Assumptions {\bf A1(a), A2, A3 } and {\bf A4} hold, and spectral clustering is applied to the subgraph; OR, 
the network is generated from the DCBM and Assumptions {\bf A1(a,b), A2, A3} and {\bf A4} hold, and regularized spectral clustering is applied to the subgraph.
    Then, for any $\tau>0$, if the constant $c_0$ in Assumption {\bf A4}  is sufficiently large, there exists an absolute positive constant $C$ such that
    \begin{equation} \label{eq:Delta1_all}
        \PP \left( m{\Delta}_{\MM} =0  \right)\geq  1- C\, m^{-\tau}.
    \end{equation}}
\end{theorem}


We conclude this section with the following remark.
\begin{remark}
{\rm
 Based on the theoretical results, we recommend setting $m$ such that $\log m \asymp \log n$, i.e., $m \asymp n^\gamma$ with $\gamma <1$. 
 Spectral clustering on the full network achieves strong consistency under the SBM when $n\alpha_n \geq c \log n$.
 In contrast, predictive assignment requires the stronger condition $m \alpha_n \geq C K^4 \log n$.
 This highlights the trade-off for scalability when using predictive assignment: achieving strong consistency requires a stricter condition.
}
\end{remark}

\section{Simulation studies}
\label{sec:sim}

We now examine the performance of predictive assignment in synthetic networks generated from the SBM and the DCBM.
We compared predictive assignment with community detection on the full network
and 
the two scalable algorithms proposed in \cite{mukherjee2021two}.

We use the following performance metrics to quantify computational cost and statistical accuracy.
For computational performance, the CPU running time and the peak RAM utilization are used to quantify
runtime and memory cost, respectively.
Note that the peak RAM utilization represents the true memory cost associated with any statistical method and it can be much larger than the size of the input dataset.
If the peak RAM value exceeds the computer's available RAM, it is impossible to execute the code.
 {The proportions of wrongly clustered/assigned nodes} $\Delta_{\MM}$, ${\Delta}_{\MM^c}$, and $\Delta$, as defined in equations \eqref{eq:d1} and \eqref{eq:d2}, are used to quantify the statistical performance for the $m$ subsampled nodes, the $(n-m)$ remaining nodes, and the  {entire set of $n$ nodes}, respectively.
 We also report $f$, the percentage of the adjacency matrix used for subgraph clustering in Step 2.
Our experiments were performed in R 4.0.2 on a state-of-the-art university high-performance research computing Linux cluster with Intel Xeon processors.
 

\subsection{Predictive assignment vs. full network under the SBM}
\label{subsec:sim_sbm}
We first compared the performance of predictive assignment to the benchmark of community detection on the full network.
We generated network data from  balanced SBMs with block probability matrix $\Omega$ such that
for $r,s\in \{1,2,\ldots,K\}$,
$$
\Omega_{rs}= (h+(K-1))^{-1}\, \alpha Kh\, I(r=s)\ \  + \ \ 
 (h+(K-1))^{-1}\, \alpha K\, I(r \neq s)
$$
where $\alpha$ is the overall expected density of the network
and $h$ is the homophily factor that determines the strength of community structure.
We set $\alpha = 0.01$, $h=3$, and considered four scenarios:
(i) $n=50000, K=15$,
(ii) $n=100000, K=20$,
(iii) $n=150000, K=20$, and
(iv) $n=200000, K=20$.
We generated $30$ random graphs under each case.

We considered two community detection methods for subgraph community detection in Step 2: spectral clustering (SC) and bias-adjusted spectral clustering (BASC).
For SC, we compute 
the $K$ orthonormal eigenvectors corresponding to the $K$ largest (in absolute value) eigenvalues of the subgraph adjacency matrix $A_{(\MM,\MM)}$, and put them in an $m\times K$ matrix. K-means clustering is applied on the matrix rows to estimate the subgraph communities \citep{rohe2011spectral, sussman2012consistent}.
BASC was proposed by \citep{lei2019}, where $K$-means clustering is carried out on 
the $K$ dominant eigenvectors of the ``bias-adjusted'' matrix ${A_{(.,\MM)}}^TA_{(.,\MM)}-D_{(\MM,\MM)}$ instead of $A_{(\MM,\MM)}$.
While BASC was proposed for multi-layer networks, we adapt it to single-layer networks in this paper, and
extend the method to rectangular (i.e., non-square) submatrices of the adjacency matrix.
The key difference between SC and BASC is that they use different portions of the adjacency matrix for subgraph community detection (see Figure \ref{fig:basc} in the Appendix for a visual illustration).
Note that $f = m^2/n^2$ for SC but  $f= \frac{2 n m - m^2}{n^2}$ for BASC, i.e., BASC uses a much larger proportion of the adjacency matrix.
We used 
$m = n^{0.85}, n^{0.9}, n^{0.95}$ for SC, 
and  
$m = n^{0.7}, n^{0.75}, n^{0.8}, n^{0.85}$ for BASC.
Following Section \ref{sec:sbm}, we used simple random sampling in Step 1 for subgraph selection and the closest community approach in Step 3 for predictive assignment.

{\spacingset{1.4}

\begin{table}[h!]
\centering
\footnotesize

 \begin{tabular}{|c|c|ccccc|}
\hline
 \multicolumn{7}{|c|}{$n=50000, K = 15$} \\
  \hline
  $\log_n m$ &
   $f$ & Mem & $\bar{\Delta}_{\MM} \pm \text{s.e.}$ & $\bar{\Delta}_{\MM^c} \pm \text{s.e.}$ & $\bar{\Delta} \pm \text{s.e.}$
   & $t$\\
   \hline
  & & \multicolumn{5}{c|}{Bias Adjusted Spectral Clustering} \\
  \hline   
   0.7 & 7.63 & 538.3 & 12.7 $\pm$ 1.9 & 9.9 $\pm$ 3.8 & 10.0 $\pm$ 3.7 & 20.61 \\ 
   0.75 & 12.93 & 595.0 &  1.0 $\pm$ 0.2 & 0.5 $\pm$ 0.2 &  0.5 $\pm$ 0.1 & 25.32 \\ 
  0.8 & 21.65 & 725.9 &  0.1 $\pm$ 0.0 & 0.1 $\pm$ 0.0 &  0.1 $\pm$ 0.0 & 37.97 \\ 
  0.85 & 35.57 & 2038.5 &  0.0 $\pm$ 0.0 & 0.0 $\pm$ 0.0 &  0.0 $\pm$ 0.0 & 70.58 \\
  \hline
  
   & & \multicolumn{5}{|c|}{Spectral Clustering} \\
  \hline  
  0.85 & 3.89 & 555.1 & 38.7 $\pm$ 3.2 &  6.5 $\pm$ 3.5 & 12.9 $\pm$ 3.4 & 194.79 \\ 
  0.9 & 11.49 & 555.1 &  2.9 $\pm$ 0.2 &  0.0 $\pm$ 0.0 &  1.0 $\pm$ 0.1 & 146.84 \\ 
  0.95 & 33.89 & 555.1 &  0.1 $\pm$ 0.0 &  0.4 $\pm$ 0.0 &  0.2 $\pm$ 0.0 & 150.46 \\ 
  \hline
  1 & 100.00 & 555.0 &  0.0 $\pm$ 0.0 &  0.0 $\pm$ 0.0 &  0.0 $\pm$ 0.0 & 233.30 \\
\hline
\end{tabular}

 \vspace{2ex}
 
\begin{tabular}{|c|c|ccccc|}
\hline
  \multicolumn{7}{|c|}{$n=100000, K = 20$} \\
  \hline
   $\log_n m$ &
   $f$ & Mem & $\bar{\Delta}_{\MM} \pm \text{s.e.}$ & $\bar{\Delta}_{\MM^c} \pm \text{s.e.}$ & $\bar{\Delta} \pm \text{s.e.}$
   & $t$\\
  \hline \hline

   & & \multicolumn{5}{c|}{Bias Adjusted Spectral Clustering} \\
  \hline 
  0.7 & 6.22 & 2175.8 & 2.9 $\pm$ 0.6 & 2.0 $\pm$ 1.0 & 2.1 $\pm$ 1.0 & 46.79 \\ 
  0.75 & 10.93 & 2212.9 & 0.0 $\pm$ 0.0 & 0.0 $\pm$ 0.0 & 0.0 $\pm$ 0.0 & 61.32 \\ 
  0.8 & 19.00 & 2556.8 & 0.0 $\pm$ 0.0 & 0.0 $\pm$ 0.0 & 0.0 $\pm$ 0.0 & 108.77 \\ 
  0.85 & 32.40 & 6634.9 & 0.0 $\pm$ 0.0 & 0.0 $\pm$ 0.0 & 0.0 $\pm$ 0.0 & 231.55\\
 \hline \hline  
  & & \multicolumn{5}{c|}{Spectral Clustering} \\
  \hline 
  0.85 & 3.16 & 1907 & 19.7 $\pm$ 1.3 & 0.0 $\pm$ 0.0 & 3.5 $\pm$ 0.2 & 328.76 \\ 
  0.9 & 10.00 & 1907 & 0.5 $\pm$ 0.0 & 0.0 $\pm$ 0.0 & 0.2 $\pm$ 0.0 & 335.14 \\ 
  0.95 & 31.62 & 1907 & 0.0 $\pm$ 0.0 & 0.0 $\pm$ 0.0 & 0.0 $\pm$ 0.0 & 437.21 \\ 
  \hline
  
  1 & 100.00 & 1907 & 0 $\pm$ 0 & 0.0 $\pm$ 0.0 & 0.0 $\pm$ 0.0 & 773.96 \\
  \hline
 \end{tabular}

\captionof{table}{\footnotesize SBM
 Case  (i)  $n=50000, K = 15$ (top panel) and 
 Case (ii) $n=100000, K = 20$ (bottom panel).
We report fraction of data used ($f$), memory cost (Mem) in Mb, error rates (mean $\pm$ standard error) in percentage, and average run-time ($t$) in seconds. 
Note that the $\log_n m = 1$ represents the full network.
}

\label{tabsbm1}
\end{table}
\begin{table}[h!]
\centering
\footnotesize

 \begin{tabular}{|c|c|ccccc|}
\hline
 \multicolumn{7}{|c|}{$n=150000, K = 20$} \\
  \hline
  $\log_n m$ &
   $f$ & Mem & $\bar{\Delta}_{\MM} \pm \text{s.e.}$ & $\bar{\Delta}_{\MM^c} \pm \text{s.e.}$ & $\bar{\Delta} \pm \text{s.e.}$
   & $t$\\
   \hline
  & & \multicolumn{5}{c|}{Bias Adjusted Spectral Clustering} \\
  \hline   
  0.7 & 5.52 & 4883.9 & 0.0 $\pm$ 0.0 & 0.0 $\pm$ 0.1 & 0.0 $\pm$ 0.1 & 78.68 \\ 
  0.75 & 9.90 & 4895.8 & 0.0 $\pm$ 0.0 & 0.0 $\pm$ 0.0 & 0.0 $\pm$ 0.0 & 113.83 \\ 
  0.8 & 17.59 & 5416.4 & 0.2 $\pm$ 1.3 & 0.2 $\pm$ 1.2 & 0.2 $\pm$ 1.2 & 207.30 \\ 
  0.85 & 30.67 & 13219.3 & 0.0 $\pm$ 0.0 & 0.0 $\pm$ 0.0 & 0.0 $\pm$ 0.0 & 483.47 \\ 
  \hline \hline
  
   & & \multicolumn{5}{|c|}{Spectral Clustering} \\
  \hline  
  0.85 & 2.80 & 4283 &  2.8 $\pm$ 0.1 &  0.0 $\pm$ 0.0 &  0.5 $\pm$ 0.0 & 354.71 \\ 
  0.9 & 9.22 & 4280 &  0.0 $\pm$ 0.0 &  0.0 $\pm$ 0.0 &  0.0 $\pm$ 0.0 & 364.09 \\ 
  0.95 & 30.37 & 4284 &  0.0 $\pm$ 0.0 &  0.0 $\pm$ 0.0 &  0.0 $\pm$ 0.0 & 629.26 \\ 
  \hline
  1 & 100.00 & 4284 &  0.2 $\pm$ 1.4 &  0.0 $\pm$ 0.0 &  0.2 $\pm$ 1.4 & 1195.43 \\
\hline
\end{tabular}

 \vspace{2ex}
 
\begin{tabular}{|c|c|ccccc|}
\hline
  \multicolumn{7}{|c|}{$n=200000, K = 20$} \\
  \hline
   $\log_n m$ &
   $f$ & Mem & $\bar{\Delta}_{\MM} \pm \text{s.e.}$ & $\bar{\Delta}_{\MM^c} \pm \text{s.e.}$ & $\bar{\Delta} \pm \text{s.e.}$
   & $t$\\
  \hline \hline

   & & \multicolumn{5}{c|}{Bias Adjusted Spectral Clustering} \\
  \hline 
  0.7 & 5.07 & 8635.2 & 0.0 $\pm$ 0.0 & 0.0 $\pm$ 0.0 & 0.0 $\pm$ 0.0 & 94.49 \\ 
  0.75 & 9.23 & 8638.5 & 0.0 $\pm$ 0.0 & 0.0 $\pm$ 0.0 & 0.0 $\pm$ 0.0 & 139.77 \\ 
  0.8 & 16.65 & 9256.9 & 0.0 $\pm$ 0.0 & 0.0 $\pm$ 0.0 & 0.0 $\pm$ 0.0 & 267.36 \\ 
 \hline 
  & & \multicolumn{5}{c|}{Spectral Clustering} \\
  \hline 
  0.85 & 2.57 & 7632 &  0.5 $\pm$ 0.0 & 0.0 $\pm$ 0.0 & 0.1 $\pm$ 0.0 & 346.41 \\ 
  0.9 & 8.71 & 7632 &  0.0 $\pm$ 0.0 & 0.0 $\pm$ 0.0 & 0.0 $\pm$ 0.0 & 464.65 \\ 
  0.95 & 29.51 & 7632 &  0.2 $\pm$ 1.3 & 0.3 $\pm$ 1.4 & 0.2 $\pm$ 1.4 & 823.02 \\ 
  \hline
  1 & 100.00 & 7632 &  0.2 $\pm$ 1.4 & 0.0 $\pm$ 0.0 & 0.2 $\pm$ 1.4 & 1698.00 \\
  \hline
  \hline
 \end{tabular}

\captionof{table}{\footnotesize SBM
 Case  (iii)  $n=150000, K = 20$ (top panel) and 
 Case (iv) $n=200000, K = 20$ (bottom).
We report fraction of data used ($f$), memory cost (Mem) in Mb, error rates (mean $\pm$ standard error) in percentage, and average run-time ($t$) in seconds. 
Note that the $\log_n m = 1$ represents the full network.
}

\label{tabsbm2}
\end{table}

}  


\noindent \textbf{Overall performance:}
     The results are reported in Tables \ref{tabsbm1} and \ref{tabsbm2}.
Note that $\log_n m = 1$ represents the baseline setting where spectral clustering is carried out on the full network, as previously reported in Table \ref{memcomp}. We observe that
     predictive assignment is 1.5 times to 18 times faster than SC on the full network.
      In most cases, predictive assignment also achieves low error rates comparable to the full network.
    Also note that the choice of $m$ affects the memory cost for BASC but not for SC.

\noindent   {\textbf{Accuracy of predictive assignment:}}
 In all cases, we observe that $\bar{\Delta}_{\MM^c} \le \bar{\Delta}_{\MM}$, meaning the predictive assignment in Step 3 is uniformly more (or equally) accurate than subgraph community detection in Step 2. In several cases, such as SC with $m=n^{0.85}$ in Table \ref{tabsbm1}, Case (ii), $\bar{\Delta}_{\MM^c}$ is \textit{much} smaller than $\bar{\Delta}_{\MM}$. Even when using a smaller $m$ that results in only a moderately accurate assignment for $i \in \MM$ (e.g., $\bar{\Delta}_{\MM} = 19.7\%$ with $m=n^{0.85}$ in Table \ref{tabsbm1}, Case (ii)), predictive assignment can still achieve perfect results ($0\%$ error) for $i \in \MM^c$.
 This is in line with our
theoretical results that predictive assignment is strongly consistent (i.e.,
$P(\Delta_{\MM^c} = 0) \rightarrow 1)$ even when subgraph community detection
is weakly consistent (i.e., ${\Delta}_{\MM} \rightarrow_P 0$). 
This highlights the key advantage of our method  --- predictive assignment is both computationally more efficient and statistically more accurate than direct community detection on the full network. 
Thus, a fast but moderately accurate community detection method in Step 2 suffices to achieve highly accurate overall results via predictive assignment.

\noindent  \textbf{SC or BASC?} 
  We next consider the choice of subgraph community detection method in Step~2. 
  BASC was proposed by \cite{lei2019} as a more accurate version of SC when applied to the full network.
  One would expect this advantage in statistical accuracy to hold for subgraph community detection as well,
   since BASC 
  uses a much higher proportion of the adjacency matrix ($f$) than SC for the same value of $m$ and $n$.
  We observe that this is indeed true; BASC is both faster and more accurate than SC for subgraph community detection. 
 However,  BASC is much more expensive than SC in terms of memory.
  Therefore, we recommend using BASC when it is feasible in terms of storage cost, and using SC otherwise.

\subsection{Comparison with existing methods}
\label{sec:compare}
We now compare the performance of our algorithm with two state-of-the-art algorithms for scalable community detection proposed in \cite{mukherjee2021two}: PACE (Piecewise Averaged Community Estimation) and GALE (Global Alignment of Local Estimates).
To make the comparison as fair as possible, we used the MATLAB code published by the authors \cite{mukherjee2021two} and the SBM model setting from their simulation study.
We built a MATLAB implementation of the predictive assignment algorithm specifically for this comparison.
Note that elsewhere we used the R implementation of our algorithm, therefore, the runtimes and storage costs reported in this subsection are different from the rest of the paper.
We consider the following model settings under the SBM:
    (i) $n=5000,K=2$, average degree $d_n=7$, and community proportions $\pi=(0.2,0.8)$ (this is the same setup as Table 5 of \cite{mukherjee2021two}) ;
    (ii) $n=10000,K=5$, average degree $d_n=100$, and balanced communities $\pi=(0.5,0.5)$; and
    (iii) $n=10000,K=8$, average degree $d_n=100$, and balanced communities $\pi=(0.5,0.5)$.
Following \cite{mukherjee2021two}, we used SC (unregularized spectral clustering) and RSC-A (regularized spectral clustering with Amini-type regularization) as the parent algorithms for PACE and GALE and for Step 2 of predictive assignment, with $m=2500$ for $n=5000$ and $m=5000$ for $n=10000$.
We used algorithmic hyperparameters recommended by \cite{mukherjee2021two} for PACE and GALE, and implemented their code in parallel in MATLAB R2019b with 18 workers. 

Table \ref{comp} reports community detection error (mean $\pm$ standard error) and average runtime from 50 networks under each model setting.
The top panel presents results from SC as the parent algorithm for PACE or GALE and as the community detection algorithm in Step 2 for predictive assignment, and the bottom panel presents results from RSC-A.
We observe that predictive assignment is much faster than both PACE and GALE, with runtime savings between 50\% and 96\%.
Predictive assignment also provides higher or similar accuracy as PACE and GALE in most cases.

{\spacingset{1.4}

\begin{table}[ht]
\centering
\footnotesize
\begin{tabular}{|l|cc|cc|cc|}
  \hline
  & \multicolumn{2}{|c|}{$n=5000,K=2$} & \multicolumn{2}{|c|}{$n=10000,K=5$}& \multicolumn{2}{|c|}{$n=10000,K=8$}\\
  \hline
 Algorithm & $\bar{\Delta}\pm$\text{s.e.}  & time & $\bar{\Delta}\pm$\text{s.e.}  & time & $\bar{\Delta}\pm$\text{s.e.}  & time \\ 
  \hline \hline
  SC+PACE & 17.06 $\pm$ 0.66 & 4.03 & 0.01 $\pm$ 0.01 & 12.29 & 2.37 $\pm$ 0.31  & 12.92\\ 
   SC+GALE & 10.50 $\pm$ 0.58 & 5.18 & 1.30 $\pm$ 0.29 & 8.51 & 74.68 $\pm$ 27.25 & 2.99\\ 
   SC+Predictive Assignment & 13.55 $\pm$ 1.97 & 0.19 & 0.03 $\pm$ 0.03 & 0.76 & 1.27 $\pm$ 0.20  & 1.00\\ 
  \hline
  \hline
     RSC-A+PACE & 17.09 $\pm$ 0.65 & 19.91 & 0.01 $\pm$ 0.01 & 24.14 & 2.11 $\pm$ 0.26 & 29.32\\ 
  RSC-A+GALE & 34.02 $\pm$ 2.85 & 19.63 & 1.26 $\pm$ 0.11 & 20.54 & 25.87 $\pm$ 19.48 & 21.05\\ 
   RSC-A+Predictive Assignment & 27.40 $\pm$ 15.52 & 3.13 & 0.02 $\pm$ 0.02 & 9.47 & 1.26 $\pm$ 0.23 & 8.98\\ 
   \hline
\end{tabular}
\captionof{table}{\footnotesize Community detection error (in percentage) and average run-times (in seconds) for PACE, GALE, and predictive assignment (our algorithm). Top panel shows results with SC and bottom panel with RSC-A as the parent algorithm, respectively.}
\label{comp}
\end{table}

} 

\subsection{Predictive assignment vs. full network under the DCBM}

\label{subsec:sim_dcbm}
Similar to Section \ref{subsec:sim_sbm}, here 
we compared predictive assignment to community detection on the full network under the DCBM.
We generated networks from the DCBM with block probability matrix $\Omega$ such that
 {
$
\Omega_{rs}= \alpha\,  I(r=s)\ \ + \ \  \alpha/h \,  I(r \neq s)
$
for $r,s\in \{1,2,\ldots,K\}$.}
As before, $\alpha$ is the sparsity parameter and $h$ is the homophily factor.
The degree parameters were generated from the Beta(1,5) distribution to ensure a positively-skewed degree distribution. 
The resultant probability matrix was scaled to make the 
networks 1\% dense in expectation.
This might make a few $P_{i,j}$'s greater than 1; while sampling edges, we simply cap such $P_{i,j}$'s to 1.
We considered two settings: 
(i) $n=100000$, $K=20$, $h=3$, and  
(ii) $n=100000$, $K=20$, $h=5$,
and generated 30 random graphs under each setting.


To implement predictive assignment, we used simple random sampling (SRS) and random walk sampling (RWS) in Step 1.
In RWS,  we first sample a node uniformly, then choose one of its neighbors at random to be included in the subgraph \citep{dasgupta2022scalable, leskovec2006sampling}.
In Step 2, we carried out community detection via regularized spectral clustering (RSC).
We compute the $K$ dominant eigenvectors of $A_{(\MM,\MM)}$ and put them in an $m\times K$ matrix, similar to SC. But unlike SC, here we first normalize the rows with respect to the respective Euclidean norms and then apply $K$-means clustering on the normalized rows to estimate the subgraph communities
\citep{lyzinski2014perfect}.
The node popularity rule \eqref{eq:dcbm3} was used in Step 3.

We used $m = n^{0.8},n^{0.85}, n^{0.9}, n^{0.95}$, and 
as before $\log_n m = 1$ represents the full network as baseline.
The results in Table \ref{tabscdcnew} are generally in line with the SBM simulation study.
We observe that
predictive assignment is much faster and requires much less memory than community detection on the full network, with little loss of accuracy.
While both SRS and RWS lead to accurate and fast community detection, RWS is generally more accurate and faster but requires more memory.
The RWS sampling method leads to denser subgraphs than SRS since higher-degree nodes are more likely to be selected,
which requires higher memory but also produces greater accuracy and lower runtime.

{\spacingset{1.4}

\begin{table}
    \centering
\footnotesize
\resizebox{\columnwidth}{!}{%
\begin{tabular}{|c|c|c|ccccc|ccccc|}
\hline
    \multicolumn{13}{|c|}{Regularized Spectral clustering}\\
    \hline 
  \hline
  & & & \multicolumn{5}{c|}{Sampling: SRS} & \multicolumn{5}{c|}{Sampling: RWS} \\
  \hline
    $h$ & 
   $\log_n m$ & $f$ & 
   Mem & $\bar{\Delta}_{\MM} \pm \text{s.e.}$ & $\bar{\Delta}_{\MM^c} \pm \text{s.e.}$ & $\bar{\Delta} \pm \text{s.e.}$ & $t$ & 
   Mem & $\bar{\Delta}_{\MM} \pm \text{s.e.}$ & $\bar{\Delta}_{\MM^c} \pm \text{s.e.}$ & $\bar{\Delta} \pm \text{s.e.}$ & $t$\\
  \hline
    3 & 0.8 & 1.00 & 877 & 75.6 $\pm$ 2.1 & 74.4 $\pm$ 2.6 & 74.6 $\pm$ 2.6 & 481.4 & 877 & 12.7 $\pm$ 0.5 & 16.8 $\pm$ 0.2 & 16.4 $\pm$ 0.2 & 120.5\\ 
  3 & 0.85 & 3.16 & 877 & 32.1 $\pm$ 1.6 & 19.1 $\pm$ 1.6 & 21.4 $\pm$ 1.6 & 318.5 & 877 &  4.7 $\pm$ 0.2 &  6.8 $\pm$ 0.1 &  6.4 $\pm$ 0.1 & 199.0\\ 
  3 & 0.9 & 10.00 & 877 & 12.1 $\pm$ 0.3 &  5.0 $\pm$ 0.1 &  7.2 $\pm$ 0.1 & 376.4 & 1288 &  1.4 $\pm$ 0.1 &  1.9 $\pm$ 0.1 &  1.8 $\pm$ 0.0 & 358.0 \\ 
  3 & 0.95 & 31.62  & 1288 &  3.1 $\pm$ 0.1 &  0.8 $\pm$ 0.0 &  2.1 $\pm$ 0.1 & 661.8 & 1781 &  0.4 $\pm$ 0.0 &  0.4 $\pm$ 0.0 &  0.4 $\pm$ 0.0 & 524.6 \\ 
  \hline
  3 & 1 & 100.00 & 2372 &  0.3 $\pm$ 0.0 &  0.0 $\pm$ 0.0 &  0.3 $\pm$ 0.0 & 927.9 & 2372 &  0.3 $\pm$ 0.0 &  0.0 $\pm$ 0.0 &  0.3 $\pm$ 0.0 & 927.9 \\ 
  \hline
  \hline
  5 & 0.8 & 1.00 & 877 & 14.1 $\pm$ 0.9 & 7.1 $\pm$ 0.6 & 7.8 $\pm$ 0.6 & 120.8 & 877 &  1.4 $\pm$ 0.1 & 2.1 $\pm$ 0.0 & 2.0 $\pm$ 0.0 & 102.65 \\ 
  5 & 0.85 & 3.16 & 877 &  4.3 $\pm$ 0.2 & 1.4 $\pm$ 0.1 & 1.9 $\pm$ 0.1 & 203.5 & 877 &  0.2 $\pm$ 0.0 & 0.3 $\pm$ 0.0 & 0.3 $\pm$ 0.0 & 143.6 \\ 
  5 & 0.9 & 10.00 & 877 &  0.6 $\pm$ 0.1 & 0.1 $\pm$ 0.0 & 0.3 $\pm$ 0.0 & 289.1 & 1288 &  0.0 $\pm$ 0.0 & 0.0 $\pm$ 0.0 & 0.0 $\pm$ 0.0 & 234.8 \\ 
  5 & 0.95 & 31.62 & 1288 &  0.0 $\pm$ 0.0 & 0.0 $\pm$ 0.0 & 0.0 $\pm$ 0.0 & 434.7 & 1781 &  0.0 $\pm$ 0.0 & 0.0 $\pm$ 0.0 & 0.0 $\pm$ 0.0 & 414.6 \\ 
   \hline
  5 & 1 & 100.00 & 2372 &  0.0 $\pm$ 0.0 & 0.0 $\pm$ 0.0 & 0.0 $\pm$ 0.0 & 696.1 & 2372 &  0.0 $\pm$ 0.0 & 0.0 $\pm$ 0.0 & 0.0 $\pm$ 0.0 & 696.1 \\ 
  \hline
\end{tabular} }
\captionof{table}{\footnotesize Results for DCBM Case (i) $n=100000$, $K=20$, $h=3$ (top panel) and  (ii) $n=100000$, $K=20$, $h=5$ (bottom panel). We report  fraction of data used ($f$), memory cost (Mem) in Mb, error rates (mean $\pm$ standard error) in percentage, and average run-time ($t$) in seconds. 
Note that $m/n=1$ represents the full network.
}
\label{tabscdcnew}
\end{table}
} 

\section{Real-data Applications: DBLP and Twitch networks}
\label{sec:data}

The DBLP 
network consists of $n=4057$ computer scientists belonging to $K=4$ communities representing research areas.
Two researchers are connected if they published at the same conference \cite{gao2009graph}.
The Twitch user network was curated from the popular streaming service \citep{sarkar2021twitch}.
An edge connects the users if they follow each other. 
 {To the extent of our knowledge, this network dataset has not been previously studied in the statistics literature.}
The dataset has 168k nodes, and the users are labeled with one of the 15 languages based on their primary language of streaming. To avoid working with heavily unbalanced data, we excluded the English-language streamers, who comprised about 122k users. 
We also removed all users with dead accounts,
users with views less than the median views for the entire network, and
users with a lifetime less than the median lifetime.
Finally, we extracted the largest connected component with $n=10983$ nodes and combined the languages into $K=5$ language groups to obtain the communities.

To implement predictive assignment on the DBLP network,
we used SC (both adjacency matrix version and Laplacian matrix version) in Step 2 to be consistent with prior SBM-based analysis of this dataset \citep{sengupta2015spectral}.
There is no prior analysis of the Twitch network in the statistics literature, and we used RSC in Step 2 given the extent of degree heterogeneity.
Thus, the two networks complement the simulation studies by providing real-world examples under the SBM and DCBM frameworks.
{For the DBLP network, we used $m =n^{0.7}, n^{0.75}, n^{0.8}$, and, for the Twitch network, we used $m =n^{0.8}, n^{0.85}, n^{0.9}$. 100 random subgraphs were generated for each value of $m$.
The results are in Table \ref{tabdata2}, where $\log_n m = 1$ represents the full network.}

For the DBLP network, SC (adjacency version) on the full network took 3.36 seconds with an error of 10.65$\%$.
Predictive assignment is approximately two times faster and achieves slightly higher accuracy.
SC (Laplacian version) on the DBLP network took 25.90 seconds with an error of 9.96$\%$.
Predictive assignment is approximately 10 times faster with comparable accuracy.
For the Twitch network, RSC on the full network took about 21 seconds with an error of 36.81$\%$ for the adjacency version of RSC and an error of 21.14$\%$ for the Laplacian version of RSC. 
Predictive assignment has similar accuracy that varies with $m$, while being 3 to 6 times faster. 

These two real-world examples attest that the proposed algorithm achieves accuracy at par with community detection on the full network, but requires runtimes that are only a small fraction of the runtime needed for community detection on the full network.

{\spacingset{1.4}

\begin{table}
\centering
 \small
\scalebox{0.9}{\begin{tabular}{|c||cc|cc||c|cc|cc|}
\hline
  \multicolumn{5}{|c||}{DBLP} & \multicolumn{5}{c|}{Twitch}\\
  \hline
  &  \multicolumn{2}{c|}{SC} & \multicolumn{2}{c||}{SC-laplacian} & & \multicolumn{2}{c|}{RSC} & \multicolumn{2}{c|}{RSC-laplacian} \\
  \hline
    $\log_n m$ & $\bar{\Delta}\pm\text{s.e.}$ & time &  $\bar{\Delta}\pm\text{s.e.}$ & time & $\log_n m$ & 
   $\bar{\Delta}\pm\text{s.e.}$ & time & 
   $\bar{\Delta}\pm\text{s.e.}$ & time \\
  \hline
   0.7 & 10.42 $\pm$ 1.40 & 1.61 & 10.41 $\pm$ 0.29 & 1.68 & 0.8 & 36.28 $\pm$ 5.59 & 3.36 & 28.52 $\pm$ 3.78 & 3.42 \\
   
   0.75 & 10.15 $\pm$ 0.26 & 1.70 & 10.38 $\pm$ 0.26 & 2.01 & 0.85 & 33.61 $\pm$ 5.77 & 4.81 & 24.99 $\pm$ 2.36 & 4.65 \\
   
   0.8 & 10.13 $\pm$ 0.24 & 1.86 & 10.35 $\pm$ 0.20 & 2.39 & 0.9 & 34.74 $\pm$ 4.35 & 6.89 & 22.11 $\pm$ 1.72 & 6.95 \\
   \hline
  
   1 & 10.65 & 3.36 & 9.96 & 22.90 & 1 & 36.81 & 21.25 & 21.14 & 21.22 \\
   \hline
\end{tabular}  }
\captionof{table}{\footnotesize Community detection errors (mean $\pm$ standard error) in percentages and average run-times in seconds for predictive assignment algorithm for different choices of $\frac{m}{n}$ for the DBLP four-area network and the Twitch network.
Note that the standard errors in this table represent the randomness arising from the subsampling in Step 1, \textit{conditional} on the observed network.
This is different from the tables in the simulation study where the standard errors represent the randomness from two sources: the data generation process and the subsampling step.
We used RWS in Step 1 for the Twitch network since the results from Section \ref{subsec:sim_dcbm} show that it leads to faster and more accurate community detection.
}
\label{tabdata2}
\end{table}   
} 

\section{Discussion}
\label{dis}
We propose the predictive assignment algorithm for scalable community detection in massive networks.
The theoretical results ascertain the statistical guarantees of the proposed method while 
the numerical experiments
demonstrate that it can substantially reduce computation costs (both runtime and memory) while producing accurate results. 
In particular, the node assignment in Step 3 has strong consistency, leading to highly accurate overall results even when the subgraph community detection in Step 2 has some errors.

We believe that the key idea of predictive assignment,
which is to replace a large-scale matrix computation with a smaller matrix computation plus a large number of vector computations,
can be used in other network inference problems beyond community detection, such as model fitting and two-sample testing.
This will be an important avenue for future research.
Future directions of research also include extending the predictive assignment method to weighted networks, heterogeneous networks, and multilayer networks.

\spacingset{1.6}
\bibliography{ref}
\bibliographystyle{apalike}



\newpage
\setcounter{page}{1}

  \setcounter{section}{0} 
  \setcounter{equation}{0}
  \renewcommand{\theequation}{A\arabic{equation}}
  \renewcommand{\thesection}{A\arabic{section}}

\begin{center}
		{\Large \bf Supplementary material for ``Scalable community detection in massive networks via predictive assignment'': Technical Proofs}
	\end{center}


\section{Proof of Theorem \ref{lem:min_max_mu_k}}  

We know that $\mu_k\sim {\rm Hypergeometric}(m, n_k, n)$ for all $k=1,2,\ldots K$. Let   $\pi_k=\frac{n_k}{n},k=1,\ldots,K$ 
and $\pi_0=\min\limits_k \pi_k$.   Corollary 1 of \citet{greene2017} yields that for $t>0$,
$\sigma_k^2=\pi_k(1-\pi_k)$, 
one has
 \begin{equation*}
    \PP( \mu_k\geq m\pi_k +mt)
    \leq \exp\left[-\frac{mt^2/2}{\sigma_k^2(1-\frac {m - 1}{n-1})+\frac{t}{3}}\right]\leq \exp\left[-\frac{mt^2/2}{\pi_k+\frac{t}{3}}\right],
\end{equation*}
 and similarly,
\begin{equation*}
    \PP( \mu_k\leq m\pi_k -mt)
    \leq \exp\left[-\frac{mt^2/2}{\sigma_k^2(1-\frac{m - 1}{n-1})+\frac{t}{3}}\right]\leq \exp\left[-\frac{mt^2/2}{\pi_k+\frac{t}{3}}\right].
\end{equation*}
%
Then, for $0<a<1$, since, under Assumption \textbf{A1(a)},  $\pi_k\geq 1/C_0K$, one has
\begin{equation*}
    \begin{split}
    \PP\left(\min_k \mu_k\geq m\min_k\pi_k (1-a) \right)
    &= \PP\left(
    \bigcap\limits_{k=1}^K \left\{ \mu_k\geq m\min_k\pi_k(1-a) \right\}\right) \\
     & \geq \PP\left(
    \bigcap\limits_{k=1}^K \left\{ \mu_k\geq m\pi_k(1-a) \right\}\right)
    \geq 1-\sum_{k=1}^{K}\exp\left[-\frac{m\pi_k^2 a^2/2}{\pi_k+\frac{\pi_k a }{3}}\right]\\
    &\geq 1-\sum_{k=1}^{K}\exp\left[-\frac{m\pi_k a^2}{4}\right] 
    \geq 1-K\exp\left[-\frac{m a^2}{4C_0K}\right].
    \end{split}
\end{equation*}
Similarly, one has
\begin{equation*}
    \PP\left(\max_k \mu_k\leq m\max_k\pi_k (1+ a ) \right)
    \geq 1-K\exp\left[-\frac{m a^2}{4C_0K}\right].
\end{equation*}
Therefore, by Assumption \textbf{A1(a)},
\begin{equation*}
    \begin{split}
    &\ \PP\left(\min_k \mu_k\geq \frac{(1- a )m}{C_0K}, \max_k \mu_k\leq \frac{(1+ a )mC_0}{K}\right)\\[.1cm]
    \geq &\ \PP\left(\min_k \mu_k\geq m\min_k\pi_k (1- a ), \max_k \mu_k\leq m\max_k\pi_k (1+ a )\right)\\[.1cm]
    \geq &\ 1 - K\exp\left[-\frac{m a^2}{4C_0K}\right] - K\exp\left[-\frac{m a^2}{4C_0K}\right] 
    \geq   1 - 2K\exp\left[-\frac{m a^2}{4C_0K}\right]
\end{split}
\end{equation*}
Due to the Assumption {\bf A3}, one has 
$K\exp\left[-\frac{m a^2}{C_0K}\right]\leq \frac{1}{m^{\tau}}$, which yields
\begin{equation}
    \PP\left(\mu_{\min}\geq \frac{(1-a)m}{C_0K}, \mu_{\max}\leq \frac{(1+a)m C_0}{K}\right)\geq 1-\frac{2}{m^{\tau}}. \label{mu0mu1}
\end{equation}


\section{Proof of Theorem \ref{lem:tmin_tmax}}  
From Eq. 2.12 of \citet{greene2017}, we have the following result.

Consider a population containing $n$ elements, $\{q_1,q_2,\ldots,q_n\},\ q_i\in \reals$. Let $1\leq i\leq m\leq n$ and let $X_i$ be the $i^{th}$ draw without replacement from this population. Let $S_m=\sum_{i=1}^m X_i$. Then,
\begin{equation}
    \PP(|S_m - m\,\bar{q}_n|>m\,t) \leq 2 \exp\left[-\frac{m\,t^2/2}{\sigma_q^2\left(1-\dfrac{m - 1}{n-1}\right)+8\,\vertiii{q}\,t}\right].
    \label{isak}
\end{equation}
where $\bar{q}_n=\dfrac{1}{n}\sum_{i=1}^n q_i,\sigma_q^2=\dfrac{1}{n}\sum_{i=1}^n(q_i - \bar{q}_n)^2,$ and $\vertiii{q} = \max_i|q_i - \bar{q}_n|$.

We will use \eqref{isak} to prove the result \eqref{lem:tmin_tmax_eq1}.
Consider the $k^{th}$ community, $k=1,\ldots,K$. Define $q_i=\theta_i\mathbb{I}(c_i=k)$. Then, due to    $\theta_i\leq 1$, and  $\vertiii{q}\leq 1$,
\begin{gather*}
    S_m=\Gamma_k,\ n\,\bar{q}_n = \sum\limits_{i=1}^n\theta_i\,\mathbb{I}(c_i=k)=t_k,\\
    \sigma_q^2 = \frac{1}{n}\sum_{i=1}^n(\theta_i\,\mathbb{I}(c_i=k) - t_k/n)^2\leq \frac{1}{n}\sum_{i=1}^n\theta_i^2\, \mathbb{I}(c_i=k)\leq \frac{t_k}{n}.
\end{gather*}

Now, using $t = t_k\, a/n$, obtain
 \begin{align*}
    &\PP( |\Gamma_k-m\,t_k/n|\leq m\,t_k\, a/n)
    \geq 1 - 2\exp\left[-\frac{m\,t_k^2\, a^2/2n^2}{2t_k(1-\frac{m - 1}{n-1})/n+8\,t_k\, a/n}\right]\\
    \geq& 1 - 2\exp\left[-m\,t_k\, a^2/20\,n\right]
    \geq 1 - 2\exp\left[-\frac{m\,a^2}{20\,C_0 K}\right],
\end{align*}
since $t_k\geq n(C_0\, K)^{-1}$ from Assumption {\bf A1(b)}. Therefore,
\begin{equation}
    \PP\left(\Gamma_{\min}\geq \frac{mt_k(1- a)}{n}, \Gamma_{\max}\leq \frac{mt_k(1+ a)}{n}\right)\geq  1 - 2K\exp\left[-\frac{m a^2}{20\, C_0\, K}\right]\geq 1-\frac{2}{m^{\tau}},
\end{equation}
due to Assumption {\bf A3(b)}. Hence, \eqref{lem:tmin_tmax_eq1} is proved.

To prove \eqref{lem:tmin_tmax_eq3}, note that, due to $t_k \leq n_k$, $k = 1, ...,K$, from Assumptions {\bf A1(a)} and {\bf A1(b)} one has
\begin{align*}
    &\Gamma_k\in\left(\frac{mt_k(1- a)}{n}, \frac{mt_k(1+ a)}{n}\right)
    \implies \Gamma_k\in\left(\frac{mt_k(1- a)}{n}, \frac{mn_k(1+ a)}{n}\right)\\
    \implies& \Gamma_k\in\left(\frac{m(1- a)}{C_0\, K}, \frac{C_0(1+ a)\, m}{K}\right).
\end{align*}
Therefore, \eqref{lem:tmin_tmax_eq1} implies that
\begin{equation}
    \PP\left(\Gamma_{\min}\geq \frac{m(1- a)}{C_0\, K}, \Gamma_{\max}\leq \frac{mC_0(1+ a)}{K}\right)\geq  1 -\frac{2}{m^{\tau}}, 
    \label{eq:gmingmax2}
\end{equation}
and \eqref{lem:tmin_tmax_eq3} holds.

\section{Proof of Theorem \ref{lem:class_weak}}



For any $1 \le k \le K$, one has
\begin{equation}
    |{\widehat{\mathcal{G}}_k}|
    \,\geq\, 
    |\mathcal{G}_k \cap\widehat{\mathcal{G}}_k| 
    \,=\,
    |\mathcal{G}_k| - |\mathcal{G}_k \cap\widehat{\mathcal{G}}^c_k| 
    \,\geq\,
    |\mathcal{G}_k|(1-\widetilde{\Delta}_{\MM})
    \,\geq\,
    \mu_{\min}(1-\widetilde{\Delta}_{\MM}).
    \label{gk1}
\end{equation}
Also, $|\mathcal{G}_k \cap\widehat{\mathcal{G}}_k^c|
    \le |\mathcal{G}_k|\,\widetilde{\Delta}_{\MM}$
 and,
 \begin{equation}
    {|\mathcal{G}_k^c \cap\widehat{\mathcal{G}}_k| } 
    \,=\,
    {\sum_{l \neq k} |\mathcal{G}_l \cap\widehat{\mathcal{G}}_k|}
        \,\leq\,
    {\sum_{l \neq k} \widetilde{\Delta}_{\MM}\,|\mathcal{G}_l| } \,=\,(m - |\mathcal{G}_k|)\widetilde{\Delta}_{\MM} \,\leq\, m\widetilde{\Delta}_{\MM}.
    \label{gk2}
\end{equation}

Note that
\begin{align}   \label{tbreak_new_weak}
  \widehat{\Theta}_{i,k}-\Theta_{i,k} = \frac{1}{|\widehat{\mathcal{G}}_k|} \sum_{j\in\widehat{\mathcal{G}}_k} (A_{(\MM^c,.)})_{i,j}\, -&\, 
  \Theta_{i, k}= \delta_{1,i,k}+\delta_{2,i,k}, \\ 
\delta_{1,i,k}  = \frac{1}{|\widehat{\mathcal{G}}_k|} \sum_{j\in \widehat{\mathcal{G}}_k} ((A_{(\MM^c,.)})_{i,j}-(P_{(\MM^c,.)})_{i,j}) ,&\quad \delta_{2,i,k} = 
\frac{1}{|\widehat{\mathcal{G}}_k|} \sum_{j\in \widehat{\mathcal{G}}_k} (P_{(\MM^c,.)})_{i,j}-
\Theta_{i, k}. 
\nonumber
\end{align}

Given $A_{(\MM,\MM)}$, $|\widehat{\mathcal{G}}_k|$ is fixed, and $\delta_{1,i,k}$ is a function of independent Bernoulli random variables $\{(A_{(\MM^c,.)})_{i,j}: j\in \widehat{\mathcal{G}}_k\}$.
Therefore, using Bernstein's inequality, for any $t>0$, derive
 \begin{equation*}
          \PP\left(\left.\left|\sum_{j\in \widehat{\mathcal{G}}_k} ((A_{(\MM^c,.)})_{i,j}-(P_{(\MM^c,.)})_{i,j})\right|
        \leq t \, \right| \, A_{(\MM,\MM)}\right) 
         \geq     1- 2\exp\left(-\frac{t^2/2}{|\widehat{\mathcal{G}}_k|\, \alpha_n + t/3 }\right).
 \end{equation*}
 Choosing
 $$t=2\sqrt{|\widehat{\mathcal{G}}_k|\,\alpha_n \log(nKm^{\tau})} + (4/3)\log(nKm^{\tau})$$
 yields that with probability at least $1- 2\,(nKm^\tau)^{-1}$,
 \begin{equation*}
          \left|\sum_{j\in \widehat{\mathcal{G}}_k} ((A_{(\MM^c,.)})_{i,j}-(P_{(\MM^c,.)})_{i,j})\right|
        \leq 2\sqrt{|\widehat{\mathcal{G}}_k|\,\alpha_n \log(nKm^{\tau})} + (4/3)\,\log(nKm^{\tau}),
 \end{equation*}
 which implies, with probability at least $1- 2\,(nKm^\tau)^{-1}$,
 \begin{equation}
     \begin{split}
    |\delta_{1,i,k}| \leq&\ 2\sqrt{\frac{\alpha_n \log(nKm^{\tau})}{|\widehat{\mathcal{G}}_k|}} + \frac{4}{3}\,\frac{\log(nKm^{\tau})}{|\widehat{\mathcal{G}}_k|}\\
    \leq&\ 2\sqrt{\frac{\alpha_n \log(nKm^{\tau})}{\mu_{\min}(1-\widetilde{\Delta}_{\MM})}} + \frac{4}{3}\,\frac{\log(nKm^{\tau})}{\mu_{\min}(1-\widetilde{\Delta}_{\MM})},
\end{split}
\label{delta1_new_weak}
\end{equation}
invoking \eqref{gk1} in the final step above.

Next, consider the expression $\delta_{2,i,k}$.
\begin{align*}
    |\delta_{2,i,k}| =&\, \left|
\frac{1}{|\widehat{\mathcal{G}}_k|} \sum_{j\in \widehat{\mathcal{G}}_k} (P_{(\MM^c,.)})_{i,j}- 
\Theta_{i, k}\right| \,=\, \left|
\frac{1}{|\widehat{\mathcal{G}}_k|} \sum_{j\in \widehat{\mathcal{G}}_k} ((P_{(\MM^c,.)})_{i,j}-
\Theta_{i, k})\right|\\
=&\, \left|
\frac{1}{|\widehat{\mathcal{G}}_k|} \sum_{j\in \widehat{\mathcal{G}}_k\cap\, \mathcal{G}_k^c} (\Theta_{i,c_j} - \Theta_{i,k})\right| \,\leq\, 
\frac{|\widehat{\mathcal{G}}_k\cap \mathcal{G}_k^c|}{|\widehat{\mathcal{G}}_k|}\,\alpha_n.
\end{align*}
Incorporating \eqref{gk1} and \eqref{gk2}, obtain
\begin{equation}
    |\delta_{2,i,k}| \leq 
\frac{m\,\widetilde{\Delta}_{\MM}\,\alpha_n}{\mu_{\min}(1-\widetilde{\Delta}_{\MM})}.
\label{delta2_new_weak}
\end{equation}
Combining \eqref{tbreak_new_weak}, \eqref{delta1_new_weak} and \eqref{delta2_new_weak}, obtain that
with probability at least $1- 2\,(nKm^\tau)^{-1}$,
\begin{equation*}
    |\widehat{\Theta}_{i,k}-\Theta_{i,k}| \leq 2\sqrt{\frac{\alpha_n \log(nKm^{\tau})}{\mu_{\min}(1-\widetilde{\Delta}_{\MM})}} + \frac{4}{3}\,\frac{\log(nKm^{\tau})}{\mu_{\min}(1-\widetilde{\Delta}_{\MM})}
+ \frac{m\,\widetilde{\Delta}_{\MM}\,\alpha_n}{\mu_{\min}(1-\widetilde{\Delta}_{\MM})}.
\end{equation*}
Therefore, with probability at least $1-2\,m^{-\tau}$,
\begin{equation}
    \max_{1\leq i\leq (n-m)}\, \max_{1\leq k\leq K} |\widehat{\Theta}_{i,k}-\Theta_{i,k}| \leq 2\sqrt{\frac{\alpha_n \log(nKm^{\tau})}{\mu_{\min}(1-\widetilde{\Delta}_{\MM})}} + \frac{4}{3}\,\frac{\log(nKm^{\tau})}{\mu_{\min}(1-\widetilde{\Delta}_{\MM})}
+ \frac{m\,\widetilde{\Delta}_{\MM}\, \alpha_n}{\mu_{\min}(1-\widetilde{\Delta}_{\MM})}.
\label{thetafinal_pt1_weak}
\end{equation}

We have assumed that 
\begin{equation}
    \PP(\widetilde{\Delta}_{\MM}\leq C_{\tau}\,\delta(n,m,K,\alpha_n)) \geq 1- \frac{C}{m^\tau},
    \label{thetafinal_pt2_weak}
\end{equation}
where $C_{\tau}\,\delta(n,m,K,\alpha_n)<1-\epsilon'$ for some $\epsilon'\in(0,1)$. Also, from Theorem \ref{lem:min_max_mu_k}, one has 
\begin{equation}
    \PP(\mu_{\min}\geq (1-a)m/(C_0\, K)) \geq 1- \frac{2}{m^\tau}.
    \label{thetafinal_pt3_weak}
\end{equation}
Plugging \eqref{thetafinal_pt2_weak}, \eqref{thetafinal_pt3_weak} into \eqref{thetafinal_pt1_weak}, one has that with probability at least $1-O(m^{-\tau})$,
\begin{align*}
    \max_{1\leq i\leq (n-m)}\, \max_{1\leq k\leq K} |\widehat{\Theta}_{i,k}-\Theta_{i,k}| \ \leq&\  2\,\sqrt{\frac{C_0\,K\,\alpha_n \log(nKm^{\tau})}{m\,(1-a)\,\epsilon'}} 
    +\,\frac{4}{3}\,\frac{C_0\,K \log(nKm^{\tau})}{m\,(1-a)\,\epsilon'}\\
    \ +&\ \frac{C_0\,C_{\tau}\,K\, \delta(n,m,K,\alpha_n)\, \alpha_n}{(1-a)\,\epsilon'}.
\end{align*}
Note that, on the right hand side above, the first term dominates the second term due to Assumption {\bf A4}, which yields \eqref{eq:max_discrepancy_weak}.

\section{Proof of Theorem \ref{lem:omega_est_weak}}
Let $e_k$ be the $k^{th}$ column of the $K\times K$ identity matrix. Then, 
$$
\widehat{\Omega}_{k,.} - \widetilde{\Omega}_{k,.} = e_k^\top (\widehat{\Omega} - \widetilde{\Omega}).
$$
Derive
\begin{equation}
    \begin{split}
    &\ \widehat{\Omega} - \widetilde{\Omega} = \widehat{M}_{(\MM,.)}^\top  A_{(\MM,\MM)} \widehat{M}_{(\MM,.)} - M_{(\MM,.)}^\top  P_{(\MM,\MM)} M_{(\MM,.)}\\
    =& \widehat{M}_{(\MM,.)}^\top  (A_{(\MM,\MM)}-P_{(\MM,\MM)}) \widehat{M}_{(\MM,.)} + (\widehat{M}_{(\MM,.)}^\top  P_{(\MM,\MM)} \widehat{M}_{(\MM,.)} - M_{(\MM,.)}^\top  P_{(\MM,\MM)} M_{(\MM,.)})\\
    =&\ \Phi_1 + \Phi_2,\ \text{say}.
    \end{split}
    \label{omegagap_break_weak}
\end{equation}

\textbf{Bounding $\umodnot{e_k^\top\Phi_1}$:}
\begin{align*}
    \umodnot{e_k^\top\Phi_1} =&\ \umodnot{e_k^\top\widehat{M}_{(\MM,.)}^\top  (A_{(\MM,\MM)}-P_{(\MM,\MM)})\, \widehat{M}_{(\MM,.)}} \\
    \leq&\ \umodnot{\widehat{M}_{(\MM,.)}\, e_k } \umodnot{A_{(\MM,\MM)}-P_{(\MM,\MM)}} \umodnot{\widehat{M}_{(\MM,.)}}\\
    =&\ \sqrt{|\widehat{\mathcal{G}}_k|}\,\umodnot{A_{(\MM,\MM)}-P_{(\MM,\MM)}} \sqrt{\max\limits_{1\leq k\leq K} |\widehat{\mathcal{G}}_k|}.
\end{align*}
The last step above follows from the definition of $\widehat{M}_{(\MM,.)}$.

Recall equations \eqref{gk1} and \eqref{gk2} from the proof of Theorem \ref{lem:class_weak}. We have
$$|\widehat{\mathcal{G}}_k| \leq |\mathcal{G}_k| + m\widetilde{\Delta}_{\MM}\leq \mu_{\max} + m\widetilde{\Delta}_{\MM},$$
Hence,
\begin{equation}
    \umodnot{e_k^\top\Phi_1}\leq (\mu_{\max} + m\widetilde{\Delta}_{\MM})\,\umodnot{A_{(\MM,\MM)}-P_{(\MM,\MM)}}.
    \label{phi1_weak}
\end{equation}

\textbf{Bounding $\umodnot{e_k^\top\Phi_2}$:}
Note that, the $(k,l)$-th element of $\Phi_2 = (\widehat{M}_{(\MM,.)}^\top  P_{(\MM,\MM)} \widehat{M}_{(\MM,.)} - M_{(\MM,.)}^\top  P_{(\MM,\MM)} M_{(\MM,.)})$ is
$$\sum\limits_{v\in \widehat{\mathcal{G}}_k}\sum\limits_{u\in \widehat{\mathcal{G}}_l}P_{v,u} - \sum\limits_{v\in \mathcal{G}_k}\sum\limits_{u\in \mathcal{G}_k}P_{v,u}.$$
\begin{align*}
    &\sum\limits_{v\in \widehat{\mathcal{G}}_k}\sum\limits_{u\in \widehat{\mathcal{G}}_l}P_{v,u}=\sum\limits_{v\in \widehat{\mathcal{G}}_k}\left(\sum\limits_{u\in \mathcal{G}_l}P_{v,u} + \sum\limits_{u\in \widehat{\mathcal{G}}_l\cap \mathcal{G}_l^c}P_{v,u} - \sum\limits_{u\in \mathcal{G}_l\cap\widehat{\mathcal{G}}_l^c}P_{v,u}\right)\\
    =&\sum\limits_{v\in \widehat{\mathcal{G}}_k}\sum\limits_{u\in \mathcal{G}_l}P_{v,u} + \sum\limits_{v\in \widehat{\mathcal{G}}_k}\left(\sum\limits_{u\in \widehat{\mathcal{G}}_l\cap \mathcal{G}_l^c}P_{v,u} - \sum\limits_{u\in \mathcal{G}_l\cap\widehat{\mathcal{G}}_l^c}P_{v,u}\right)\\
    =&\sum\limits_{u\in \mathcal{G}_l}\sum\limits_{v\in \widehat{\mathcal{G}}_k}P_{v,u} + \sum\limits_{v\in \widehat{\mathcal{G}}_k}\left(\sum\limits_{u\in \widehat{\mathcal{G}}_l\cap \mathcal{G}_l^c}P_{v,u} - \sum\limits_{u\in \mathcal{G}_l\cap\widehat{\mathcal{G}}_l^c}P_{v,u}\right)\\
    =&\sum\limits_{u\in \mathcal{G}_l}\left(\sum\limits_{v\in \mathcal{G}_k}P_{v,u} + \sum\limits_{v\in \widehat{\mathcal{G}}_k\cap \mathcal{G}_k^c}P_{v,u} - \sum\limits_{v\in \mathcal{G}_k\cap\widehat{\mathcal{G}}_k^c}P_{v,u}\right) + \sum\limits_{v\in \widehat{\mathcal{G}}_k}\left(\sum\limits_{u\in \widehat{\mathcal{G}}_l\cap \mathcal{G}_l^c}P_{v,u} - \sum\limits_{u\in \mathcal{G}_l\cap\widehat{\mathcal{G}}_l^c}P_{v,u}\right)\\
    =&\sum\limits_{u\in \mathcal{G}_l}\sum\limits_{v\in \mathcal{G}_k}P_{v,u} + \sum\limits_{u\in \mathcal{G}_l}\left(\sum\limits_{v\in \widehat{\mathcal{G}}_k\cap \mathcal{G}_k^c}P_{v,u} - \sum\limits_{v\in \mathcal{G}_k\cap\widehat{\mathcal{G}}_k^c}P_{v,u}\right) + \sum\limits_{v\in \widehat{\mathcal{G}}_k}\left(\sum\limits_{u\in \widehat{\mathcal{G}}_l\cap \mathcal{G}_l^c}P_{v,u} - \sum\limits_{u\in \mathcal{G}_l\cap\widehat{\mathcal{G}}_l^c}P_{v,u}\right).
\end{align*}

This implies,
\begin{align*}
    &\,\left|\sum\limits_{v\in \widehat{\mathcal{G}}_k}\sum\limits_{u\in \widehat{\mathcal{G}}_l}P_{v,u} - \sum\limits_{v\in \mathcal{G}_k}\sum\limits_{u\in \mathcal{G}_l}P_{v,u}\right|\\
    =&\,\left|\sum\limits_{u\in \mathcal{G}_l}\left(\sum\limits_{v\in \widehat{\mathcal{G}}_k\cap\, \mathcal{G}_k^c}P_{v,u} - \sum\limits_{v\in \mathcal{G}_k\cap\, \widehat{\mathcal{G}}_k^c}P_{v,u}\right) + \sum\limits_{v\in \widehat{\mathcal{G}}_k}\left(\sum\limits_{u\in \widehat{\mathcal{G}}_l\cap\, \mathcal{G}_l^c}P_{v,u} - \sum\limits_{u\in \mathcal{G}_l\cap\,\widehat{\mathcal{G}}_l^c}P_{v,u}\right)\right|\\
    \leq&\, \alpha_n\, |\mathcal{G}_l|\,(|\widehat{\mathcal{G}}_k\cap \mathcal{G}_k^c|+|\mathcal{G}_k\cap\widehat{\mathcal{G}}_k^c|) + \alpha_n\,|\widehat{\mathcal{G}}_k|\, (|\widehat{\mathcal{G}}_l\cap \mathcal{G}_l^c| + |\mathcal{G}_l\cap\widehat{\mathcal{G}}_l^c|),\ \text{since }P_{v,u}\leq\alpha_n\\
    \leq&\, \alpha_n\, |\mathcal{G}_l|\,(|\widehat{\mathcal{G}}_k\cap \mathcal{G}_k^c|+|\mathcal{G}_k\cap\widehat{\mathcal{G}}_k^c|) + \alpha_n\,(|\mathcal{G}_k| + |\widehat{\mathcal{G}}_k\cap \mathcal{G}_k^c|)\, (|\widehat{\mathcal{G}}_l\cap \mathcal{G}_l^c| + |\mathcal{G}_l\cap\widehat{\mathcal{G}}_l^c|)\\
    \leq&\, \alpha_n\, \mu_{\max}\,m\widetilde{\Delta}_{\MM} + \alpha_n\, (\mu_{\max}+m\widetilde{\Delta}_{\MM})\, m \widetilde{\Delta}_{\MM},\\
    &\text{since }|\mathcal{G}_l\cap\widehat{\mathcal{G}}_l^c| \leq |\mathcal{G}_l|\widetilde{\Delta}_{\MM},\ \text{and }|\widehat{\mathcal{G}}_l\cap\mathcal{G}_l^c|\leq (m-|\mathcal{G}_l|)\widetilde{\Delta}_{\MM}\text{ from }\eqref{gk2}\\
    =&\, 2\,\alpha_n\, \mu_{\max}\, m \widetilde{\Delta}_{\MM} + \alpha_n\, m^2 \widetilde{\Delta}_{\MM}^2.
\end{align*}

Hence
\begin{equation}
    \umodnot{e_k^\top \Phi_2} = \sqrt{\sum_{l=1}^K\left(\sum\limits_{v\in \widehat{\mathcal{G}}_k}\sum\limits_{u\in \widehat{\mathcal{G}}_l}P_{v,u} - \sum\limits_{v\in \mathcal{G}_k}\sum\limits_{u\in \mathcal{G}_l}P_{v,u}\right)^2} \leq 2\,\sqrt{K}\,\alpha_n\, \mu_{\max}\, m \widetilde{\Delta}_{\MM} + \sqrt{K}\,\alpha_n\, m^2\widetilde{\Delta}_{\MM}^2.
    \label{phi2_weak}
\end{equation}

\textbf{Conclusion:} Combining \eqref{omegagap_break_weak}, \eqref{phi1_weak} and \eqref{phi2_weak}, obtain
\begin{equation}
    \umodnot{e_k^\top (\widehat{\Omega} - \widetilde{\Omega})} \leq (\mu_{\max} + m\widetilde{\Delta}_{\MM})\,\umodnot{A_{(\MM,\MM)}-P_{(\MM,\MM)}} + 2\,\sqrt{K}\,\alpha_n\, \mu_{\max}\, m \widetilde{\Delta}_{\MM} + \sqrt{K}\,\alpha_n\, m^2\widetilde{\Delta}_{\MM}^2.
    \label{omegagap}
\end{equation}
We can construct probability bound for the quantity on the right-hand side above as follows.

First, we have assumed that 
\begin{equation*}
    \PP(\widetilde{\Delta}_{\MM}\leq C_{\tau}\,\delta(n,m,K,\alpha_n)) \geq 1- \frac{C}{m^\tau}.
    \label{omegagap_pt1}
\end{equation*}
Next, from Theorem \ref{lem:min_max_mu_k}, one has 
\begin{equation*}
    \PP(\mu_{\max}\leq (1+a)\, m C_0/K) \geq 1- \frac{2}{m^\tau}.
    \label{omegagap_pt2}
\end{equation*}
Finally, Theorem 5.2 of \cite{lei2015consistency} yields that under Assumption {\bf A4},
\begin{equation*}
    \PP(\umodnot{A_{(\MM,\MM)}-P_{(\MM,\MM)}}\leq C_{\tau}\sqrt{m\alpha_n})\geq 1-O(m^{-\tau}).
    \label{omegagap_pt3}
\end{equation*}
Combining together, one has that with probability at least $1-O(m^{-\tau})$,
\begin{align*}
    \umodnot{e_k^\top (\widehat{\Omega} - \widetilde{\Omega})}
    \leq&\ C_{\tau}\,\bigg[\bigg(\dfrac{m}{K} + m\,\delta(n,m,K,\alpha_n)\bigg)\,\sqrt{m\alpha_n} \ + \\
    &\quad\quad\ \ \dfrac{m^2\alpha_n}{\sqrt{K}}\, \delta(n,m,K,\alpha_n) + \sqrt{K}\, m^2\,\alpha_n\,\delta(n,m,K,\alpha_n)^2\bigg]\\[.2cm]
    =&\ C_{\tau}\left(\dfrac{m^{3/2}\sqrt{\alpha_n}}{K} + \dfrac{m^2\alpha_n}{\sqrt{K}}\, \delta(n,m,K,\alpha_n)\right)(1 + K\,\delta(n,m,K,\alpha_n)),
\end{align*}
for some positive constant ${C}_{\tau}$ depending on $\tau$.
\\

\medskip
Now, we prove the second part of Theorem~\ref{lem:omega_est_weak}.    
We expand $(\widehat{\Omega} - \widetilde{\Omega})$ as
\begin{equation}
    \begin{split}
    &\ \widehat{\Omega} - \widetilde{\Omega} = \widehat{M}_{(\MM,.)}^\top  A_{(\MM,\MM)} \widehat{M}_{(\MM,.)} - M_{(\MM,.)}^\top  P_{(\MM,\MM)} M_{(\MM,.)}\\
    =&\ (\widehat{M}_{(\MM,.)}^\top  A_{(\MM,\MM)} \widehat{M}_{(\MM,.)} - M_{(\MM,.)}^\top  A_{(\MM,\MM)} M_{(\MM,.)}) + M_{(\MM,.)}^\top  (A_{(\MM,\MM)}-P_{(\MM,\MM)}) M_{(\MM,.)}.
    \end{split}
    \label{omegagap_break}
\end{equation}
Consider the set $\mathscr{E} = \left\{\omega: \Delta_{\MM} =0   \right\}$ in the sample space  with $\PP(\mathscr{E}) \geq 1 - C\, m^{-\tau}$.
Note that  for $\omega \in \mathscr{E}$ one has correct community assignments for all nodes in $\MM$, so that, for any $1 \le k \le K$, one has
$\widehat{\mathcal{G}}_k =  {\mathcal{G}}_k$ and $\widehat{M}_{(\MM,.)} = {M}_{(\MM,.)}$, which implies that
\begin{equation}
    \umodnot{e_k^\top(\widehat{M}_{(\MM,.)}^\top  A_{(\MM,\MM)} \widehat{M}_{(\MM,.)} - M_{(\MM,.)}^\top  A_{(\MM,\MM)} M_{(\MM,.)})} = 0.
    \label{omegagap_break.pt1}
\end{equation}

Now, concerning the second term on the right-hand side of \eqref{omegagap_break},
\begin{equation}
    \begin{split} 
    &\ \umodnot{e_k^\top ({M}_{(\MM,.)}^\top  (A_{(\MM,\MM)}-P_{(\MM,\MM)}) {M}_{(\MM,.)})} 
    \ =\ \sqrt{\sum_{l=1}^K \left(\sum_{i\in\mathcal{G}_k} \sum_{j\in\mathcal{G}_l} (A_{i,j} - P_{i,j})\right)^2}. 
\end{split}
\label{omegagap_break.pt2}
\end{equation}   
Using Hoeffding's inequality, we have
\begin{equation*}
    \PP\left(\left|\sum_{i\in\mathcal{G}_k} \sum_{j\in\mathcal{G}_l} (A_{i,j} - P_{i,j}) \right|\leq \sqrt{\frac{|\mathcal{G}_k|\, |\mathcal{G}_l|\log(K\,m^{\tau})}{2}}\right)\geq 1-2\,K^{-1}m^{-\tau}.
\end{equation*}
Applying Theorem \ref{lem:min_max_mu_k} and taking the union bound, we obtain
\begin{equation*}
    \PP\left(\bigcap_{k=1}^K\left\{\umodnot{e_k^\top ({M}_{(\MM,.)}^\top  (A_{(\MM,\MM)}-P_{(\MM,\MM)}) {M}_{(\MM,.)})} \leq \frac{mC_0\,(1+a)}{\sqrt{K}}\,\sqrt{\frac{\log(K m^{\tau})}{2}}\right\} \right)\geq 1-O(m^{-\tau}).
\end{equation*}

Combining terms, we conclude
\begin{equation}
    \PP\left(\max_{1\leq k\leq K}\umodnot{\widehat{\Omega}_{k,.} - \widetilde{\Omega}_{k,.}}  \leq \frac{mC_0\,(1+a)}{\sqrt{K}}\,\sqrt{\frac{\log(K m^{\tau})}{2}}\right) \geq 1-O(m^{-\tau}),
\end{equation}
which completes the proof. 


\section{Proof of Theorem \ref{th:step2_weak}}   

\noindent
\textbf{Case 1: Closest community approach under the SBM}

Recall that $a_j$ and $p_j$ are the $j^{th}$ columns of $A_{(\MM^c,.)}$ and $P_{(\MM^c,.)}$ respectively. Then,  
for the closest community approach, ${\Delta}_{\MM^c}$ can be written as,
$$
    {\Delta}_{\MM^c} = \frac{1}{n-m}\sum\limits_{j \in \MM^c}  \mathbb{I}\left(\min_{l \neq c_j}\, \left\{ \umodnot{a_j-\widehat{\Theta}_{.,l}}^2 \leq \umodnot{a_j-\widehat{\Theta}_{.,c_j}}^2 
    \right\}\right)
$$
Consider a node $j\in \MM^c$.
Then $p_j=\Theta_{.,c_j}$, 
and the node is correctly clustered if
\begin{equation}
   \umodnot{a_j-\widehat{\Theta}_{.,c_j}}^2 \leq \min_{l \neq k}\, \umodnot{a_j-\widehat{\Theta}_{.,l}}^2 
    \label{event_weak}
\end{equation}
Fix some $j \in\MM^c$ and $l \neq c_j$. Then,
\begin{equation*}
    \begin{split}
     \umodnot{a_j-\widehat{\Theta}_{.,l}}^2 & -\umodnot{a_j-\widehat{\Theta}_{.,c_j}}^2 \geq 
    \umodnot{{\Theta}_{.,c_j}-\widehat{\Theta}_{.,l}}^2 -\umodnot{{\Theta}_{.,c_j}-\widehat{\Theta}_{.,c_j}}^2-2\langle a_j-{\Theta}_{.,c_j},\widehat{\Theta}_{.,l}-\widehat{\Theta}_{.,c_j}\rangle\\
    \geq\,
    & \dfrac{1}{2}\,\umod{{\Theta}_{.,c_j}-{\Theta}_{.,l}}^2 -2\,\umodnot{{\Theta}_{.,l}-\widehat{\Theta}_{.,l}}^2 -\umodnot{{\Theta}_{.,c_j}-\widehat{\Theta}_{.,c_j}}^2-2\langle a_j-{\Theta}_{.,c_j},\widehat{\Theta}_{.,l}-\widehat{\Theta}_{.,c_j}\rangle,
    \end{split}
    \end{equation*}
since $\|a+b\|^2 \leq 1/2\, \|a\|^2 + 2\, \|b\|^2$ for any $a$ and $b$. 
Denoting $\epsilon_j=a_j-{\Theta}_{.,c_j}=a_j-p_j$, obtain 
  \begin{equation}
  \begin{split}
  &\ \umodnot{a_j-\widehat{\Theta}_{.,l}}^2 -\umodnot{a_j-\widehat{\Theta}_{.,c_j}}^2\\
  \geq&\ \frac{1}{2}\,\min_{k\neq l} \umodnot{{\Theta}_{.,k}-{\Theta}_{.,l}}^2-3\,n\max_{i,\, k}(\widehat{\Theta}_{i,k}-\Theta_{i,k})^2-2\,\langle \epsilon_j,\widehat{\Theta}_{.,l}-\widehat{\Theta}_{.,c_j}\rangle.
     \end{split}
 \label{thetabreak2_weak}
\end{equation}

\textbf{Bounding $\min\limits_{k\neq l}\umod{\Theta_{\cdot,k} - \Theta_{\cdot,l}}^2$:}
We establish a lower bound for $\min\limits_{k\neq l}\umod{\Theta_{\cdot,k} - \Theta_{\cdot,l}}^2$. Let $\Lambda_{ns}={M_{(\MM^c,.)}}^\top M_{(\MM^c,.)}$ be the diagonal matrix of the true community sizes for the sub-graph $A_{(\MM^c,\MM^c)}$. Now,
\begin{equation*}
    \begin{split}
    \umod{\Theta_{\cdot,k} - \Theta_{\cdot,l}}^2  
     =&\,  \alpha_n^2\left((\Omega_0)_{\cdot,k} - {(\Omega_0})_{\cdot,l}\right)^\top\Lambda_{ns}\left((\Omega_0)_{\cdot,k} - ({\Omega_0})_{\cdot,l}\right)\\
    \geq&\,  \alpha_n^2\lambda_{\min}(\Lambda_{ns})\umod{(\Omega_0)_{\cdot,k} - (\Omega_0)_{\cdot,l}}^2 
    \,\geq\,   \alpha_n^2\, \lambda_{\min}(\Lambda_{ns})\, \lambda_{\min}^2(\Omega_0)\\
    %
 \geq&\, \alpha_n^2\lambda^2\ \underset{k}{\min}(n_k-\mu_k),\ \text{by Assumption {\bf A2}}. 
    \end{split}
\end{equation*}
%
Now, due to $n\geq 2\,(1+a)\,m$, 
$$ 
\underset{k}{\min}(n_k-\mu_k)\geq \underset{k}{\min}\left(\frac{n\,C_0}{K}-\frac{(1+a)\,m\,C_0}{K}\right) \geq \frac{n\,C_0}{2K},
$$
with probability at least $1-2\, m^{-\tau}$, by Assumption \textbf{A1(a)} and Theorem \ref{lem:min_max_mu_k}. 
Hence, for some positive constant $C_1$, one has
 %
%
\begin{equation}
    \PP\left(\min\limits_{k\neq l}\umod{\Theta_{\cdot,k} - \Theta_{\cdot,l}}^2\geq  C_1\, K^{-1}\, n\alpha_n^2\right)\geq 1- 2\, m^{-\tau}.
    \label{2nd_weak}
\end{equation}

\textbf{Bounding $\left|\langle \epsilon_j,\widehat{\Theta}_{.,l}-\widehat{\Theta}_{.,c_j}\rangle\right|$:} Let us examine the expression 
$\langle \epsilon_j,\widehat{\Theta}_{.,l}-\widehat{\Theta}_{.,c_j}\rangle$. 
It is easy to see that given $A_{(.,\MM)}$, vectors $\widehat{\Theta}_{.,l}, \widehat{\Theta}_{.,c_j}$ are constants and the term is only a function of 
$a_j=((A_{(\MM^c,.)})_{1,j},
\ldots,(A_{(\MM^c,.)})_{n-m,j})^\top$. Then, 
$$\langle \epsilon_j,\widehat{\Theta}_{.,l}-\widehat{\Theta}_{.,c_j}\rangle=\sum\limits_{i=1}^{n-m}((A_{(\MM^c,.)})_{i,j}-(P_{(\MM^c,.)})_{i,j})(\widehat{\Theta}_{i,l}-\widehat{\Theta}_{i,c_j}).$$ 
 The $i$-th summand is bounded above by $|\widehat{\Theta}_{i,l} - \widehat{\Theta}_{i,c_j}|$. 
Also, note that
 \begin{align*}
     &\,\sum\limits_{i=1}^{n-m} \EE\left(\left.\left(((A_{(\MM^c,.)})_{i,j}-(P_{(\MM^c,.)})_{i,j})(\widehat{\Theta}_{i,l}-\widehat{\Theta}_{i,c_j})\right)^2\right| \, A_{(.,\MM)}\right)\\
     =&\, \sum\limits_{i=1}^{n-m} (P_{(\MM^c,.)})_{i,j}(1-(P_{(\MM^c,.)})_{i,j}) (\widehat{\Theta}_{i, l}-\widehat{\Theta}_{i,c_j})^2\, \leq\, \alpha_n \umodnot{\widehat{\Theta}_{.,l}-\widehat{\Theta}_{.,c_j}}^2.
 \end{align*}
 Therefore, using Bernstein's inequality, for any $t>0$, derive
 \begin{equation*}
          \PP\left(\left.\left|\langle \epsilon_j,\widehat{\Theta}_{.,l}-\widehat{\Theta}_{.,c_j}\rangle\right|
        \leq t \, \right| \, A_{(.,\MM)}\right) 
         \geq     1- 2\exp\left(-\frac{t^2/2}{\alpha_n\umodnot{\widehat{\Theta}_{.,l}-\widehat{\Theta}_{.,c_j}}^2 + (t/3)\,\umodnot{\widehat{\Theta}_{.,l}-\widehat{\Theta}_{.,c_j}}_{\infty} }\right).
 \end{equation*}
 Choosing
 $$t=2\,\sqrt{\alpha_n}\,\umodnot{\widehat{\Theta}_{.,l}-\widehat{\Theta}_{.,c_j}}\sqrt{\log(nKm^{\tau})} + (4/3) \,\umodnot{\widehat{\Theta}_{.,l}-\widehat{\Theta}_{.,c_j}}_{\infty}\log(nKm^{\tau})$$
 yields that, with probability at least $1- 2(nKm^\tau)^{-1}$,  one has
 \begin{equation*}
          \left|\langle \epsilon_j,\widehat{\Theta}_{.,l}-\widehat{\Theta}_{.,c_j}\rangle\right|
        \ \leq\ 2\,\sqrt{\alpha_n}\, \umodnot{\widehat{\Theta}_{.,l}-\widehat{\Theta}_{.,c_j}}\sqrt{\log(nKm^{\tau})} + (4/3)\, \umodnot{\widehat{\Theta}_{.,l}-\widehat{\Theta}_{.,c_j}}_{\infty}\log(nKm^{\tau}).
 \end{equation*}
 Note that,
 \begin{equation*}
     \begin{split}
        \umodnot{\widehat{\Theta}_{.,l}-\widehat{\Theta}_{.,c_j}}_\infty&\leq \umodnot{\widehat{\Theta}_{.,l}-{\Theta}_{.,l}}_\infty + \umodnot{\widehat{\Theta}_{.,c_j}-{\Theta}_{.,c_j}}_\infty + \umodnot{{\Theta}_{.,c_j}-{\Theta}_{.,l}}_\infty\\
        &\leq 2\max_{i,\, k}|\widehat{\Theta}_{i,k}-{\Theta}_{i,k}|  + \alpha_n,
     \end{split}
\end{equation*}
and, 
\begin{equation*}
     \umodnot{\widehat{\Theta}_{.,l}-\widehat{\Theta}_{.,c_j}} \leq \sqrt{n}\,\umodnot{\widehat{\Theta}_{.,l}-\widehat{\Theta}_{.,c_j}}_\infty  
        \leq  2\,\sqrt{n}\,\max_{i,\, k}|\widehat{\Theta}_{i,k}-{\Theta}_{i,k}| + \sqrt{n}\,\alpha_n.
\end{equation*}
Therefore, 
with probability at least $1 - 2\,m^{-\tau}$, one has
\begin{align*}
&\,\max_{j\in\MM^c,\,l\neq c_j }\,
2\left|\langle \epsilon_j,\widehat{\Theta}_{.,l}-\widehat{\Theta}_{.,c_j}\rangle\right|\\
\leq&\,  8\,\sqrt{n\alpha_n}\, \max_{i,\, k}|\widehat{\Theta}_{i,k}-{\Theta}_{i,k}| \sqrt{\log(nKm^{\tau})} + 
4\,\sqrt{n\alpha_n}\,\alpha_n\, \sqrt{\log(nKm^{\tau})}\\ +&\, \frac{16}{3} \max_{i,\, k}|\widehat{\Theta}_{i,k}-{\Theta}_{i,k}|\log(nKm^{\tau}) + \frac{8}{3} \alpha_n\log(nKm^{\tau})\\
=&\, 8\,\left(\sqrt{n\alpha_n} + \frac{2}{3} \sqrt{\log(nKm^{\tau})}\right)\, \max_{i,\, k}|\widehat{\Theta}_{i,k}-{\Theta}_{i,k}| \sqrt{\log(nKm^{\tau})}\\
+&\, 
4\,\left(\sqrt{n\alpha_n}+ \frac{2}{3} \sqrt{\log(nKm^{\tau})}\right)\,\alpha_n\, \sqrt{\log(nKm^{\tau})} \\
\leq&\, C_{\tau}\left(\max_{i,\, k}|\widehat{\Theta}_{i,k}-{\Theta}_{i,k}| 
+\alpha_n\right) \sqrt{n\alpha_n} \sqrt{\log n},
\end{align*}
since $\log (n K m^{\tau}) \leq (\tau + 2) \log n$, and $\log n$ is dominated by $n\alpha_n$ due to Assumption {\bf A4}.

\textbf{Conclusion:} Combination of the above inequality and \eqref{thetabreak2_weak} yields
\begin{align}
     \PP\left(\bigcap_{j \in \MM^c} \bigcap_{l\neq {c_j}} \right. & \bigg\{\umodnot{a_j-\widehat{\Theta}_{.,l}}^2 -\umodnot{a_j-\widehat{\Theta}_{.,c_j}}^2\geq \frac{1}{2} \min_{k\neq l}\umod{{\Theta}_{.,k}-{\Theta}_{.,l}}^2 - \left. 3\, n\max_{i,\, k}(\widehat{\Theta}_{i,k}-\Theta_{i,k})^2 -\right.\nonumber\\ 
     & \left.C_{\tau}\left(\max_{i,\, k}|\widehat{\Theta}_{i,k}-{\Theta}_{i,k}| 
    +\alpha_n\right) \sqrt{n\alpha_n} \sqrt{\log n}\bigg\} \right) \geq 1 - O(m^{-\tau}).
\label{penultimate_weak}
\end{align}
Now, it follows from \eqref{2nd_weak} and Theorem \ref{lem:class_weak} that, 
with probability at least $1 - O(m^{-\tau})$, one has 
\begin{equation*}
    \begin{split}       
        \bigcap_{j \in \MM^c} \bigcap_{l\neq {c_j}} \bigg\{ \umodnot{a_j-\widehat{\Theta}_{.,l}}^2 -\umodnot{a_j-\widehat{\Theta}_{.,c_j}}^2 \geq \frac{C_1 \, n\alpha_n^2}{K} -
         C_{\tau}\,   \bigg( \frac{K n \alpha_n\log n}{m} + K^2 n \alpha_n^2\,\delta(n,m,K,\alpha_n)^2 \,+\\ \left( \sqrt{\frac{K\alpha_n\log n}{m}} + K \alpha_n\,\delta(n,m,K,\alpha_n)  
        +\alpha_n  \right) \sqrt{n\alpha_n} \sqrt{\log n}\bigg)\bigg\} .
    \end{split}
\label{ultimate}
\end{equation*}
Here, the right-hand side of the inequality is bounded below by 
\begin{align*}
    \frac{C_1 \, n\alpha_n^2}{K}  \bigg[1 - \frac{C_\tau}{C_1}\bigg(\frac{K^2 \log n}{m \alpha_n} \ +&\ K^3\,\delta(n,m,K,\alpha_n)^2 + \frac{K^{3/2} \log n}{\sqrt{m n}\alpha_n}\\
    \ +&\ (1+ K \delta(n,m,K,\alpha_n))\,\frac{K \sqrt{\log n}}{\sqrt{n \alpha_n}} 
    \bigg)  \bigg]. 
    %
\end{align*}
Therefore, if $K^3\,\delta(n,m,K,\alpha_n)^2 \rightarrow 0$ and the constant $c_0$ in Assumption {\bf A4} is sufficiently large, then the quantity above is positive, and one has
\begin{equation*}
    \PP({\Delta}_{\MM^c}=0) =  \PP\left(\bigcap\limits_{j \in \MM^c} \bigcap\limits_{l \neq c_j}\left( \left\{ \umodnot{a_j-\widehat{\Theta}_{.,l}}^2 > \umodnot{a_j-\widehat{\Theta}_{.,c_j}}^2
    \right\}\right)\right)\\
     \geq 1- O(m^{-\tau}),
\end{equation*}
which completes the proof.

\noindent
\textbf{Case 2: Node popularity approach under the DCBM}

For the node popularity approach, ${\Delta}_{\MM^c}$ can be written as,
$$
    {\Delta}_{\MM^c} = \frac{1}{n-m}\sum\limits_{i\in\MM^c}  \mathbb{I}\left(\min_{l \neq c_i}\, \left\{ \umod{\widetilde{N}_{i,.}-\frac{\widehat{\Omega}_{l,.}}{\sum\limits_r \widehat{\Omega}_{l,r}}} \leq \umod{\widetilde{N}_{i,.}-\frac{\widehat{\Omega}_{c_i,.}}{\sum\limits_r \widehat{\Omega}_{c_i,r}}}
    \right\}\right).
$$
The $i^{th}$ node is correctly clustered if
\begin{equation}
   \umod{\widetilde{N}_{i,.}-\frac{\widehat{\Omega}_{c_i,.}}{\sum\limits_r \widehat{\Omega}_{c_i,r}}} \leq \min_{l \neq c_i}\, \umod{\widetilde{N}_{i,.}-\frac{\widehat{\Omega}_{l,.}}{\sum\limits_r \widehat{\Omega}_{l,r}}}.
    \label{event2}
\end{equation}
Fix some $i \in\MM^c$ and $l \neq c_i$. Then, by repeated application of the triangle inequality, we obtain
\begin{align}
     &\ \umod{\widetilde{N}_{i,.}-\frac{\widehat{\Omega}_{l,.}}{\sum\limits_r \widehat{\Omega}_{l,r}}}  -\umod{\widetilde{N}_{i,.}-\frac{\widehat{\Omega}_{c_i,.}}{\sum\limits_r \widehat{\Omega}_{c_i,r}}}
    \ \geq\ \umod{\frac{\widehat{\Omega}_{l,.}}{\sum\limits_r \widehat{\Omega}_{l,r}} - \frac{\widehat{\Omega}_{c_i,.}}{\sum\limits_r \widehat{\Omega}_{c_i,r}}}  -2\umod{\widetilde{N}_{i,.}-\frac{\widehat{\Omega}_{c_i,.}}{\sum\limits_r \widehat{\Omega}_{c_i,r}}}\nonumber\\
    \geq
    &\ \umod{\frac{\widetilde{\Omega}_{c_i,.}}{\sum\limits_r \widetilde{\Omega}_{c_i,r}} - \frac{\widetilde{\Omega}_{l,.}}{\sum\limits_r \widetilde{\Omega}_{l,r}}} -\umod{\frac{\widehat{\Omega}_{c_i,.}}{\sum\limits_r \widehat{\Omega}_{c_i,r}} - \frac{\widetilde{\Omega}_{c_i,.}}{\sum\limits_r \widetilde{\Omega}_{c_i,r}}} -\umod{\frac{\widehat{\Omega}_{l,.}}{\sum\limits_r \widehat{\Omega}_{l,r}} - \frac{\widetilde{\Omega}_{l,.}}{\sum\limits_r \widetilde{\Omega}_{l,r}}} -2\umod{\widetilde{N}_{i,.}-\frac{\widehat{\Omega}_{c_i,.}}{\sum\limits_r \widehat{\Omega}_{c_i,r}}}\nonumber\\
    \geq
    &\ \min_{k\neq l}\umod{\frac{\widetilde{\Omega}_{k,.}}{\sum\limits_r \widetilde{\Omega}_{k,r}} - \frac{\widetilde{\Omega}_{l,.}}{\sum\limits_r \widetilde{\Omega}_{l,r}}} -2\max_{k}\umod{\frac{\widehat{\Omega}_{k,.}}{\sum\limits_r \widehat{\Omega}_{k,r}} - \frac{\widetilde{\Omega}_{k,.}}{\sum\limits_r \widetilde{\Omega}_{k,r}}} -2\umod{\widetilde{N}_{i,.}-\frac{\widehat{\Omega}_{c_i,.}}{\sum\limits_r \widehat{\Omega}_{c_i,r}}}\nonumber\\
    \geq
    &\ \min_{k\neq l}\umod{\frac{\widetilde{\Omega}_{k,.}}{\sum\limits_r \widetilde{\Omega}_{k,r}} - \frac{\widetilde{\Omega}_{l,.}}{\sum\limits_r \widetilde{\Omega}_{l,r}}} -4\max_{k}\umod{\frac{\widehat{\Omega}_{k,.}}{\sum\limits_r \widehat{\Omega}_{k,r}} - \frac{\widetilde{\Omega}_{k,.}}{\sum\limits_r \widetilde{\Omega}_{k,r}}} -2\umod{\widetilde{N}_{i,.}-\frac{\widetilde{\Omega}_{c_i,.}}{\sum\limits_r \widetilde{\Omega}_{c_i,r}}}.\label{main_break_weak}
\end{align}
\textbf{Bounding $\min\limits_{k\neq l}\umod{\frac{\widetilde{\Omega}_{k,.}}{\sum\limits_r \widetilde{\Omega}_{k,r}} - \frac{\widetilde{\Omega}_{l,.}}{\sum\limits_r \widetilde{\Omega}_{l,r}}}$:}
First, we establish a lower bound for $\min\limits_{k\neq l}\umod{\frac{\widetilde{\Omega}_{k,.}}{\sum\limits_r \widetilde{\Omega}_{k,r}} - \frac{\widetilde{\Omega}_{l,.}}{\sum\limits_r \widetilde{\Omega}_{l,r}}}$.
\\

Recall that, for any $k\in [K]$,
$$\frac{\widetilde{\Omega}_{k,.}}{\sum\limits_r \widetilde{\Omega}_{k,r}}=\left(\frac{\Omega_{k,1}\,\Gamma_1}{\sum\limits_r \Omega_{k,r}\,\Gamma_r}, \ldots, \frac{\Omega_{k,K}\,\Gamma_K}{\sum\limits_r \Omega_{k,r}\,\Gamma_r}\right).$$
Define $(K\times K)$ diagonal matrices $D_1$ and $D_2$ such that
$$D_1=\diag\left(\frac{1}{\sum\limits_r \Omega_{1,r}\Gamma_r},\ldots,\frac{1}{\sum\limits_r \Omega_{K,r}\Gamma_r}\right),\ D_2=\diag(\Gamma_1,\ldots,\Gamma_K).$$
Let $e_k$ be the $k^{th}$ column of the $K\times K$ identity matrix.
Then, $\frac{\widetilde{\Omega}_{k,.}}{\sum\limits_r \widetilde{\Omega}_{k,r}}$ can be written as $e_k^\top D_1\,\Omega\, D_2$.
For $k\neq l$, obtain
\begin{align*}
    &\ \umod{\frac{\widetilde{\Omega}_{k,.}}{\sum\limits_r \widetilde{\Omega}_{k,r}} - \frac{\widetilde{\Omega}_{l,.}}{\sum\limits_r \widetilde{\Omega}_{l,r}}}^2
    =\umod{e_k^\top D_1\,\Omega\, D_2 - e_l^\top D_1\,\Omega\, D_2}^2= \umod{(e_k - e_l)^\top D_1\,\Omega\, D_2}^2\\
    =&\ (e_k - e_l)^\top D_1\,\Omega\, D_2^2\,\Omega\, D_1(e_k - e_l)
    \ \geq\ \Gamma_{\min}^2(e_k - e_l)^\top D_1\,\Omega^2\, D_1(e_k - e_l)\\
    \geq&\ \Gamma_{\min}^2\,\lambda^2\,\alpha_n^2\, (e_k - e_l)^\top D_1^2\,(e_k - e_l) \\
    \geq&\ 2\,\Gamma_{\min}^2\,\alpha_n^2\,\lambda^2\, \min_k\left(\frac{1}{\sum\limits_r \Omega_{k,r}\Gamma_r}\right)^2
    \ \geq\ 2\,\Gamma_{\min}^2\, \alpha_n^2\, \lambda^2 \,\frac{1}{\Gamma_{\max}^2\, K^2\, \alpha_n^2}
    \,=\,\frac{2\,\Gamma_{\min}^2\,\lambda^2}{\Gamma_{\max}^2\, K^2}.
\end{align*}
since $\Omega\succeq \lambda\,\alpha_n\,I_K$ by Assumption {\bf A2}.
Therefore,
$$
\min_{k\neq l}\umod{\frac{\widetilde{\Omega}_{k,.}}{\sum\limits_r \widetilde{\Omega}_{k,r}} - \frac{\widetilde{\Omega}_{l,.}}{\sum\limits_r \widetilde{\Omega}_{l,r}}} 
\geq \frac{\sqrt{2}\,\Gamma_{\min}\,\lambda}{\Gamma_{\max}\, K} \geq \frac{\sqrt{2}\,(1-a)\,\lambda}{C_0^2\,(1+a)\,K},
$$
with probability at least $1-2\,m^{-\tau}$, by Theorem \ref{lem:tmin_tmax}.
Hence, for some positive constant $C_1$, one has
 %
%
\begin{equation}
    \PP\left(\min_{k\neq l}\umod{\frac{\widetilde{\Omega}_{k,.}}{\sum\limits_r \widetilde{\Omega}_{k,r}} - \frac{\widetilde{\Omega}_{l,.}}{\sum\limits_r \widetilde{\Omega}_{l,r}}} \geq \frac{C_1}{K}\right)\geq 1- \frac{2}{m^{\tau}}.
    \label{1st_dcbm_weak}
\end{equation}

Next, we establish upper bounds for $\max\limits_{k}\umod{\frac{\widehat{\Omega}_{k,.}}{\sum\limits_r \widehat{\Omega}_{k,r}} - \frac{\widetilde{\Omega}_{k,.}}{\sum\limits_r \widetilde{\Omega}_{k,r}}}$ and $\umod{\widetilde{N}_{i,.}-\frac{\widetilde{\Omega}_{c_i,.}}{\sum\limits_r \widetilde{\Omega}_{c_i,r}}}$.

\textbf{Bounding $\max\limits_{k}\umod{\frac{\widehat{\Omega}_{k,.}}{\sum\limits_r \widehat{\Omega}_{k,r}} - \frac{\widetilde{\Omega}_{k,.}}{\sum\limits_r \widetilde{\Omega}_{k,r}}}$:} 
\begin{align*}
    &\ \umod{\frac{\widehat{\Omega}_{k,.}}{\sum\limits_r \widehat{\Omega}_{k,r}} - \frac{\widetilde{\Omega}_{k,.}}{\sum\limits_r \widetilde{\Omega}_{k,r}}}
    \leq \umod{\frac{\widehat{\Omega}_{k,.}}{\sum\limits_r \widehat{\Omega}_{k,r}} - \frac{\widetilde{\Omega}_{k,.}}{\sum\limits_r \widehat{\Omega}_{k,r}}} + \umod{\frac{\widetilde{\Omega}_{k,.}}{\sum\limits_r \widehat{\Omega}_{k,r}} - \frac{\widetilde{\Omega}_{k,.}}{\sum\limits_r \widetilde{\Omega}_{k,r}}}\\[.2cm]
    =&\ \frac{\umodnot{\widehat{\Omega}_{k,.} - \widetilde{\Omega}_{k,.}}}{\sum\limits_r \widehat{\Omega}_{k,r}} + \frac{\umodnot{\widetilde{\Omega}_{k,.}}}{\sum\limits_r \widehat{\Omega}_{k,r}\sum\limits_r \widetilde{\Omega}_{k,r}} \left|\sum\limits_r \widehat{\Omega}_{k,r} - \sum\limits_r \widetilde{\Omega}_{k,r}\right|\\[.2cm]
    \leq&\ \frac{\umodnot{\widehat{\Omega}_{k,.} - \widetilde{\Omega}_{k,.}}}{\sum\limits_r \widehat{\Omega}_{k,r}} + \frac{\sqrt{K}\,\umodnot{\widetilde{\Omega}_{k,.}}\umodnot{\widehat{\Omega}_{k,.} - \widetilde{\Omega}_{k,.}}}{\sum\limits_r \widehat{\Omega}_{k,r}\sum\limits_r \widetilde{\Omega}_{k,r}},\ \text{using Cauchy-Schwarz inequality}\\[.2cm]
    \leq&\ \frac{2\sqrt{K}\,\umodnot{\widehat{\Omega}_{k,.} - \widetilde{\Omega}_{k,.}}}{\sum\limits_r \widehat{\Omega}_{k,r}},\ \text{since }\umodnot{\widetilde{\Omega}_{k,.}}\leq \sum\limits_r \widetilde{\Omega}_{k,r}\\
    \leq&\ \frac{2\sqrt{K}\,\umodnot{\widehat{\Omega}_{k,.} - \widetilde{\Omega}_{k,.}}}{\sum\limits_r \widetilde{\Omega}_{k,r} - \sqrt{K}\,\umodnot{\widehat{\Omega}_{k,.} - \widetilde{\Omega}_{k,.}}},\ \text{using Cauchy-Schwarz inequality}\\[.2cm]
    \leq&\ \frac{2\sqrt{K}\,\umodnot{\widehat{\Omega}_{k,.} - \widetilde{\Omega}_{k,.}}}{\Gamma_{\min}^2\, \lambda\, \alpha_n - \sqrt{K}\umodnot{\widehat{\Omega}_{k,.} - \widetilde{\Omega}_{k,.}}},\ \text{since }\sum\limits_r \widetilde{\Omega}_{k,r}\geq\widetilde{\Omega}_{k,k}\geq \Gamma_{\min}^2\,\lambda\,\alpha_n.
\end{align*}
Leveraging Theorems \ref{lem:tmin_tmax} and \ref{lem:omega_est_weak}, one has that with probability at least $1-O(m^{-\tau})$,
\begin{align*}
    &\ \max\limits_{k}\umod{\frac{\widehat{\Omega}_{k,.}}{\sum\limits_r \widehat{\Omega}_{k,r}} - \frac{\widetilde{\Omega}_{k,.}}{\sum\limits_r \widetilde{\Omega}_{k,r}}}  
    \ \leq\ \frac{2\sqrt{K}\max\limits_{k}\umodnot{\widehat{\Omega}_{k,.} - \widetilde{\Omega}_{k,.}}}{\Gamma_{\min}^2\lambda\,\alpha_n - \sqrt{K}\max\limits_{k}\umodnot{\widehat{\Omega}_{k,.} - \widetilde{\Omega}_{k,.}}}\\[.2cm]
    \leq&\ \dfrac{2\sqrt{K}\,C_{\tau}\left(\dfrac{m^{3/2}\sqrt{\alpha_n}}{K} + \dfrac{m^2\alpha_n}{\sqrt{K}}\, \delta(n,m,K,\alpha_n)\right)(1 + K\,\delta(n,m,K,\alpha_n))}{\dfrac{m^2(1-a)^2}{{C_0}^2 K^2}\lambda\,\alpha_n - \sqrt{K}\,C_{\tau}\left(\dfrac{m^{3/2}\sqrt{\alpha_n}}{K} + \dfrac{m^2\alpha_n}{\sqrt{K}}\, \delta(n,m,K,\alpha_n)\right)(1 + K\,\delta(n,m,K,\alpha_n))}\\[.2cm]
    =&\ \dfrac{2\,\dfrac{{C_0}^2\, C_{\tau}}{(1-a)^2\lambda} \, \left(\dfrac{K^{3/2}}{\sqrt{m\alpha_n}} + K^2\,\delta(n,m,K,\alpha_n)\right)(1 + K\,\delta(n,m,K,\alpha_n))}{1 - \dfrac{{C_0}^2\, C_{\tau}}{(1-a)^2\lambda} \, \left(\dfrac{K^{3/2}}{\sqrt{m\alpha_n}} + K^2\,\delta(n,m,K,\alpha_n)\right)(1 + K\,\delta(n,m,K,\alpha_n))}\\[.2cm]
    \leq&\ \dfrac{2\,\dfrac{{C_0}^2\, C_{\tau}}{(1-a)^2\lambda} \, \left(\dfrac{1}{\sqrt{c_0\,K \log n}} + K^2\,\delta(n,m,K,\alpha_n)\right)(1 + K\,\delta(n,m,K,\alpha_n))}{1 - \dfrac{{C_0}^2\, C_{\tau}}{(1-a)^2\lambda} \, \left(\dfrac{1}{\sqrt{c_0\,K \log n}} + K^2\,\delta(n,m,K,\alpha_n)\right)(1 + K\,\delta(n,m,K,\alpha_n))},
\end{align*}
\\
where the final step follows from by Assumption {\bf A4}. 

Now, if $K=O(\log n)$, and $K^3\,\delta(n,m,K,\alpha_n)\rightarrow 0$, then for all sufficiently large $n$, the numerator on the right-hand side above is upper-bounded by $2\,C_{1\tau}/K\sqrt{c_0}$ for some positive constant $C_{1\tau}$. If
 $c_0$ satisfies $C_{1\tau}/K\sqrt{c_0}< 1/2$, then one has
\begin{equation}
    \PP\left(\max\limits_{k}\umod{\frac{\widehat{\Omega}_{k,.}}{\sum\limits_r \widehat{\Omega}_{k,r}} - \frac{\widetilde{\Omega}_{k,.}}{\sum\limits_r \widetilde{\Omega}_{k,r}}} 
    \leq \dfrac{4\,C_{1\tau}}{K\sqrt{c_0}} \right) \
    \geq 1-O(m^{-\tau}).
    \label{2nd_dcbm_weak}
\end{equation}

\textbf{Bounding $\umod{\widetilde{N}_{i,.}-\frac{\widetilde{\Omega}_{c_i,.}}{\sum\limits_r \widetilde{\Omega}_{c_i,r}}}$:}
\begin{equation*}
    \umod{\widetilde{N}_{i,.}-\frac{\widetilde{\Omega}_{c_i,.}}{\sum\limits_r \widetilde{\Omega}_{c_i,r}}} = \sqrt{\sum\limits_k \left(\frac{\sum\limits_{u \in \widehat{\mathcal{G}}_k} A_{i,u} }{\sum\limits_{u \in \MM} A_{i,u}} - \frac{\widetilde{\Omega}_{c_i,k}}{\sum\limits_r \widetilde{\Omega}_{c_i,r}}\right)^2}.
\end{equation*}
Observe that
\begin{equation*}
\frac{\sum\limits_{u \in \mathcal{G}_k} P_{i,u} }{\sum\limits_{u \in \MM} P_{i,u}}= \frac{\sum\limits_{u \in \mathcal{G}_k} \theta_i\,\Omega_{c_i,k}\, \theta_u}{\sum\limits_r\sum\limits_{u \in \mathcal{G}_r} \theta_i\,\Omega_{c_i,r}\, \theta_u}
 =\frac{\Omega_{c_i,k}\,\Gamma_k}{\sum\limits_r\Omega_{c_i,r}\,\Gamma_r}=\frac{\widetilde{\Omega}_{c_i,k}}{\sum\limits_r \widetilde{\Omega}_{c_i,r}} \text{ for } k=1, \ldots, K.
\end{equation*}
So, we have
\begin{equation}
    \umod{\widetilde{N}_{i,.}-\frac{\widetilde{\Omega}_{c_i,.}}{\sum\limits_r \widetilde{\Omega}_{c_i,r}}} = \sqrt{\sum\limits_{k=1}^K \left(\frac{\sum\limits_{u \in \widehat{\mathcal{G}}_k} A_{i,u} }{\sum\limits_{u \in \MM} A_{i,u}} - \frac{\sum\limits_{u \in \mathcal{G}_k} P_{i,u} }{\sum\limits_{u \in \MM} P_{i,u}}\right)^2}.
    \label{ni_minus_omega_weak}
\end{equation}

\begin{align}
    &\ \left|\frac{\sum\limits_{u \in \widehat{\mathcal{G}}_k} A_{i,u} }{\sum\limits_{u \in \MM} A_{i,u}} - \frac{\sum\limits_{u \in \mathcal{G}_k} P_{i,u} }{\sum\limits_{u \in \MM} P_{i,u}}\right|\nonumber\\
    \leq&\ \left|\frac{\sum\limits_{u \in \widehat{\mathcal{G}}_k} A_{i,u} }{\sum\limits_{u \in \MM} A_{i,u}} - \frac{\sum\limits_{u \in \widehat{\mathcal{G}}_k} A_{i,u} }{\sum\limits_{u \in \MM} P_{i,u}}\right|
    + \left|\frac{\sum\limits_{u \in \widehat{\mathcal{G}}_k} A_{i,u} }{\sum\limits_{u \in \MM} P_{i,u}} - \frac{\sum\limits_{u \in \widehat{\mathcal{G}}_k} P_{i,u} }{\sum\limits_{u \in \MM} P_{i,u}}\right|
    +\left|\frac{\sum\limits_{u \in \widehat{\mathcal{G}}_k} P_{i,u} }{\sum\limits_{u \in \MM} P_{i,u}} - \frac{\sum\limits_{u \in \mathcal{G}_k} P_{i,u} }{\sum\limits_{u \in \MM} P_{i,u}}\right|\nonumber\\
    =&\ 
    \frac{\sum\limits_{u \in \widehat{\mathcal{G}}_k} A_{i,u} \left|\sum\limits_{u \in \MM} (A_{i,u} - P_{i,u})\right|}{\sum\limits_{u \in \MM} A_{i,u}\sum\limits_{u \in \MM} P_{i,u}}
    + \frac{\left|\sum\limits_{u \in \widehat{\mathcal{G}}_k} (A_{i,u} - P_{i,u})\right|}{\sum\limits_{u \in \MM} P_{i,u}}
    + \frac{\left|\sum\limits_{u \in \widehat{\mathcal{G}}_k} P_{i,u} - \sum\limits_{u \in \mathcal{G}_k} P_{i,u}\right|}{\sum\limits_{u \in \MM} P_{i,u}}
    \nonumber\\
    \leq&\ \frac{\left|\sum\limits_{u \in \MM} (A_{i,u} - P_{i,u})\right|}{\sum\limits_{u \in \MM} P_{i,u}} 
    + \frac{\left|\sum\limits_{u \in \widehat{\mathcal{G}}_k} (A_{i,u} - P_{i,u})\right|}{\sum\limits_{u \in \MM} P_{i,u}}
    + \frac{\left|\sum\limits_{u \in \widehat{\mathcal{G}}_k} P_{i,u} - \sum\limits_{u \in \mathcal{G}_k} P_{i,u}\right|}{\sum\limits_{u \in \MM} P_{i,u}}.\label{ai_minus_pi_weak}
\end{align}

{\bf Bounding the first term on the right-hand side of \eqref{ai_minus_pi_weak}:} 
Using Bernstein's inequality, one has
\begin{equation*}
    \PP\left(\left|\sum\limits_{u \in \MM} (A_{i,u} - P_{i,u})\right|\leq t\right)\geq 1-2\exp\left(\frac{-t^2/2}{\sum_{u \in \MM} P_{i,u} + t/3}\right),
\end{equation*}
since $\text{Var}(\sum_{u \in \MM} A_{i,u})\leq\sum_{u \in \MM} P_{i,u}$. Choosing 
$$ t = 2\sqrt{\sum_{u \in \MM} P_{i,u} \log(n m^{\tau})} + 4/3\, \log(n m^{\tau})$$
yields that, with probability at least $1- 2\, (nm^\tau)^{-1}$,
 \begin{equation*}
          \left|\sum\limits_{u \in \MM} (A_{i,u} - P_{i,u})\right|
        \leq 2\sqrt{\sum_{u \in \MM} P_{i,u} \log(n m^{\tau})} + 4/3\, \log(n m^{\tau}).
 \end{equation*}
Now, for all $i\in \MM^c$,
\begin{equation}
\sum\limits_{u \in \MM} P_{i,u} \ =\  \theta_i\sum\limits_k\Omega_{c_i,k}\,\Gamma_k \ \geq\  \theta_i\,\Gamma_{\min}\,\lambda\, \alpha_n\ \geq\  \frac{\theta_i\,\lambda\, m\alpha_n\,(1 - a)}{C_0\,K},
\label{sumpi.lbound_weak}
\end{equation}
with probability at least $1-2\,m^{-\tau}$, using Theorem \ref{lem:tmin_tmax}.
Noting that $\theta_i\, m\alpha_n/K$ dominates $\log n$ by Assumption {\bf A4}, one has
\begin{equation}
    \PP\left(\bigcap_{i\in\MM^c} \left\{\left|\sum\limits_{u \in \MM} (A_{i,u} - P_{i,u})\right|\leq C_{2\tau}\sqrt{\sum_{u \in \MM} P_{i,u} \log n}\right\}\right)\geq 1 - \frac{4}{m^{\tau}},
    \label{npvec_p1_weak}
\end{equation}
where $C_{2\tau}$ is a positive constant depending on $\tau$.

Combining \eqref{sumpi.lbound_weak} and \eqref{npvec_p1_weak}, obtain
\begin{equation*}
    \PP\left(\bigcap_{i\in\MM^c} \left\{\frac{\left|\sum\limits_{u \in \MM} (A_{i,u} - P_{i,u})\right|}{\sum\limits_{u \in \MM} P_{i,u}} \leq C_{2\tau}\sqrt{\frac{C_0 K\log n}{\theta_i\,\lambda\, m\alpha_n\,(1 - a)}} \right\}\right)\geq 1 - \frac{6}{m^{\tau}}.
\end{equation*}
Applying Assumption {\bf A4}, one has
\begin{equation}
    \PP\left(\bigcap_{i\in\MM^c} \left\{\frac{\left|\sum\limits_{u \in \MM} (A_{i,u} - P_{i,u})\right|}{\sum\limits_{u \in \MM} P_{i,u}} \leq C_{2\tau}\sqrt{\frac{C_0}{c_0\,K^3\,\lambda\,(1 - a)}} \right\}\right)\geq 1 - \frac{6}{m^{\tau}}.
    \label{npvec_p2_weak}
\end{equation}

{\bf Bounding the second term on the right-hand side of \eqref{ai_minus_pi_weak}:} 
Note that $\widehat{\mathcal{G}}_k$ is independent of $A_{i,u}$, since $i\in\MM^c$. Using Bernstein's inequality, one has
\begin{equation*}
    \PP\left(\left.\left|\sum\limits_{u \in \widehat{\mathcal{G}}_k} (A_{i,u} - P_{i,u})\right|\leq t\right|\widehat{\mathcal{G}}_k\right)\geq 1-2\exp\left(\frac{-t^2/2}{\sum_{u \in \MM} P_{i,u} + t/3}\right),
\end{equation*}
since $\text{Var}(\sum_{u \in \widehat{\mathcal{G}}_k} A_{i,u}|\widehat{\mathcal{G}}_k)\leq \sum_{u \in \widehat{\mathcal{G}}_k} P_{i,u}\leq \sum_{u \in \MM} P_{i,u}$. Choosing 
$$ t = 2\sqrt{\sum_{u \in \MM} P_{i,u} \log(n K m^{\tau})} + 4/3\, \log(n K m^{\tau}),$$
one can follow the same argument as before to obtain
\begin{equation}
    \PP\left(\bigcap_{i\in\MM^c} \bigcap_{k=1}^K\left\{\frac{\left|\sum\limits_{u \in \widehat{\mathcal{G}}_k} (A_{i,u} - P_{i,u})\right|}{\sum\limits_{u \in \MM} P_{i,u}} \leq C_{3\tau}\sqrt{\frac{C_0}{c_0\,K^3\,\lambda\,(1 - a)}} \right\}\right)\geq 1 - \frac{6}{m^{\tau}},
    \label{npvec_p3_weak}
\end{equation}
where $C_{3\tau}$ is a positive constant depending on $\tau$.

{\bf Bounding the third term on the right-hand side of \eqref{ai_minus_pi_weak}:} Recall equations \eqref{gk1} and \eqref{gk2} from the proof of Theorem \ref{lem:class_weak}. 
\begin{equation*}
    \left|\sum\limits_{u \in \widehat{\mathcal{G}}_k} P_{i,u} - \sum\limits_{u \in \mathcal{G}_k} P_{i,u} \right|
    \,\leq\, \theta_i\,\alpha_n(|\widehat{\mathcal{G}}_k\cap\mathcal{G}_k^c| + |\mathcal{G}_k\cap\widehat{\mathcal{G}}_k^c|)\,\leq\, \theta_i\,\alpha_n\,m\widetilde{\Delta}_{\MM}.
\end{equation*}
We have assumed that 
\begin{equation}
    \PP(\widetilde{\Delta}_{\MM}\leq C_{\tau}\,\delta(n,m,K,\alpha_n)) \geq 1- \frac{C}{m^\tau},
    \label{comdet_subgraph_weak}
\end{equation}
which implies
\begin{equation*}
    \left|\sum\limits_{u \in \widehat{\mathcal{G}}_k} P_{i,u} - \sum\limits_{u \in \mathcal{G}_k} P_{i,u} \right|
    \leq C_{\tau}\,\theta_i\,m\alpha_n\,\delta(n,m,K,\alpha_n).
\end{equation*}
Combining the above inequality with \eqref{sumpi.lbound_weak}, one has
\begin{equation*}
    \PP\left(\bigcap_{i\in\MM^c} \bigcap_{k=1}^K\left\{\frac{\left|\sum\limits_{u \in \widehat{\mathcal{G}}_k} P_{i,u} - \sum\limits_{u \in \mathcal{G}_k} P_{i,u} \right|}{\sum\limits_{u \in \MM} P_{i,u}} \leq \dfrac{C_{\tau}\,C_0}{\lambda\,(1-a)}\,K\,\delta(n,m,K,\alpha_n) \right\}\right)\geq 1 - O(m^{-\tau}),
\end{equation*}
Note that we did not take an union bound to obtain the above result, as the probability inequality in \eqref{sumpi.lbound_weak}  considers all $i\in\MM^c$ and is free from $k\in[K]$, and the probability inequality in \eqref{comdet_subgraph_weak} is free from both $i\in\MM^c$ and $k\in[K]$.

By $K^3\,\delta(n,m,K,\alpha_n)\rightarrow\infty$, for all sufficiently large $n$, one has
\begin{equation}
    \PP\left(\bigcap_{i\in\MM^c} \bigcap_{k=1}^K\left\{\frac{\left|\sum\limits_{u \in \widehat{\mathcal{G}}_k} P_{i,u} - \sum\limits_{u \in \mathcal{G}_k} P_{i,u} \right|}{\sum\limits_{u \in \MM} P_{i,u}} \leq \dfrac{C_{\tau}\,C_0}{\lambda\,(1-a)K^2\sqrt{c_0}} \right\}\right)\geq 1 - O(m^{-\tau}),
    \label{npvec_p5_weak}
\end{equation}
Therefore, from \eqref{ai_minus_pi_weak}, \eqref{npvec_p2_weak}, \eqref{npvec_p3_weak} and \eqref{npvec_p5_weak}, one has that for all $i\in \MM^c$,
\begin{align*}
    \left|\frac{\sum\limits_{u \in \widehat{\mathcal{G}}_k} A_{i,u} }{\sum\limits_{u \in \MM} A_{i,u}} - \frac{\sum\limits_{u \in \mathcal{G}_k} P_{i,u} }{\sum\limits_{u \in \MM} P_{i,u}}\right|
    \leq&\ (C_{2\tau} + C_{3\tau})\,\sqrt{\frac{C_0}{c_0\,K^3\,\lambda\,(1 - a)}} +\dfrac{C_{\tau}\,C_0}{\lambda\,(1-a)K^2\sqrt{c_0}} \leq \frac{C_{4\tau}}{\sqrt{c_0\,K^3}},
\end{align*}
with probability at least $1-O(m^{-\tau})$, for some positive constant $C_{4\tau}$ depending on $\tau$.
Then, from \eqref{ni_minus_omega_weak}, one has
\begin{equation}
    \PP\left(\bigcap\limits_{i \in \MM^c}\umod{\widetilde{N}_{i,.}-\frac{\widetilde{\Omega}_{c_i,.}}{\sum\limits_r \widetilde{\Omega}_{c_i,r}}}\leq  \frac{C_{4\tau}}{K\sqrt{c_0}}\right)\geq 1-O(m^{-\tau}).
    \label{3rd_dcbm_weak}
\end{equation}

\textbf{Conclusion:} Combining \eqref{main_break_weak}, \eqref{1st_dcbm_weak}, \eqref{2nd_dcbm_weak}, and \eqref{3rd_dcbm_weak}, we have that for all $i\in \MM^c$ and $l\neq c_i$,
\begin{align*}
    \umod{\widetilde{N}_{i,.}-\frac{\widehat{\Omega}_{l,.}}{\sum\limits_r \widehat{\Omega}_{l,r}}}  -\umod{\widetilde{N}_{i,.}-\frac{\widehat{\Omega}_{c_i,.}}{\sum\limits_r \widehat{\Omega}_{c_i,r}}}
    \geq&\ \frac{C_1}{K} - \dfrac{16\,C_{1\tau}}{K\sqrt{c_0}} -\frac{2\,C_{4\tau}}{K\sqrt{c_0}} \\
    =&\ \frac{1}{K}\left(C_1 - \dfrac{16\,C_{1\tau}}{K\sqrt{c_0}} -\frac{2\,C_{4\tau}}{K\sqrt{c_0}}\right) > 0,
\end{align*}
with probability at least $1-O(m^{-\tau})$, provided $c_0$ is large enough. Hence,
\begin{equation*}
    \PP({\Delta}_{\MM^c}=0) =  \PP\left(\bigcap\limits_{i \in \MM^c} \bigcap\limits_{l \neq c_i}\left( \left\{  \umod{\widetilde{N}_{i,.}-\frac{\widehat{\Omega}_{l,.}}{\sum\limits_r \widehat{\Omega}_{l,r}}}  >\umod{\widetilde{N}_{i,.}-\frac{\widehat{\Omega}_{c_i,.}}{\sum\limits_r \widehat{\Omega}_{c_i,r}}}\right\}\right)\right)\\
     \geq 1- O(m^{-\tau}),
\end{equation*}
which completes the proof.


\section{Proof of Theorem \ref{th:step1}}   

Define $U=M\Lambda^{-\frac{1}{2}},U_s=M_{(\MM,.)}\Lambda_s^{-\frac{1}{2}}$, so  that $U^\top U={U_s}^\top  U_s=I_K$, the identity matrix.

\noindent
\textbf{Case 1: Spectral clustering under the SBM}\\
%
Let $V, \widehat{V}$ be $m\times K$ 
matrices consisting of the $K$ leading eigenvectors of $P_{(\MM,\MM)}$ and $A_{(\MM,\MM)}$ respectively, 
and let $\lambda_K(P_{(\MM,\MM)})$ be the smallest non-zero eigenvalue of $P_{(\MM,\MM)}$.
We first establish a high-probability lower bound on $\lambda_K(P_{(\MM,\MM)})$:
\begin{align*}        
    P_{(\MM,\MM)}&= \alpha_n\,M_{(\MM,.)}\,\Omega_0\, {M_{(\MM,.)}}^\top = \alpha_n\, {U}_s\,\Lambda_s^{\frac{1}{2}}\, \Omega_0\, \Lambda_s^{\frac{1}{2}}\,{U}_s^\top \succeq \lambda\,\alpha_n\, {U}_s\,\Lambda_s\, {U}_s^\top 
     \succeq \lambda\,\mu_{\min}\,\alpha_n\, {U}_s U_s^\top,
\end{align*}
using the fact that $\Omega_0\succeq \lambda\,I$ by Assumption {\bf A2}.

By Theorem \ref{lem:min_max_mu_k}, obtain that with probability at least $1-O(m^{-\tau})$,
\begin{equation}
    \lambda_K(P_{(\MM,\MM)})  \,\geq\, \lambda\,\mu_{\min}\alpha_n \,\geq\, \dfrac{\lambda(1-a)\,m\alpha_n}{C_0 K}.
    \label{zetam}
\end{equation}

Thereofore, by Corollary 4.1 of \citet{xie2021entrywise}, we obtain that under Assumption {\bf A4}, there exists a $(K\times K)$ orthogonal matrix $W$ such that
\begin{equation}
    \umodnot{\widehat{V} - VW}_{2,\infty} \ \leq\ C\,\dfrac{\sqrt{m\alpha_n \log m}}{\lambda_K(P_{(\MM,\MM)})}\,\umod{V}_{2,\infty},
    \label{twotoinf1}
\end{equation}
with probability at least $1-O(m^{-\tau})$.

Next, note that if $P_{(\MM,\MM)}$ is rank $K$, we can consider $V=U_s\,Q$ for a $K\times K$ orthogonal matrix $Q$, since $P_{(\MM,\MM)}= \alpha_n\, {U}_s\,(\Lambda_s^{\frac{1}{2}}\, \Omega_0\, \Lambda_s^{\frac{1}{2}})\,{U}_s^\top$, and $(\Lambda_s^{\frac{1}{2}}\, \Omega_0\, \Lambda_s^{\frac{1}{2}})$ is a $K\times K$ matrix with full rank. So,
\begin{equation*}
    \umod{V}_{2,\infty}\,=\, \umod{U_s}_{2,\infty} \,\leq\, \dfrac{1}{\sqrt{\mu_{\min}}}.
\end{equation*}
Again, applying Theorem \ref{lem:min_max_mu_k}, obtain that with probability at least $1-O(m^{-\tau})$,
\begin{equation}
    \umod{V}_{2,\infty}\,\leq\, \sqrt{\dfrac{C_0 K}{m(1-a)}}.
    \label{twotoinf2}
\end{equation}

Combining \eqref{zetam}, \eqref{twotoinf1} and \eqref{twotoinf2}, obtain that with probability at least $1-O(m^{-\tau})$,
\begin{equation}
    \umodnot{\widehat{V} - VW}_{2,\infty} \leq C\,\dfrac{K^{3/2}}{\sqrt{m}}\, \sqrt{\dfrac{\log m}{m\alpha_n}}.
    \label{twotoinf3}
\end{equation}
Note that, the probability statement in Theorem \ref{lem:min_max_mu_k} concerns with the randomly chosen subsample $\MM$ which is independent of $A$, and hence, independent of $\widehat{V}$.

Now, by Lemma 1 of \citet{pensky2024davis}, the number of misclassified nodes obtained from estimating 
the row clusters of $VW$ by an $(1+\beta)$-approximate solution to K-means problem with input $\widehat{V}$, 
i.e. the number $m\Delta_{\MM}$ of misclustered nodes in step 2, is bounded above by 
\begin{equation}
    m\Delta_{\MM} \,\leq\, \#\{i: \umodnot{\widehat{V}_{i,.} - (VW)_{i,.}} > \gamma/2 - \delta\},
\end{equation}
provided there exists $\delta\in(0,\gamma/2)$ such that
\begin{equation}
    \lVert\widehat{V}-VW\rVert_F\leq \dfrac{\delta\,\sqrt{\mu_{min}}}{\sqrt{K}(1+\sqrt{1+\beta})}.
    \label{delta_cond}
\end{equation}
Here, $\gamma$ is the minimum pairwise Euclidean norm separation among the $K$ distinct rows of $VW$, which   doesn't depend on $W$.
Note that by Theorem~\ref{lem:min_max_mu_k}  and \eqref{twotoinf3}, one has, with probability at least $1 - O(m^{-\tau})$
\begin{equation}
\label{high.prob.event}
\begin{split}
    \mu_{\min} \geq (C_0 K)^{-1}\, (1-a)\,m,&\quad \mu_{\max} \leq K^{-1}\, (1+a)\,m C_0,\\
    \umodnot{\widehat{V} - VW}_{2,\infty} &\,\leq\, C\,\dfrac{K^{3/2}}{\sqrt{m}}\, \sqrt{\dfrac{\log m}{m\alpha_n}}.    
\end{split}
\end{equation}
Then, on the same set, due to Assumption {\bf A4}, one has
\begin{gather*}
    \gamma \,\geq\, \min\limits_{k\neq l}\sqrt{\frac{1}{\mu_k}+\frac{1}{\mu_l}}\,\geq\, \sqrt{\frac{2}{\mu_{\max}}}\,\geq\, \sqrt{\frac{2K}{(1+a)\, m\,C_0}},\\
    \umodnot{\widehat{V} - VW}_{F} \,\leq\, \sqrt{m}\,\umodnot{\widehat{V} - VW}_{2,\infty} \,\leq\, C\,K^{3/2}\, \sqrt{\dfrac{\log m}{m\alpha_n}} \,\leq\,\dfrac{C}{\sqrt{c_0\,K}}.
\end{gather*}
Note that, when $\mu_{\min} \geq (C_0 K)^{-1}\, (1-a)\,m$, inequality \eqref{delta_cond} holds if 
$$
\lVert\widehat{V}-VW\rVert_F\leq \dfrac{\delta\,\sqrt{(1-a)\,m}}{\sqrt{C_0} K(1+\sqrt{1+\beta})}.
$$
Hence, if the constant $c_0$ in Assumption {\bf A4(a)} is large enough, one can choose $\delta=\gamma/4$, so that \eqref{delta_cond} holds. 
Therefore, the events in \eqref{high.prob.event} imply that
\begin{equation}
    m\Delta_{\MM} \,\leq\, \#\{i: \umodnot{\widehat{V}_{i,.} - (VW)_{i,.}} > \gamma/4\}.
\end{equation}
Now, since $\umodnot{\widehat{V} - VW}_{2,\infty} \leq C\,(K^{3/2}/\sqrt{m})\, \sqrt{\log m/m\alpha_n}$ and $\gamma\geq \sqrt{2K/((1+a)\, m\, C_0)}$, 
there will be no $i$ such that $\umodnot{\widehat{V} - VW}_{2,\infty}>\gamma/4$, if the constant $c_0$ in Assumption {\bf A4} is large enough, 
and perfect clustering is guaranteed, that is, $m\Delta_{\MM}=0$.
Finally, we note that by Theorem \ref{lem:min_max_mu_k} and \eqref{twotoinf3}, the events in \eqref{high.prob.event} hold with probability at least $1-O(m^{-\tau})$. 
This concludes the proof of the first part of the theorem.

\medskip    


\noindent
\textbf{Case 2: Regularized spectral clustering under the DCBM}\\
Let $\Xi=\diag(\theta)$ be the diagonal matrix containing the degree parameters $(\theta_i)$ as its diagonal elements. 
Define $\widetilde{\Lambda}_s = M_{(\MM,.)}^\top\,\Xi_{(\MM,\MM)}^2\, M_{(\MM,.)}$ and $\widetilde{U}_s=\Xi_{(\MM,\MM)}\,M_{(\MM,.)}\,\widetilde{\Lambda}_s^{-\frac{1}{2}}$,
and observe that matrix $\widetilde{\Lambda}_s$ is diagonal and  $\widetilde{U}_s^\top  \widetilde{U}_s=I_K$, the identity matrix.
Note that, the $k$-th diagonal element of $\widetilde{\Lambda}_s$ is 
$$
\widetilde{\mu}_k=\sum_{j\in\mathcal{G}_k} \theta_j^2,\quad k=1,\ldots, K.
$$
Also define
$$
\widetilde{\mu}_{\max}=\max_{k}\widetilde{\mu}_k , \quad \widetilde{\mu}_{\min}=\min_{k}\widetilde{\mu}_k.
$$
Let $V$ and $\widehat{V}$ be $(m\times K)$ matrices, consisting of the $K$ leading eigenvectors of $P_{(\MM,\MM)}$ and $A_{(\MM,\MM)}$ respectively, 
and $\Psi,\widehat{\Psi}$ be the corresponding diagonal matrices consisting of the $K$ leading eigenvalues. 
Let $\lambda_K(P_{(\MM,\MM)})$ be the smallest non-zero eigenvalue of $P_{(\MM,\MM)}$.
We first establish a high-probability lower bound on $\lambda_K(P_{(\MM,\MM)})$:
\begin{align*}        
    P_{(\MM,\MM)}&= \alpha_n\,\Xi_{(\MM,\MM)}\,M_{(\MM,.)}\,\Omega_0\, {M_{(\MM,.)}}^\top\Xi_{(\MM,\MM)} = \alpha_n\, {\widetilde{U}}_s\,\widetilde{\Lambda}_s^{\frac{1}{2}}\, \Omega_0\, \widetilde{\Lambda}_s^{\frac{1}{2}}\,{\widetilde{U}}_s^\top\\
    &\succeq \lambda\,\alpha_n\, {\widetilde{U}}_s\,\widetilde{\Lambda}_s\, {\widetilde{U}}_s^\top 
     \succeq \lambda\,\widetilde{\mu}_{\min}\,\alpha_n\, {\widetilde{U}}_s \widetilde{U}_s^\top,
\end{align*}
due to the fact that $\Omega_0\succeq \lambda\,I_K$ by Assumption {\bf A2}.

By Theorems \ref{lem:min_max_mu_k} and \ref{lem:tmin_tmax}, obtain that with probability at least $1-O(m^{-\tau})$,
 \begin{equation*}
     \widetilde{\mu}_{\min}\geq \min_k\frac{1}{|\mathcal{G}_k|} \left(\sum\limits_{i\in\mathcal{G}_k}\theta_i\right)^2 
     \geq\frac{\Gamma_{\min}^2}{\mu_{\max}}\geq \frac{m\,(1-a)^2}{C_0^3\, (1+a)\, K},
 \end{equation*}
 which implies
\begin{equation}
     \lambda_K(P_{(\MM,\MM)})  \,\geq\, \dfrac{\lambda(1-a)^2\,m\alpha_n}{C_0^3\, (1+a)\, K}.
    \label{zetam_dcbm}
\end{equation}

Let $W$ be the $(K\times K)$ orthogonal matrix for which $\umodnot{\widehat{V}-VW}$ is minimized, and let $\widehat{V}_r$ and $V_r$ be the regularized versions of $\widehat{V}$ and $VW$ respectively, that is, 
$$
(\widehat{V}_r)_{i,.}=\frac{\widehat{V}_{i,.}}{\umodnot{\widehat{V}_{i,.}}}, \quad  ({V}_r)_{i,.}=\frac{(VW)_{i,.}}{\umodnot{(VW)_{i,.}}},\ 1\leq i\leq m.
$$
Note that if $P_{(\MM,\MM)}$ is of rank $K$, we can write $V=\widetilde{U}_s\,Q$ with  a $(K\times K)$ orthogonal matrix $Q$, since $P_{(\MM,\MM)}= \alpha_n\, {\widetilde{U}}_s\,(\widetilde{\Lambda}_s^{\frac{1}{2}}\, \Omega_0\, \widetilde{\Lambda}_s^{\frac{1}{2}})\,{\widetilde{U}}_s^\top$, and 
$(\widetilde{\Lambda}_s^{\frac{1}{2}}\, \Omega_0\, \widetilde{\Lambda}_s^{\frac{1}{2}})$ is a $(K\times K)$ matrix of full rank. 
Choosing $V=\widetilde{U}_sQ$ yields that $V_r=M_{(\MM,.)}Q\,W$, so that, for nodes within the same community, the corresponding rows of matrix $V_r$ will be the same.

Now, by Lemma 1 of \citet{pensky2024davis}, the number  $m\Delta_{\MM}$ of misclustered nodes in step 2,  obtained from estimating 
the row clusters of $V_r$ by an $(1+\beta)$-approximate solution to K-means problem with input $\widehat{V}_r$, 
is bounded above by 
\begin{equation}
    m\Delta_{\MM} \,\leq\, \#\{i: \umodnot{(\widehat{V}_r)_{i,.} - (V_r)_{i,.}} > \gamma/2 - \delta\},
\end{equation}
provided there exists $\delta\in(0,\gamma/2)$ such that
\begin{equation}
    \lVert\widehat{V}_r-V_r\rVert_F\leq \dfrac{\delta\,\sqrt{\mu_{min}}}{\sqrt{K}(1+\sqrt{1+\beta})}.
    \label{delta_cond_dcbm}
\end{equation}
Here $\gamma$ is the minimum pairwise Euclidean norm separation between the $K$ distinct rows of $V_r$. 
Observe that 
\begin{align*}
    \umodnot{\widehat{V}_r - V_r}_{2,\infty} =&\, \max_{1\leq i\leq m}\umod{\frac{\widehat{V}_{i,.}}{\umodnot{\widehat{V}_{i,.}}} -
    \frac{(VW)_{i,.}}{\umodnot{(VW)_{i,.}}}} \\
    \leq&\, \max_{1\leq i\leq m}\left(\frac{\umodnot{\widehat{V}_{i,.}-(VW)_{i,.}}}{\umodnot{\widehat{V}_{i,.}}}
    + \umodnot{(VW)_{i,.}}\umod{\frac{1}{\umodnot{\widehat{V}_{i,.}}} -
    \frac{1}{\umodnot{(VW)_{i,.}}}}\right) \\
    =&\, \max_{1\leq i\leq m}\left(\frac{\umodnot{\widehat{V}_{i,.}-(VW)_{i,.}}}{\umodnot{\widehat{V}_{i,.}}}
    + \frac{|\umodnot{\widehat{V}_{i,.}} - \umodnot{(VW)_{i,.}}|}{\umodnot{\widehat{V}_{i,.}}}\right) \\
    \leq&\, 2 \max_{1\leq i\leq m}\frac{\umodnot{\widehat{V}_{i,.}-(VW)_{i,.}}}{\umodnot{\widehat{V}_{i,.}}}\,
    \leq\, 2 \max_{1\leq i\leq m}\frac{\umodnot{\widehat{V}_{i,.}-(VW)_{i,.}}}{\umodnot{(VW)_{i,.}} - \umodnot{\widehat{V}_{i,.}-(VW)_{i,.}}}
\end{align*}
Defining $e_i$ as the $m$-dimensional unit vector with the $i$-th element equal to 1, derive
\begin{equation}
    \umodnot{\widehat{V}_r - V_r}_{F} 
    \leq \sqrt{m}\,\umodnot{\widehat{V}_r - V_r}_{2,\infty}\leq 2\sqrt{m}\,\max_{1\leq i\leq m}\frac{\umodnot{e_i^\top(\widehat{V}-VW)}}{\umodnot{V_{i,.}} - \umodnot{e_i^\top(\widehat{V}-VW)}}.
    \label{eigengap_dcbm}
\end{equation}

Note that $(\widehat{V}-VW)$ can be expanded as
\begin{equation*}
    \widehat{V}-VW\,=\, (I-VV^{\top})(A_{(\MM,\MM)}-P_{(\MM,\MM)})\widehat{V}\widehat{\Psi}^{-1} + V(V^{\top} \widehat{V} - W).
\end{equation*}
Therefore, 
\begin{align*}
    &\,\umodnot{e_i^\top(\widehat{V}-VW)} \\
    \leq&\, \umodnot{e_i^\top(A_{(\MM,\MM)}-P_{(\MM,\MM)})\widehat{V}\widehat{\Psi}^{-1}}
    + \umodnot{e_i^\top VV^{\top}(A_{(\MM,\MM)}-P_{(\MM,\MM)})\widehat{V}\widehat{\Psi}^{-1}} + \umodnot{e_i^\top V(V^{\top} \widehat{V} - W)}\\
    \leq&\, \umodnot{e_i^\top(A_{(\MM,\MM)}-P_{(\MM,\MM)})\widehat{V}}\umodnot{\widehat{\Psi}^{-1}}
    + \umodnot{e_i^\top V}\umodnot{A_{(\MM,\MM)}-P_{(\MM,\MM)}}\umodnot{\widehat{\Psi}^{-1}} + \umodnot{e_i^\top V}\umodnot{V^{\top} \widehat{V} - W}\\
    \leq&\, \dfrac{\umodnot{e_i^\top(A_{(\MM,\MM)}-P_{(\MM,\MM)})\widehat{V}}}{\lambda_K(P_{(\MM,\MM)}) - \umodnot{A_{(\MM,\MM)}-P_{(\MM,\MM)}}}
    + \umodnot{V_{i,.}}\dfrac{\umodnot{A_{(\MM,\MM)}-P_{(\MM,\MM)}}}{\lambda_K(P_{(\MM,\MM)}) - \umodnot{A_{(\MM,\MM)}-P_{(\MM,\MM)}}} + \umodnot{V_{i,.}}\umodnot{V^{\top} \widehat{V} - W}.
\end{align*}
In the final step above, we used the fact that $\umodnot{\widehat{\Psi}^{-1}} = 1/\lambda_K(A_{(\MM,\MM)})$ and $|\lambda_K(A_{(\MM,\MM)})-\lambda_K(P_{(\MM,\MM)})|\leq \umodnot{A_{(\MM,\MM)}-P_{(\MM,\MM)}}$ by Weyl's inequality.

Consider the following events:
\begin{equation}
\label{high.prob.event_dcbm}
\begin{split}
    \mathscr{E}_1 &=\, \{\mu_{\min} \geq (C_0 K)^{-1}\, (1- a)m,\quad \mu_{\max} \leq K^{-1}\, (1+ a)m C_0\}\\
    \mathscr{E}_2 &=\,\{\Gamma_{\min} \geq (C_0 K)^{-1}\, (1- a)m,\quad \Gamma_{\max} \leq K^{-1}\, (1+ a)m C_0\}\\
    \mathscr{E}_3 &=\,\left\{\umodnot{A_{(\MM,\MM)}-P_{(\MM,\MM)}}\,\leq\, C_{\tau}\sqrt{m\alpha_n}
    \right\}\\
    \mathscr{E}_4 &=\, \left\{\umodnot{e_i^\top(A_{(\MM,\MM)}-P_{(\MM,\MM)})\widehat{V}}\leq C_{\tau}\,\sqrt{\theta_i\,m \alpha_n\log m}\, \umodnot{V}_{2,\infty}\right\}.    
\end{split}
\end{equation}

The event $\mathscr{E}_1 \cap \mathscr{E}_2$ implies that \eqref{zetam_dcbm} holds, that is $\lambda_K(P_{(\MM,\MM)})K/m\alpha_n$ is bounded away from zero. 
From Lemma C.3 of \cite{agterberg2024jointspectralclusteringmultilayer},
\begin{equation}
    \umodnot{V^{\top} \widehat{V} - W} \leq C\,\dfrac{\umodnot{A_{(\MM,\MM)}-P_{(\MM,\MM)}}^2}{\lambda_K(P_{(\MM,\MM)})^2},
\end{equation}
provided that $\sqrt{m\alpha_n\log n}/\lambda_K(P_{(\MM,\MM)})$ and $\log n/\theta_{\min}\,m\alpha_n$ are bounded above by a constant. 
Both of these conditions hold due to Assumption {\bf A4} and since $\lambda_K(P_{(\MM,\MM)})K/m\alpha_n$ is bounded away from zero.
Since $V=\widetilde{U}_s\,Q$,  one has
\begin{equation*}
    \umod{V}_{2,\infty}\,=\, \umodnot{\widetilde{U}_s}_{2,\infty} \,\leq\, \dfrac{1}{\sqrt{\widetilde{\mu}_{\min}}}.
\end{equation*}
So, $\mathscr{E}_1$ and $\mathscr{E}_2$ together also imply that
\begin{equation}
    \umod{V}_{2,\infty}\,\leq\, \sqrt{\frac{C_0^3\,(1+ a)K}{m\,(1- a)^2}}.
    \label{twotoinf_dcbm2}
\end{equation}
Therefore, the event  $\mathscr{E} = \mathscr{E}_1 \cap \mathscr{E}_2 \cap \mathscr{E}_3 \cap \mathscr{E}_4$ implies that
\begin{equation}
    \begin{split}
        \umodnot{e_i^\top(\widehat{V}-VW)}\leq&\, \dfrac{C\,(1 + \sqrt{\theta_i\,m \alpha_n\log m}) \,\umodnot{V}_{2,\infty}}{m\alpha_n/K - C\sqrt{m\alpha_n}}\\
    +&\, \umodnot{V_{i,.}}\dfrac{C\sqrt{m\alpha_n}}{m\alpha_n/K - C\sqrt{m\alpha_n}} + \umodnot{V_{i,.}}\,C\, \dfrac{m\alpha_n}{(m\alpha_n/K)^2}\\[.2cm]
    \leq&\, \dfrac{C\,K^{3/2}\,(1 + \sqrt{\theta_i\,m \alpha_n\log m})}{\sqrt{m}\,m\alpha_n} +\umodnot{V_{i,.}}\,C\,\dfrac{K}{\sqrt{m\alpha_n}},
    \end{split}
    \label{eigengap_dcbm_pt1}
\end{equation}
since  $\sqrt{m\alpha_n}/K \to \infty$   under Assumption {\bf A4}.

Since  $V=\widetilde{U}_sQ$, we have $\umodnot{V_{i,.}}\geq\dfrac{\theta_i}{\sqrt{\widetilde{\mu}_{\max}}}$. Noting that $\widetilde{\mu}_{\max}\leq \Gamma_{\max}$ since $\theta_j\leq 1$ for all $j$, the event $\mathscr{E}_2$ implies that
\begin{equation}
    \umodnot{V_{i,.}}\,\geq\, C\,\theta_i\sqrt{\dfrac{K}{m}}.
    \label{eigengap_dcbm_pt2}
\end{equation}
Plugging \eqref{eigengap_dcbm_pt1} and \eqref{eigengap_dcbm_pt2} into \eqref{eigengap_dcbm}, we obtain
\begin{align*}
    \umodnot{\widehat{V}_r - V_r}_{F} 
 \leq&\, 2\sqrt{m}\,\max_{1\leq i\leq m} \dfrac{\dfrac{C\,K^{3/2}\,(1 + \sqrt{\theta_i\,m \alpha_n\log m})}{\sqrt{m}\,m\alpha_n} +\umodnot{V_{i,.}}\,C\,\dfrac{K}{\sqrt{m\alpha_n}}}{\umodnot{V_{i,.}}-\dfrac{C\,K^{3/2}\,(1 + \sqrt{\theta_i\,m \alpha_n\log m})}{\sqrt{m}\,m\alpha_n} -\umodnot{V_{i,.}}\,C\,\dfrac{K}{\sqrt{m\alpha_n}}}\\[.2cm]
 \leq&\, C\sqrt{m}\,\max_{1\leq i\leq m}\left(\dfrac{1}{\theta_i}\sqrt{\dfrac{m}{K}}\dfrac{K^{3/2}\,(1 + \sqrt{\theta_i\,m \alpha_n\log m})}{\sqrt{m}\,m\alpha_n} + C\,\dfrac{K}{\sqrt{m\alpha_n}}\right)\\
 =&\, C\sqrt{m}\,\max_{1\leq i\leq m}\left(\dfrac{K}{\theta_i\,m\alpha_n} + \dfrac{K}{\sqrt{\theta_i}}\sqrt{\dfrac{\log m}{m\alpha_n}} + C\,\dfrac{K}{\sqrt{m\alpha_n}}\right)\\
 \leq&\,C\,\dfrac{\sqrt{m}}{\sqrt{c_0}\,K},
\end{align*}
due to  Assumption {\bf A4}.

Note that under the event $\mathscr{E}_1$, \eqref{delta_cond_dcbm} holds if 
$$
\lVert\widehat{V}_r-V_r\rVert_F\leq \delta\,\sqrt{(1- a)\,m}/(\sqrt{C_0} K(1+\sqrt{1+\beta})).
$$
Also, Since $V_r = M_{(\MM,.)}QW$, we have $\gamma\geq \sqrt{2}$.
Therefore, if the constant $c_0$ in Assumption {\bf A4} is large enough, we can choose $\delta=\gamma/4$ so that \eqref{delta_cond_dcbm} holds. 
Hence, the event  $\mathscr{E}$, which constitutes the intersection of all events in  in \eqref{high.prob.event}, implies that
\begin{equation}
    m\Delta_{\MM} \,\leq\, \#\{i: \umodnot{(\widehat{V}_r)_{i,.} - (V_r)_{i,.}} > \gamma/4\}.
\end{equation}
Now, since $\umodnot{\widehat{V}_r - V_r}_{2,\infty} \leq C/\sqrt{c_0}\,K$ and $\gamma\geq \sqrt{2}$, there will not be any $i$ such that $\umodnot{\widehat{V}_r - V_r}_{2,\infty}>\gamma/4$, if the constant $c_0$ in Assumption {\bf A4(b)} is large enough. Then, we have perfect clustering, that is, $m\Delta_{\MM}=0$.

Finally, it remains to show that the event  $\mathscr{E}$ occurs with high probability.
It follows from Theorems \ref{lem:min_max_mu_k} and \ref{lem:tmin_tmax} that the events $\mathscr{E}_1$ and $\mathscr{E}_2$ occur with probability at least $1-O(m^{-\tau})$.
By Theorem 5.2 of \cite{lei2015consistency}, the event $\mathscr{E}_3$ holds with probability at least $1-O(m^{-\tau})$ under Assumption {\bf A4}.
To obtain the lower probability bound for the event $\mathscr{E}_4$, first apply Lemma C.5 of \cite{agterberg2024jointspectralclusteringmultilayer} , which states that  with probability at least $1-O(m^{-{\tau}})$ one has
\begin{equation*}
    \umodnot{e_i^\top(A_{(\MM,\MM)}-P_{(\MM,\MM)})\widehat{V}}\leq C\,\sqrt{\theta_i\,m \alpha_n\log m}\, \umodnot{\widehat{V}}_{2,\infty}.
\end{equation*}
Next, note that $\umodnot{\widehat{V}}_{2,\infty}\leq \umodnot{\widehat{V}-VW}_{2,\infty} + \umodnot{V}_{2,\infty}$, and from Lemma C.6 of \cite{agterberg2024jointspectralclusteringmultilayer}, $\umodnot{\widehat{V}-VW}_{2,\infty}\leq C\sqrt{m\alpha_n\log n}/\lambda_K(P_{(\MM,\MM)})\,\umodnot{V}_{2,\infty}$ with probability at least $1-O(m^{-{\tau}})$.
Finally, the quantity $\sqrt{m\alpha_n\log n}/\lambda_K(P_{(\MM,\MM)})$ is bounded above by a constant, as argued earlier. 
Thus, the event $\mathscr{E}_4$ also holds with probability at least $1-O(m^{-{\tau}})$.
This concludes the proof.

\newpage
\section{Schematic for predictive assignment under the SBM using BASC for subgraph community detection}

\begin{figure}[h!]
	\centering
\includegraphics[width=\linewidth]{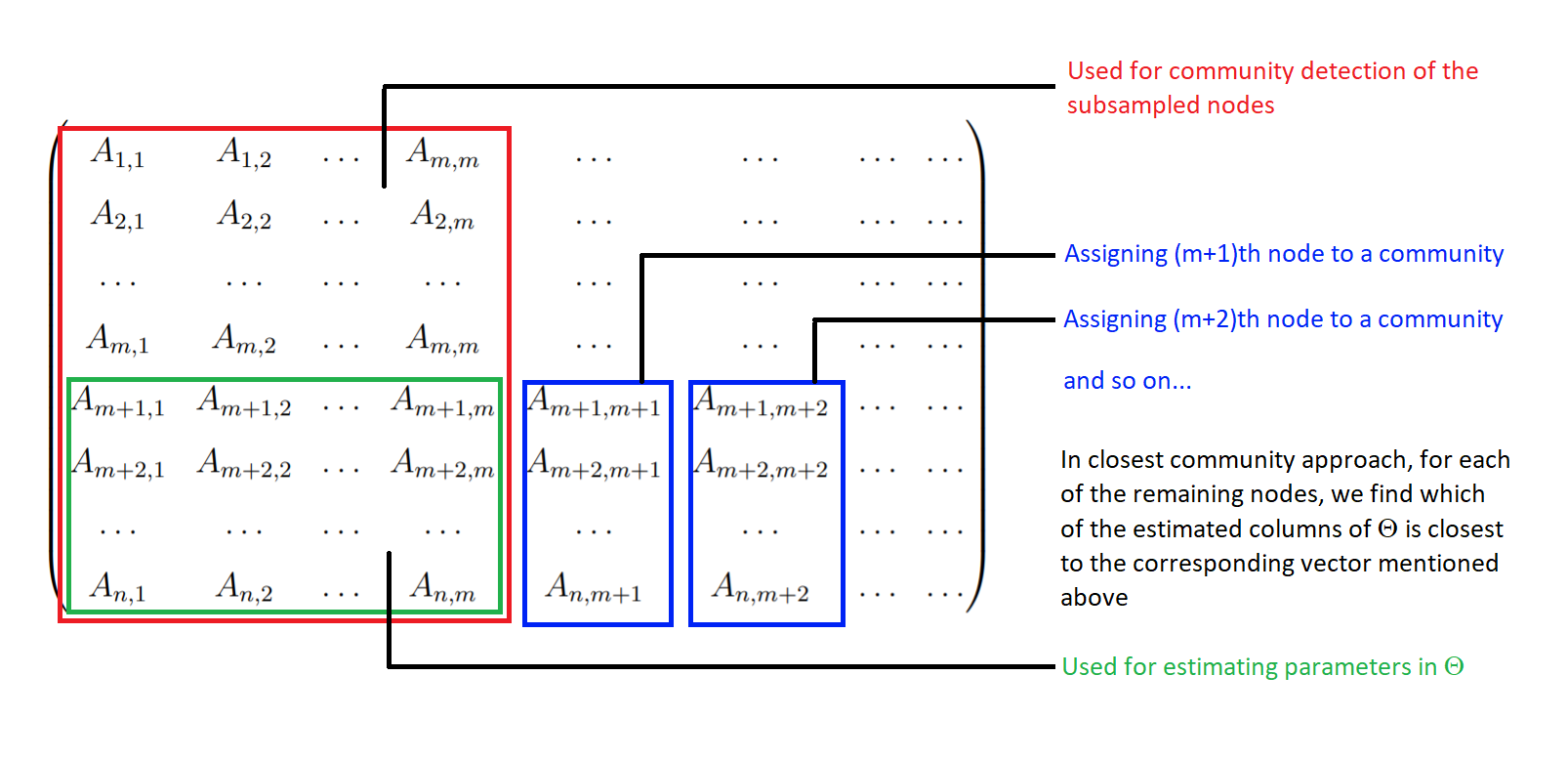}

\captionof{figure}{Use of the different sections of the adjacency matrix using BASC under SBM for community detection in Step 2.
Here we have assumed, for the sake of simplicity, that $\MM = \{1, \ldots, m\}$.
The submatrices
 $A_{(.,\MM)}$ (red border) and $A_{(\MM^c,\MM)}$ (green border) are utilized for subgraph community detection and estimation of $\Theta$, respectively.
The blue-bordered vectors represent $a_j$  and are used to assign nodes to communities one by one in Step 3.
\label{fig:basc}}
\end{figure}


\end{document}